\documentclass[aps,prd, onecolumn,superscriptaddress,nofootinbib,10pt,floatfix]{revtex4-2}

\usepackage[english]{babel}
\usepackage{graphicx}
\usepackage[dvipsnames,svgnames]{xcolor}
\usepackage{hyperref}
\hypersetup{colorlinks=true, linkcolor=Blue, citecolor=Blue,urlcolor=Blue}
\usepackage{physics}
\usepackage{amsmath}
\usepackage{amsfonts}
\usepackage{amssymb}
\usepackage{slashed}
\usepackage{dsfont}
\definecolor{verde}{HTML}{006400}

\usepackage{enumitem}
\usepackage{float}
\usepackage[normalem]{ulem}
\newcommand{\bs}{\boldsymbol}

\newcommand{\beq}{\begin{equation}}
\newcommand{\eeq}{\end{equation}}
\newcommand{\barr}{\begin{eqnarray}}
\newcommand{\earr}{\end{eqnarray}}
\newcommand{\mbf}{\mathbf}

\newcommand{\mbb}{\mathbb}

\begin{document}

\title{  Cherenkov radiation in isotropic chiral matter: unlocking threshold-free emission  }

\author{R. Martínez von Dossow}
\email{ricardo.martinez@correo.nucleares.unam.mx}
\affiliation{Instituto de Ciencias Nucleares, Universidad Nacional Aut\'{o}noma de M\'{e}xico, 04510 Ciudad de M\'{e}xico, M\'{e}xico}

\author{Eduardo Barredo-Alamilla}
\email{eduardo.barredo@metalab.ifmo.ru}
\affiliation{School of Physics and Engineering, ITMO University, 197101 Saint  Petersburg, Russia}

\author{Maxim A. Gorlach}
\email{m.gorlach@metalab.ifmo.ru}
\affiliation{School of Physics and Engineering, ITMO University, 197101 Saint Petersburg, Russia}

\author{L. F. Urrutia}
\email{urrutia@nucleares.unam.mx}
\affiliation{Instituto de Ciencias Nucleares, Universidad Nacional Aut\'{o}noma de M\'{e}xico, 04510 Ciudad de M\'{e}xico, M\'{e}xico}

\begin{abstract}

We investigate Cherenkov radiation in isotropic chiral matter using Carroll-Field-Jackiw electrodynamics, with an axion angle linear in time, to describe a charge moving at constant velocity. By solving the modified Maxwell's equations in cylindrical coordinates and in the space-frequency domain, we derive closed expressions for the circularly polarized electromagnetic fields contributing  independently to the radiation. The dispersion relations are obtained by imposing causality at a cylindrical surface at infinity, ensuring outgoing waves. Contrary to initial suppositions, each spectral energy distribution is gauge-invariant and positive, describing radiation at a characteristic angle. We characterize the angles and identify frequency ranges that allow for zero, one, or two Cherenkov cones. Notably, one sector of the model enables threshold-free Cherenkov radiation from slowly moving charges. Our results agree with partial findings in the nonrelativistic limit of earlier iterative analysis and clarify the regimes in which Cherenkov radiation arises in isotropic chiral matter.
\end{abstract}
\maketitle

 \section{ Introduction:}  
 \label{INTRO}
  Since its discovery \cite{Cherenkov:1934ilx, 
 vavilov1934cr} 
 and subsequent understanding in terms of Maxwell's equations \cite{Frank:1937fk}, Cherenkov radiation (CHR) has been instrumental in physics, with applications spanning  Cherenkov
 detectors \cite{Ypsilantis:1993cp, E598:1974sol, TibetASgamma:2021tpz}, light sources \cite{adamo2009light, liu2012surface,liu2017integrated}, and more recently medical imaging \cite{hachadorian2020imaging, alexander2021color, shaffer2017utilizing}, 
and  photodynamic therapy \cite{wang2022cherenkov, kotagiri2015breaking, kamkaew2016cerenkov}. 
Research in this area remains vibrant, fitting within the broader theme of radiation engineered via structured environments. Key settings for current exploration include two-dimensional materials, metamaterials, photonic crystals, and external fields.
For a detailed review, see, for example Refs. 
\cite{zrelov1970cherenkov,hu2021free}. 
The Cherenkov threshold imposes a fundamental restriction, requiring the charge velocity to exceed the light velocity in the medium. This typically necessitates high-energy particles, often only available in high-energy accelerators. In medical applications, the energy range is typically on the order of MeV, whereas GeV-range energies are required for charged particle identification in accelerators. To overcome this limitation, a significant research effort has focused on generating Cherenkov radiation with low-speed charges, effectively creating threshold-free Cherenkov radiation \cite{liu2017integrated, zhang2022tunable, hu2020nonlocality,  gong2023interfacial}. 

A further challenge in harnessing Cherenkov radiation is the collinear propagation of the radiation and the particles that produce it, making it difficult to isolate the Cherenkov cone. This limitation has spurred research into reversed Cherenkov radiation (RCHR)  \cite{skryabin2017backward, genevet2015controlled, galyamin2009reversed},  a concept theoretically proposed by Veselago  \cite{veselago1967electrodynamics} using materials with negative refractive index, also known as left-handed materials. 
 The experimental realization of such metamaterials \cite{shelby2001experimental} has enabled several observations of reversed Cherenkov radiation  \cite{xi2009experimental,
 zhang2009flipping,lu2019generation,duan2017observation}. 
 Challenging earlier assumptions that negative refractive index materials were required, recent studies have shown that RCHR can also occur in natural materials with positive refractive indices. 
 A key step was taken in  Ref. \cite{Franca:2019twk}, which demonstrated the existence of RCHR when a particle is incident perpendicularly to the interface between vacuum and a topological insulator (a magnetoelectric medium described by axion electrodynamics). This phenomenon can be classified as an interfacial process, following the terminology of Ref.  \cite{gong2023interfacial}.    Additionally, Ref. \cite{chen2025gain} showed that RCHR can emerge in a positive-index isotropic slab with optical gain.
  
Carroll-Field-Jackiw (CFJ) electrodynamics \cite{Carroll:1989vb},
a specific case of    axion electrodynamics \cite{Sikivie:1983ip,Wilczek:1987mv}, has been extensively explored at least  in  two main contexts: high-energy physics models with Lorentz symmetry breaking  \cite{Adam:2001ma,Kostelecky:2002ue,Lehnert:2004hq, Lehnert:2004be,Kaufhold:2005vj,
Colladay:2016rmy, Schreck:2017isa, Lisboa-Santos:2023pwc, OConnor:2023izw},  and condensed matter physics models dealing with the electromagnetic response of magnetoelectric media \cite{Franca:2019twk, Silva:2020dli, 
Silva:2021fzh, Franca:2021svc,Franca:2021irg,
 Barredo-Alamilla:2023xdt,Silva:2023ffk,Franca:2024fav}. 
 
The axionic contribution is part of the photon sector of the Standard–Model Extension (SME) \cite{PhysRevD.55.6760, PhysRevD.58.116002}, where it represents a minimal Lorentz- and CPT-violating modification of electrodynamics. It alters photon dispersion and allows forbidden processes such as vacuum Cherenkov radiation and other exotic scattering phenomena \cite{Kostelecky:2002ue,Lehnert:2004hq,Lehnert:2004be,Charneski:2012py}.  This contribution  can also be radiatively induced from the fermion sector of the SME \cite{Jackiw:1999yp,MartinezvonDossow:2025mxv} and  establishes a direct link between fundamental field-theoretic extensions of the Standard Model and the effective Hamiltonians that capture the electromagnetic response of topological quantum materials like Weyl semimetals \cite{Zyuzin2012Sep,Zyuzin2012Apr,Goswami:2012db}.

CFJ-electrodynamics provides the effective electromagnetic response of chiral matter and serves as a framework to describe diverse systems such as Weyl semimetals~\cite{Zyuzin2012Apr,Zyuzin2012Sep}, quark–gluon plasmas~\cite{Kharzeev2008Oct}, and dark matter~\cite{Donghan2024Sep}. Recently, it has been shown that such a response can be realized in realistic photonic systems whose magnetization rapidly varies in time 
\cite{2rrr-glyn}. The distinctive {optical} and radiative properties of such systems not only offer a playground to study the chiral 
anomaly~\cite{Huang:2018hgk}, but also allow exotic phenomena as anomalous reflectance exceeding unity at a finite frequency range for specific circularly polarized light~\cite{Tatsuya2023,
Yusuke2023,silva2025anomalous}, and mechanisms to generate terahertz (THz) circularly polarized light via Cherenkov radiation \cite{Hansen:2024kvc} and ultrafast photocurrents~\cite{gao2020chiral}, with potential applications in spectroscopy~\cite{Kaushik:2018tjj}.

We focus on the electrodynamics of isotropic chiral matter, a specific instance of CFJ electrodynamics, which extends standard electrodynamics by introducing a magnetoelectric coupling term in Ampère's law, proportional to $\sigma\mbb{B}$. We stress that the term ``chiral'' here does not refer to optical activity in reciprocal bi-isotropic (Pasteur) media described by the constitutive relations $\mbb{D} = \epsilon \mbb{E} + i \xi_\text{c} \mbb{B}$ and $\mbb{H} = \mu^{-1} \mbb{B} + i \xi_\text{c} \mbb{E}$ ~\cite{lindell1994electromagnetic,serdyukov2001electromagnetics}, 
which has been extensively studied both historically and in modern contexts~\cite{Fernandez-Corbaton2016,voronin2024chiral,Totful2024Nov,
Ribeiro:2025byd,Costa:2024bxw,
dyakov2025strong}. Instead, we focus on media exhibiting the chiral anomaly, where the macroscopic manifestation appears directly in Maxwell’s equations as a Lorentz-violating term that produces an effective magnetic current $\sigma\mbb{B}$, without requiring explicit frequency dispersion or spatial nonlocality, though such extensions are possible. For monochromatic plane waves, isotropic CFJ electrodynamics admits an equivalent description in terms of effective chiral response with the identification $\sigma = 2 \xi_\text{c}\omega$. Standard chiral media require $\xi_\text{c} \to 0$ as $\omega \to 0$, whereas the magnetic conductivity $\sigma$ can persist in the static limit. Important differences include: (i) bi-isotropic chiral media are stable, while isotropic chiral matter may admit unstable modes  \cite{qiu2017electrodynamics}; (ii) while reciprocal bi-isotropic chiral media produce polarization rotation upon transmission but not upon reflection~\cite{jaggard1990recent}, isotropic CFJ-type response can induce nonreciprocal effects such as cross-polarized transmission and reflection at time-interfaces~\cite{2rrr-glyn}.

We investigate radiation produced by a charge moving at constant velocity $v$ through isotropic chiral matter, a topic previously explored in Refs. \cite{Altschul:2014bba,Schober:2015rya, Altschul:2017xzx,DeCosta:2018nyf, 
Tuchin:2018sqe,Tuchin:2018mte,Tuchin:2020gtz, Hansen:2020irw,Hansen:2023wzp}. This problem is complicated by the lack of gauge invariance in both local energy density and local Poynting vector, unlike standard electrodynamics. Additionally, the local energy density is not positive definite, and plane wave solutions exhibit runaway modes with imaginary frequencies in the dispersion relations.
To address these challenges, we adopt a fresh perspective, recognizing that plane waves are not the propagating normal modes in our case. Instead, we focus on determining the actual normal modes to properly assess the problem. The unusual features of isotropic chiral electrodynamics raise questions about its observable consequences, particularly regarding radiation. Notably, there is ongoing debate about ``vacuum'' Cherenkov radiation, i.e., $n=1, \sigma\neq 0$, with some studies suggesting instabilities may prevent it due to cancellations between positive and negative energy modes \cite{Altschul:2014bba,Schober:2015rya}, while others report observing high-energy radiation in this regime \cite{Tuchin:2018sqe,Hansen:2020irw}.
Motivated by this controversy, but going one step further, we present an alternative derivation for Cherenkov radiation in isotropic chiral materials with $\epsilon \geq 1$. Our approach yields exact results for arbitrary $\sigma$ and $v$, providing a realistic description of CHR. To address the issue of imaginary frequencies, we employ a standard method in electrodynamics: solving Maxwell's equations in the space-frequency domain \cite{Jackson:1998nia, schwinger2019classical}. This involves performing a time Fourier transform on all fields and operators, using only real measurable frequencies. Our calculation closely follows the original method proposed in Ref. \cite{Frank:1937fk} for standard CHR.

\begin{figure}[h!]
  \centering
\includegraphics[scale=0.1]{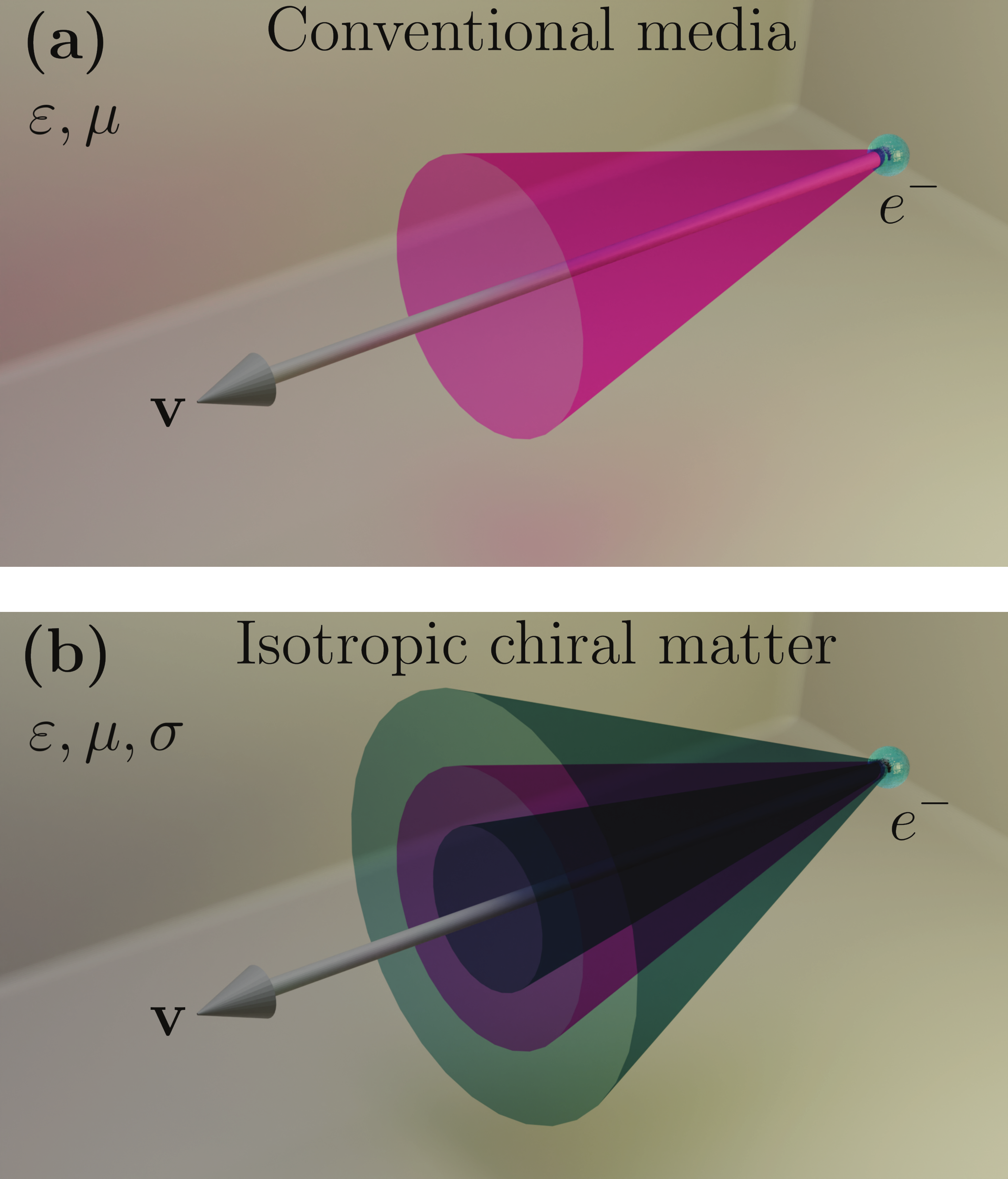} 
  \caption{Cherenkov radiation in conventional media and in isotropic chiral matter. (a) Standard Cherenkov radiation for $n=\sqrt{\mu\varepsilon}>1$ represented by the magenta cone. (b) Splitting of the Cherenkov radiation in isotropic chiral matter with the same material parameters $\varepsilon,\mu$ as in the conventional medium. The outer and inner cones have different circular polarization. The Standard Cherenkov cone (magenta) is overlaid for comparison.}
  \label{CONES}
\end{figure}

We observe novel phenomena which leads to either a shift or a splitting of the standard Cherenkov cone, where each cone corresponds to a distinct circular polarization (see Fig. \ref{CONES}). Also, we show that  threshold-free Cherenkov radiation is possible in isotropic chiral matter. Notably, this encompasses the phenomenon of the so-called ``vacuum'' Cherenkov radiation.

A clarification on notation is needed. In studies examining Lorentz symmetry breaking  in high energy physics, such as  Refs. \cite{Lehnert:2004hq,Lehnert:2004be}, for example, modifications to electrodynamics like CFJ electrodynamics are attributed to the effects of Lorentz invariance violation in the standard  vacuum. Although our work involves chiral matter, in the case $n=1$ and $\sigma\neq 0$ we adopt the term "vacuum" Cherenkov radiation for convenience, acknowledging this terminology might be somewhat imprecise for our specific case.

This work presents a comprehensive and detailed expansion of our recent letter \cite{MartnezvonDossow2025}, with the goal of definitively resolving the question of Cherenkov radiation in chiral matter.

The paper is organized as follows. Section \ref{MAXEQSSUB} presents Maxwell's equations for isotropic chiral matter in CFJ electrodynamics, along with the corresponding energy and momentum densities. Although these local quantities are gauge-dependent, we demonstrate that their integrated expressions are gauge-independent, particularly for the total radiated energy. This section makes contact with the Appendix \ref{BIREFMEDIA}, which briefly discusses the main differences between CFJ electrodynamics and bi-isotropic electrodynamics, a well-established area of research in electromagnetism.
Section \ref{EMFCHM} focuses on determining the dispersion relations, which yield two polarization modes and provide explicit solutions for the electromagnetic fields and potentials in cylindrical coordinates. The details are collected in Appendix \ref{SOLMAXWEQ}.
Section \ref{RADAPP} describes the radiation process, concentrating on the determination of the total energy radiated per unit frequency and per unit length flowing across a cylindrical surface at infinity. We determine the asymptotic behavior of the fields, demanding outgoing waves across this surface to preserve causality, which in turn determines the conditions for non-zero CHR. The phase velocity of each mode and the characteristic angle for each Cherenkov cone are also determined, and various possibilities for the output radiation are discussed based on the setup parameters. The calculations are detailed in Appendix \ref{Ap-En}.
The non-relativistic limit of our exact results is presented in Section \ref{NRLIMIT}. The primary goal is to validate these results by comparing them with those previously obtained in the literature using an iterative method that starts from the non-relativistic approximation for the fields of a particle moving at constant velocity. We also examine the $z$-symmetry, a fundamental tool that underpins the results obtained using this iterative method. A detailed expansion of the electromagnetic fields is included in the Appendix \ref{Ap-Exp}. Some Fourier transforms required in this section are collected in the Appendix \ref{APPENDIXF}. In Section \ref{APPLIC} we explore the key features of CHR emitted by an electron traversing a material with magneto-optic properties similar to a  Weyl semimetal, a prime example of chiral matter. The paper closes with a summary and conclusions in Section \ref{conclusion}. Our notation for the electromagnetic fields, vector potentials and currents is: for  vectors in the space-time domain 
we use 
blackboard bold letters: $\mathbb{V}=\mathbb{V}(\mathbf{x},t)$, with components $\mathbb{V}_i$ $(i=1,2,3)$. The Fourier  transforms in time, i.e. the vectors in the space-frequency domain are denoted  by a tilde over the blackboard bold symbol: $\tilde{\mathbb{V}}=\tilde{\mathbb{V}}(\mathbf{x}, \omega)$, with components $\tilde{\mathbb{V}}_i$. In cylindrical coordinates with axial symmetry we introduce the further splitting $\tilde{\mathbb{V}}_i(\mathbf{x}, \omega)=e^{i\frac{\omega}{v}z} V_i(\rho, \omega)$. With these conventions we can avoid writing explicitly the dependence either on $( \mathbf{x},t)$ or $( \mathbf{x}, \omega )$ in each case, unless confusion arises.

\section{Maxwell's equations in isotropic chiral matter}

\label{MAXEQSSUB}

 We consider isotropic chiral matter with permittivity $\epsilon$, permeability $\mu=1$ and magnetoelectric susceptibility $\sigma$. This model is directly connected with the low–energy description of Weyl semimetals (WSMs). In fact, the effective electrodynamics of WSMs is governed by axion electrodynamics (also known as Carroll–Field–Jackiw (CFJ) electrodynamics). Thus, isotropic chiral matter provides a simplified but physically motivated framework to capture the main features of the electromagnetic response of WSMs.
At low energies, WSMs are described by fermionic quasiparticles coupled to an axial background four–vector $b_\mu=(b_0,\mathbf{b})$, according to the Dirac-like action
\begin{equation}
	S_{\text{WSM}} = \int d^4x \, \bar{\psi}\left(i\gamma^\mu (\partial_\mu -{ie A_\mu}) - b_\mu  \gamma^5 \gamma^\mu \right)\psi,
	\label{WSM-action}
\end{equation}
where $b_\mu$ parametrizes the separation of the Weyl nodes, with $\mathbf{b}$ giving their separation in momentum space and $b_0$ their separation in energy \cite{Zyuzin2012Sep,Zyuzin2012Apr,Goswami:2012db}. This action corresponds to a particular case of the Standard–Model Extension (SME) \cite{PhysRevD.55.6760, PhysRevD.58.116002}, where $b_\mu$ is one of the coefficients that break Lorentz and CPT invariance.
Integrating out the fermions in (\ref{WSM-action}) induces the effective CFJ action
\begin{equation}
	S_{\text{CFJ}} = {S_0} + \frac{e^2}{32\pi^2}\int d^4x \, \theta(x)\, \epsilon^{\mu\nu\rho\sigma}F_{\mu\nu}F_{\rho\sigma}, 
	\qquad \theta(x)=x_\mu b^\mu = b_0 t - \mathbf{b}\cdot \mathbf{r}
	\label{eq:CFJ},
\end{equation}
{where $S_{0}$ is the standard Maxwell's action.
Here, $\theta(x)$ is the so-called axion angle which 
 encapsulates  the microscopic parameters of the WSM and governs anomalous transport phenomena such as the anomalous Hall effect and the chiral magnetic effect.
A more complete description of realistic WSMs requires including anisotropies and tilting of the cones. This can be achieved by considering additional SME coefficients, leading to the generalized action
\begin{align}
	S = \int d ^{4} x \, \bar{\Psi} \left( \Gamma ^{\mu} i \partial _{\mu} - M
	- e \Gamma ^{\mu} A _{\mu} \right) \Psi,  \label{ACTION1}
\end{align}
with
\begin{equation}
	\Gamma ^{\mu }={ \gamma^\mu}+c^{\mu }{}_{\nu }\gamma ^{\nu }+d^{\mu
	}{}_{\nu }\gamma ^{5} \gamma ^{\nu },\qquad 
	M=a_{\mu }\gamma ^{\mu }+b_{\mu}\gamma ^{5} \gamma ^{\mu }.
	\label{GAMMAMU}
\end{equation}
The set of SME coefficients $\{a_\mu, b_\mu, c^\mu{}_\nu, d^\mu{}_\nu\}$ can be mapped onto the macroscopic parameters characterizing Weyl semimetals, such as tilt and anisotropy of the Weyl cones, as discussed in Ref.~\cite{Gomez:2023jyl}.
 Integrating out the fermions in this more general case again produces a CFJ–type effective action, but now the coefficient is an effective four–vector $\mathcal{B}_\mu=({\mathcal B}_0, \bs{\mathcal{B}})$, which is a function of the SME parameters \cite{Gomez:2023jyl}. Thus, both the ideal case (only $b_\mu$) and the anisotropic/tilted models, with $b_\mu$  replaced by   $\mathcal{B}_\mu$, share the same CFJ structure, differing only in the effective parameter controlling the axion angle. 
 In the following, we focus on the isotropic situation where only the temporal component contributes. The axion angle then reduces to
\begin{equation}
\theta(x)=\sigma t, 
\qquad \sigma \equiv b_0 \quad \text{(ideal case)} 
\quad \text{or} \quad \sigma \equiv \mathcal{B}_0 \quad \text{(general case)}.
\label{THETACHM1}
\end{equation}
Both identifications lead to the same axion electrodynamics, so $\sigma$ can be regarded as the chiral  magnetic conductivity governing the response of the medium. 

From the  action (\ref{eq:CFJ}), {together with the choice in (\ref{THETACHM1})},  we find that the electromagnetic response is described by the following modified Maxwell's equations in Gaussian units
\begin{eqnarray}
		&&\boldsymbol{\nabla}\cdot\epsilon\mbb{E}=4\pi\bar{\rho}, \qquad \qquad \quad \boldsymbol{\nabla}\cdot \mbb{B}=0,  \label{MAXW1}
		\\
&&\boldsymbol{\nabla}\times\mbb{E}=-\frac{1}{c} \frac{\partial \mbb{B} }{\partial t}, 	\qquad 	\qquad \boldsymbol{\nabla}\times \mbb{B}=\frac{4\pi}{c}\mbb{J} + \frac{\epsilon}{c}\frac{\partial \mbb{E} }{\partial t} + \frac{\sigma}{c} \mbb{B},
\label{MAXW2}
\end{eqnarray}
where $c$ is the speed of light in vacuum and we take $\mu=1$. The speed of light in the medium is $c/n$, with refractive index $n=\sqrt{\epsilon}$. We  assume $\epsilon$ to be constant within a finite frequency range where this approximation holds, allowing us to isolate the effects of the magnetoelectric parameter $\sigma$, while noting that in real materials, permittivity and permeability typically exhibit frequency dependence. As a matter of notation, we call  isotropic chiral matter  those materials whose electromagnetic response is governed by Eqs. (\ref{MAXW1}) and (\ref{MAXW2}), i. e when the only additional parameter in Maxwell equations, besides  $\epsilon$ and $\mu$, is the magnetoelectric conductivity $\sigma$. 

It is interesting to remark that the inhomogeneous equations in (\ref{MAXW1}) and (\ref{MAXW2}) can be derived from the standard Maxwell equations in media, in terms of $\mbb{D}$ and $\mbb{H}$, by inserting the constitutive relations 
\beq
\mbb{D}=  \epsilon \mbb{E} + \theta(x) \mbb{B}, \qquad \mbb{H}=  \mbb{B} - \theta(x) \mbb{E}, \label{CONSTRELTHETA}
\eeq
with $\theta(x)$ given by  Eq. (\ref{THETACHM1}).

In Eqs. (\ref{MAXW1}) and (\ref{MAXW2}),
we take the source to be a charge $q$ moving along the $z$-axis with constant velocity $v$, with charge density $\bar{\rho}= q \, \delta(x)\delta(y)\delta(z-vt)$ and current density $\mbb{J}=\bar{\rho} \, v \, \mbf{\hat{k}}$. As usual, it is convenient to introduce 
the potentials $\mbb{A}$ and $\Phi$ such that $\mbb{B}= \boldsymbol{\nabla}\times \mbb{A}$ and $\mbb{E}=-\boldsymbol{\nabla} \Phi -\frac{1}{c}
\frac{\partial \mbb{A}}{\partial t}$.

Since we aim at discussing radiation in this medium, we write the corresponding local energy density ${U}$ and the local energy flux ${\mbb S}$
\beq
    {U} = \frac{1}{8\pi}(\epsilon \mbb{E}^2+\mbb{B}^2) -\frac{1}{8\pi} \frac{\sigma}{c}\mbb{A} \cdot \mbb{B}, \qquad \quad 
    {\mbb{S}}  = \frac{c}{4\pi} \mbb{E} \times \mbb{B} + \frac{\sigma}{8\pi}\left(\mbb{A} \times \mbb{E} - \Phi \mbb{B}\right)\label{CONS}, 
    \eeq
satisfying the conservation equation $\partial_t {U}+\boldsymbol{\nabla}\cdot \mbb{ S}=0$ outside the sources. 

Although the expressions (\ref{CONS}) have two drawbacks - a non-positive-definite local energy density and lack of gauge invariance - at this stage, we verify that the relevant physical quantities are indeed gauge-invariant. We will provide detailed expressions for these quantities after obtaining the explicit solutions for the electromagnetic fields and potentials.

A brief clarification on notation is in order. We use the term ``local'' to describe the energy density and energy flux, which are related to the local energy-momentum tensor $T^{\mu\nu}$   as $U=T^{00}$ and  $ S^i= T^{i0}$. These quantities can be derived via the Noether theorem, yielding a local conservation equation $\partial_\mu T^{\mu\nu}=0$ outside the sources. While local densities like the energy-momentum tensor lack physical meaning on their own, the associated conserved charges are physically relevant.  This is evident from the fact that the energy-momentum tensor can be redefined as  
${\tilde T}^{\mu\nu}
$  such that
\beq {
{\tilde T}^{\mu\nu}=T^{\mu\nu} + \partial_\rho S^{\mu \rho \nu},} 
\eeq
with $S^{\mu \rho \nu}=-S^{\rho \mu \nu}$
  being an arbitrary tensor, antisymmetric in its first two indices. This ensures that the modified tensor ${\tilde T}^{\mu\nu}$ is also conserved, with the same conserved charges as the original tensor $T^{\mu\nu}$. This property enables the construction of the symmetric energy-momentum tensor via the Belinfante method and allows for a manifestly gauge-invariant formulation of the canonical energy-momentum tensor. Notably, a direct calculation of the latter, via Noether's theorem, yields non-gauge-invariant results \cite{Jackson:1998nia}, but an appropriate choice of $S^{\mu \rho \nu}$  can produce the standard gauge-invariant expression. In other words, even if the local energy-momentum tensor is not gauge-invariant, we require the corresponding charge,  the total energy  
\beq
{ E = \int d^3 x \, T^{00}},
 \label{1}
\eeq
to be gauge-invariant. This can be directly verified in our case. Starting from Eq. (\ref{CONS}), we focus on the potentially problematic term and perform an infinitesimal gauge transformation $\delta_G  \mbb{A}=\nabla \delta \Lambda$, yielding
\beq 
\delta_G E= -\frac{\sigma}{8\pi c}\int d^3 x \, (\mbb{B}\cdot\nabla \delta \Lambda )= -\frac{\sigma}{8\pi c}\int d^3 x \, \delta \Lambda \, \nabla \cdot \, \mbb{B} =0,
\label{GT0}
\eeq 
after integrating by parts and using $\boldsymbol{\nabla} \cdot \mbb{B}=0$.

Considering now the radiation regime, we introduce the spectral distribution of the total radiated energy per unit length, ${\cal E}$, which originates from the total energy ${E}$ flowing across a closed surface ${\cal E}$ at infinity. This quantity is the relevant gauge-invariant observable. We have 
\beq
{ E}= \int_{-\infty}^{+\infty} dt \oint_{\cal S} d{ \cal S} \,  \hat{\boldsymbol{n}}\cdot 
{\mbb{S}}.
\label{TOTALRE}
\eeq
Let us  examine the conflicting term in the local energy flux $\mbb{S}$, given in Eq. (\ref{CONS}), which depends on the potentials. Applying the full gauge transformation $\delta_G \mbb{A}=\bs{\nabla} \delta \Lambda, \, \delta_G \Phi=-\frac{1}{c}\partial_t \delta \Lambda $ to its contribution  
 $\tilde E$ to the total radiated energy, we obtain
\beq
\delta_G \tilde{E}=\frac{\sigma}{8\pi} \int_{-\infty}^{+\infty} dt \oint_{\cal S} \Big((\boldsymbol{\nabla} \delta \Lambda) \times \mbb{E} + \frac{1}{c} (\partial_t \delta \Lambda)\, \mbb{B}\Big)\cdot \hat{\mbf{n}}\,  d {\cal{S}}
\label{GT1}.
\eeq 

Assuming $(\delta \Lambda  \mbb{B})$ vanishes at $t=\pm \infty $, we integrate  the second term on the right-hand side of Eq. (\ref{GT1}) by parts in time and apply Faraday's law along with Gauss's theorem, obtaining
\barr
&& \delta_G \tilde{E}=\frac{\sigma}{8\pi} \int_{-\infty}^{+\infty} dt \oint_{\cal S} \Big((\boldsymbol{\nabla} \delta \Lambda) \times \mbb{E} + \delta \Lambda \boldsymbol{\nabla}\times \mbb{E} \Big)\cdot \hat{\mbf{n}} d {\cal{S}} 
=\frac{\sigma}{8\pi} \int_{-\infty}^{+\infty} dt \int_V d^3 x \, \boldsymbol{\nabla} \cdot \mbb{V}, 
\label{GT22}
\earr 
where we denoted by $\mbb{V}$ the vector in round brackets in the second term of Eq. (\ref{GT22}). Using standard vector identities, it is a direct matter to show that $\boldsymbol{\nabla} \cdot \mbb{V}=0$, which proves  the gauge invariance of $\tilde E$. 

In our case, we choose to work in cylindrical coordinates $\rho,\, \phi,\, z$ and consider the surface ${\cal S}$ as an infinite cylinder with axis along the $z$-direction and radius $\rho \rightarrow \infty$. Starting from standard electrodynamics, i.e., $\sigma=0$, where we have $ E =  \rho \int_{0}^{+\infty} \frac{d\omega}{2\pi} \int_{-\infty}^{+\infty} dz\, \hat{\boldsymbol{\rho}}\cdot \, {\rm Re}( c \,  \tilde{\mbb{E}}^*\times \tilde{\mbb{B}})$, we need to include the additional contributions of the local energy flux in (\ref{CONS}). Let us notice that we have just changed from the $(\mbf{x},t)$ domain to the $(\mbf{x},\omega)$  domain, in cylinrical coordinates.
After doing this, we can read the spectral distribution of the total radiated energy per unit length 
\beq
{\cal E}= \frac{d^2 E}{d \omega dz},
\label{SED0}
\eeq
to be discussed in detail in Section \ref{RADAPP}. Summarizing, we have shown that the total energy flowing across a closed surface at infinity is gauge-invariant, despite the local energy flux $\mbb{S}$ being gauge-dependent.

A word of caution is in order to distinguish our case from the well-studied electrodynamics in bi-isotropic chiral media, as mentioned earlier in the Introduction. We elaborate on this point further in the Appendix \ref{BIREFMEDIA}.

\section{The electromagnetic fields in chiral matter }

\label{EMFCHM}

In addition to our choice of cylindrical coordinates $\rho, \phi, z $, we work in the space-frequency domain, where the sources are
\beq
{\bar \rho}(\rho, z, \omega) = \frac{q}{ v} \frac{\delta(\rho)}{2 \pi \rho} e^{i\frac{ \omega}{v}z}, \qquad    \tilde{\mbb{J}}_x =\tilde{\mbb{J}}_y=0, \qquad    
\tilde{\mbb{J}}_z=  v \bar{\rho},
\label{SOURCES0}
\eeq 
as appropriate to the axial symmetry of the system.
Before going on, let us clarify the origin of the $\delta(\rho)/2\pi \rho$ in (\ref{SOURCES0}).  The  source in the space-time domain 
is a charge $q$ moving at constant velocity $v$ along the $z$-axis, such that the charge density  is $\rho(\mbf{x},t)=q \delta(x)\delta(y) \delta(z-vt)$. This yields the time-Fourier transform ${\bar \rho}(\mbf{x},\omega)=\frac{q}{c} e^{i\frac{\omega}{v}z} \delta(x)\delta(y)$. Now, recalling that we have axial symmetry along the $z$-axis we have $dx dy= 2\pi \rho d \rho$, since all fields are independent of the angular variable $\phi$. This leads to the equivalence  $\delta^2(\mbf{x}_\perp)=\delta(x) \delta(y)= \frac{\delta(\rho)}{2\pi \rho}$, such that we obtain the result $1$ when integrating $\delta^2(\mbf{x}_\perp)$ over the whole $(x,y)$ space in cylindrical coordinates. Then we recover the first expression in (\ref{SOURCES0}).  The appearence of the factor $e^{i\frac{\omega}{v}z}$ in all the sources of Maxwell equations in the space-frequency space motivates to implement  the separation of variables
\beq
\tilde{\mbb V}_i(\mathbf{x}, \omega)=e^{ikz} V_i(x,y, \omega)= e^{ikz} V_i(\rho, \omega), \qquad k=\frac{\omega}{v}, 
\label{FACT_CONV}
\eeq
for all vector components. Incidentally, this yields $\partial_z= ik$ in our space.

Going back to the space-time domain we have
\beq
{\mbb V}_i(\mbf{x},t)=\int_{-\infty}^{+\infty} \frac{d \omega}{2\pi} e^{-i\omega t} \,  \tilde{\mbb V}_i(\mathbf{x}, \omega)= \int_{-\infty}^{+\infty} \frac{d \omega}{2\pi} e^{-i \frac{\omega}{v}(z-v t)} V_i(\rho, \omega),
\eeq
so that the resulting functions in the space-time domain depend upon $\rho$ and the combination $(z-vt)$.

Next we deal with Maxwell's equations (\ref{MAXW1}) and (\ref{MAXW2}) in this Fourier space,  where we  have $
\partial_t= -i\omega$ and  $\partial_z=+ik$.
In the following we take advantage of the factorization of the fields in the form 
highlighted in Eq. (\ref{FACT_CONV}), by displaying  only  the contribution ${V}_i(\rho, \omega) $  for each component, to be denoted by $V_a$, and implicitly assuming the dependence in  $(\rho,\omega)$, unless confusion arises. In cylindrical coordinates $i: \rho, \, \phi, \, z$. 
The corresponding solution of  Maxwell's equations  are discussed in detail in  the Appendix \ref{SOLMAXWEQ}. Here we present only the results

The dispersion relation yielding  the allowed magnitude of the momenta $Q$ in the $\rho$-direction in the radiation approximation  is
\begin{eqnarray}
Q^4-(\alpha^2+\gamma^2)Q^2+\alpha^2\gamma^2-\frac{n^2 \omega^2 \sigma^2}{c^4}&=&0,
\label{DISPREL0}
\end{eqnarray}
which  is a quartic equation with solutions
 \begin{equation}
	Q^2_\nu(\omega)=\frac{\omega^2}{v^2}-\frac{n^2 \omega^2}{c^2}-\frac{\sigma^2}{2c^2}+\nu \frac{\sigma}{2c^2} \sqrt{\sigma^2+4n^2 \omega^2}, \qquad \nu= \pm.
	\label{DISPREL1}
\end{equation}
The modes $\nu= + \, (\nu=-) $ describe  circular polarization in the plane perpendicular to the wave vector. The conventions are: 
$\nu=+$ correspond to negative  helicity waves (right circular polarization in the optics convention) and
$\nu=-$ correspond to  positive   helicity waves (left circular polarization in the optics convention). This is discussed in detail in the Appendix \ref{POL}.

The electromagnetic fields are calculated in detail in the Appendices \ref{EMFOUTS}  and \ref{BOUNDC}, yielding 
\begin{eqnarray}
E_\rho&=&\frac{\omega\Gamma}{c}\left( Q_+ \Omega_+ K_1(Q_+\rho)-Q_- \Omega_-  K_1(Q_-\rho)\right),\label{ERHO} \\
	E_\phi&=&-\frac{i\omega\Gamma}{c}\left(Q_+ K_1(Q_+ \rho) - Q_- K_1(Q_- \rho)\right), \label{EPHI}\\
	E_z&=&-\frac{iv\Gamma}{c}\left(\Omega_+ Q^2_+ K_0(Q_+ \rho)-\Omega_- Q^2_-  K_0(Q_- \rho)\right),\label{EZ}\\
	B_\rho&=&\frac{i\omega\Gamma}{v}\left(Q_+ K_1(Q_+ \rho)-Q_- K_1(Q_- \rho)\right) \label{BRHO} \\
	B_\phi&=&\frac{\Gamma}{v\omega}\left( Q_+ \Omega_+ \left( \omega^2  - v^2Q^2_+ \right)K_1(Q_+ \rho)- Q_- \Omega_- \left( \omega^2  - v^2Q^2_- \right)K_1(Q_- \rho)\right), \label{BPHI} \\
	B_z&=&\Gamma\left( Q^2_+ K_0(Q_+\rho)-  Q^2_- K_0(Q_-\rho)\right).
	\label{BZ}
\end{eqnarray}
 The notation is 
 \beq
\alpha^2 = k^2 - \frac{\epsilon \omega^2}{c^2} - \frac{\sigma^2}{c^2}, \quad  \gamma^2 = k^2 - \frac{\epsilon \omega^2}{c^2}, \quad  \Gamma={2q}/{\sqrt{\sigma^2+4n^2 \omega^2}}, \quad \Omega_\nu=
{k \sigma}/{(Q^2_\nu-\gamma^2)c}.
\label{CONV0}
 \eeq
The electromagnetic potentials that enter in the calculation of the spectral distribution of the total radiated energy per unit length ${\cal E}$ are also obtained  in the Appendix \ref{EMPOT}.  First they were calculated in the Lorenz gauge in Eqs. (\ref{AZ1}), but an additional gauge transformation allowed to obtain the simpler result $\mbf{A}=-\frac{ic}{\omega} \mbf{E}$,  with $\Phi=0$. The  explicit  final expressions are 
\begin{eqnarray}
A_\rho &=& -i\Gamma \sum_{\nu=\pm } \nu \, Q_\nu \Omega_\nu K_1(Q_\nu \rho), \qquad 
A_\phi = -\Gamma  \sum_{\nu=\pm } \nu \, Q_\nu K_1(Q_\nu \rho), \nonumber \\
A_z &=& - \frac{v\Gamma}{\omega} \sum_{\nu=\pm } \nu \, \Omega_\nu Q^2_\nu K_0(Q_\nu \rho). 
\label{TOTALA1}
\end{eqnarray}
The vector potential in Eq. (\ref{TOTALA1}) satisfies $\bs{\nabla}\cdot \mbf{A}=0$ when $\rho >0$, so that it is written in  the Coulomb gauge outside the sources.

\section{The radiation zone}
\label{RADAPP}

In this section we deal with the main subject of the paper: the characterization of the Cherenkov radiation. As shown in Section \ref{MAXEQSSUB} the relevant physical quantity is total energy $E$ flowing across a closed surface $\cal S$ at infinity,  given by Eq. (\ref{TOTALRE}), and 
which is gauge-invariant.    Taking   the surface ${\cal S}$ as  that of an infinite cylinder with axis along the $z$-direction and radius $\rho \rightarrow \infty$, we have 
\beq E =  \rho \int_{0}^{+\infty} \frac{d\omega}{2\pi} \int_{-\infty}^{+\infty} dz\, \hat{\boldsymbol{\rho}}\cdot \, {\rm Re}\left( c \,  \mbf{E}^*\times \mbf{B}+\frac{\sigma}{2}(\mbf{A}^*\times \mbf{E}+ \Phi \mbf{B}^*)\right).
\label{TOTALRADE0}
\eeq
From the above equation  we read  the spectral distribution of the total radiated energy per unit length, ${\cal E}= \frac{d^2 E}{d \omega dz}$, yielding 
\barr
&&{\cal E}=\lim_{\rho \to \infty}  \frac{\rho}{2\pi}  \mathrm{Re} \Big[ c E^*_{ \phi} B_z -  c E^*_zB_\phi 
+   \frac{\sigma}{2}(A_\phi E^*_z-A_z E^*_\phi -\Phi\, B^*_\rho) \Big], 
\label{FINALSED}
\earr
which is finally written in the space-frequency domain, as a function of the fields (\ref{ERHO})-(\ref{BZ}) and (\ref{TOTALA1})

Before delving into the details, we further verify gauge invariance by demonstrating that the value of $\cal{E}$ in terms of electromagnetic fields is identical in both the Coulomb gauge (CG) and the Lorenz gauge (LG). Specifically, in the Coulomb gauge ($\Phi=0, \, \mbf{A}=-\frac{ic}{\omega} \mbf{E}$), we have 
\barr
{\cal E}_{\rm CG}&=&\lim_{\rho \to \infty}  \frac{\rho}{2\pi}  \mathrm{Re} \Big[ c E^*_{ \phi} B_z -  c E^*_zB_\phi 
 -\frac{ic}{\omega} \frac{\sigma}{2}( E_\phi E^*_z- E_z E^*_\phi \Big], \\
 &=& \lim_{\rho \to \infty}  \frac{\rho}{2\pi}  \mathrm{Re} \Big[ c E^*_{ \phi} B_z -  c E^*_zB_\phi 
 -\frac{ic \sigma}{\omega} E_\phi E^*_z].
 \label{FINALCALE}
\earr
In the Lorenz gauge $\Phi\neq 0$,  and the relevant relations are $ A_\phi=-\frac{ic}{\omega} E_\phi$ and $A_z=-\frac{ic}{\omega} E_z+ \frac{c}{v} \Phi$ according to Eqs. (\ref{APHI1}), respectively. Adding the remaining piece proportional to $\Phi$  we obtain
\beq
{\cal E}_{\rm LG}= {\cal E}_{\rm CG}  
-\Phi \, \left(\frac{c}{v}  E_\phi^* + B_\rho^* \right).
\label{GT3}
\eeq
The first component of Faraday's law in Eq. (\ref{B}), which is evident in the solutions (\ref{EPHI}) and (\ref{BRHO}), yields $B_\rho=-\frac{c}{v} E_\phi$. This relation renders the term proportional to $\Phi$ equal to zero, confirming the equivalence .

To examine the conditions for non-zero radiation, we examine the fields in the asymptotic region $\rho \rightarrow \infty$. Specifically, we aim to determine the frequency range that permits radiation for a given set of parameters $\sigma, n , v$ and to identify the contribution of each  polarization mode  $\nu$. The fields in question are expressed in terms of modified Bessel functions $K_{0,1}(Q\rho)$, which exhibit the  asymptotic behavior
\beq
\lim_{\rho \to \infty}K_{0,1}(Q_\nu\rho)=
\sqrt{\frac{\pi}{2 Q_\nu \rho}}e^{-Q_\nu \rho},
\label{BESSELASYMT}
\eeq
with $Q_\nu^2$ given by Eq. (\ref{DISPREL1}).

We then enforce causality by imposing outgoing waves at the cylindrical surface as  $\rho \to \infty$, which requires choosing
$
Q_\nu=-i{\cal Q}_\nu,
$
 with ${\cal Q}_\nu $ real and positive. Going back to the dispersion relation (\ref{DISPREL1}) we factor a minus sign and take ${\cal Q}_\nu= \sqrt{q_\nu}$, with 
\begin{equation}
	{\cal Q}_\nu^2=q_\nu\equiv \Big(\frac{n^2 \omega^2}{c^2}+\frac{\sigma^2}{2c^2}-\nu \frac{\sigma}{2c^2} \sqrt{\sigma^2+4n^2 \omega^2}\Big) -\frac{\omega^2}{v^2} \geq 0.	
	\label{POSDEF1}
\end{equation}
The final relation $q_\nu \geq 0$ determines the conditions under which a particular polarization mode $\nu$ contributes to the radiation and constitutes one of the main results of the manuscript. At the risk of being repetitive we emphasize that the existence of radiation depends fundamentally on $q_\nu$ being positive. For example, if we start from ${\cal Q}_\nu^2$ in Eq.(\ref{POSDEF1}), set $\sigma=0$, and also  neglect the term $\frac{n^2\omega^2}{c^2}$ we will end up with $q_\nu$ being negative, thus forbidding radiation.

Before obtaining a simplified expression for  the radiation condition (\ref{POSDEF1}), we first assess the contribution to the spectral energy distribution of the two polarization modes,  $\nu=\pm 1 $.  Our goal is to investigate the potential interference between these modes by computing the quadratic products in the fields given by Eq. (\ref{FINALSED}). From the general structure of the spectral energy density 
${\cal E}$ we  expect the result $
{\cal E}={\cal E}_+ + {\cal E}_- + {\cal E}_{(+,-) }$, 
where each  ${\cal E}_{\pm}$ can be read  from the expression (\ref{FINALSED})  after  substituting the respective fields and potentials for each polarization $\nu= \pm $. The term   ${\cal E}_{(+,-) }$  represents the interference between the two polarization modes.
These contributions are calculated in the Appendix \ref{Ap-En}, using the asymptotic field expressions. Notably, the interference term vanishes exactly, indicating that the spectral energy distribution is simply presented  as the sum  of the two polarization modes. Since this is another important result of this paper we provide a detailed algebraic proof  of this cancellation in the Appendix.  Given our earlier proof of gauge invariance, which holds separately for each polarization, the cancellation of ${\cal E}_{(+,-)}$ implies that we can effectively consider each polarization's contribution independently, i. e. we can effectively write $\mbb{S}=\mbb{S}_+ + \mbb{S}_- $.  Specifically, the radiation from each polarization makes an independent physical contribution to the spectral energy distribution.  As shown in the following, each polarization is associated with a specific emission angle making them experimentally distinguishable by varying the observation angle, and  resulting in the appearance of one cone, two cones, or no cones in the spectrum. The calculation of ${\cal E}_\nu$ is straightforward but tedious (see Appendix \ref{Ap-En} ), 
yielding the following result 

\begin{equation}
	{\cal E}_\nu = \frac{ q^2 \omega}{2c^2}\left( 1-\frac{1}{n^2 \beta^2}-\nu\frac{\sigma}{\sqrt{\sigma^2+4n^2 \omega^2}}\left(1+\frac{1}{n^2 \beta^2}\right)\right) \equiv \frac{q^2 \, \omega}{c^2}\, \Omega,
	\label{ELAMBDA2}
\end{equation}
where $\beta=v/c$ is the ratio of the velocity of the moving  charge over the velocity of the light in vacuum. Note that the total spectral energy distribution ${\cal E} = {\cal E}_+ + {\cal E}_-$ becomes independent of the chiral parameter $\sigma$. However, this property does not extend to the Cherenkov spectrum, where the output of each cone in ${\cal E}_+$ or ${\cal E}_-$ must be considered separately, and each of these contributions is clearly dependent on $\sigma$.

As shown at the end of the Appendix \ref{Ap-En}, equation (\ref{ELAMBDA2}) can be presented in the alternative form
\begin{equation}
{\cal E}_\nu=\frac{ q^2 \omega}{2 c^2} \frac{{\cal F}_\nu}{\sqrt{1+\frac{ \Sigma^2}{4}}} 
\left[1-\frac{1}{ n^2 \beta^2 {{\cal F}_\nu}^2}\right],
\label{ELAMBDA}
\end{equation} 
which will prove useful in investigating the positivity properties of ${\cal E}_\nu$.  We introduce  the notation  $\Sigma = \frac{\sigma}{n \omega} > 0$ together with
\beq
    {\cal F}_\nu= \sqrt{1+\frac{\Sigma^2}{4}}-\nu\frac{\Sigma}{2}= 
    \frac{1}{2 n \omega}\Big(\sqrt{\sigma^2+ 4 n^2 \omega^2} -\nu \sigma\Big), 
    \label{FNU}
\eeq
    where  ${\cal F}_\nu$,  being of the form $\sqrt{1+ z^2} \pm z$ with $z$ real, is always positive.

Now we calculate the phase velocity and the Cherenkov angle at which radiation is emitted. In the radiation zone, the fields in each polarization mode are proportional to
$e^{i \mbf{K}_\nu \cdot \bs{r}} = e^{i (\mathcal{Q}_\nu \rho + k z)}$, so that the wave vector is
$
 {\mbf K}_\nu= {\cal Q}_\nu \, \hat{\bs{\rho}} + k \hat{\bs{z}},
$
 with 
 \barr
&& |{\mbf K}_\nu|=\sqrt{{\cal Q}^2_\nu+ k^2}
 = \frac{1}{2c}\Big(\sqrt{\sigma^2+4n^2 \omega^2}-\nu \sigma\Big)
  = \frac{n \omega}{c} {\cal F}_\nu,
  \label{MODK}
 \earr
where we use ${\cal Q}^2_\nu=q_\nu$ from Eq.~(\ref{POSDEF1}), together with ${\cal F}_\nu>0$. 
The phase velocity of each mode $\nu$ is given by
\beq
v_{\mathrm{ph},\nu}(\omega)\equiv \frac{\omega}{|{\mbf K}_\nu(\omega)|}
= \frac{c}{n\,{\cal F}_\nu(\omega)},
\label{VPHASE0}
\eeq
and, correspondingly, the angle of the emitted radiation with respect to the $z$-axis is obtained as
\beq
 \cos \Theta_\nu=\frac{k}{|{\mbf K}_\nu|}= \frac{1}{n \beta {\cal F}_\nu},
 \label{THETAC}
\eeq
where $k=\omega/v=\omega/(\beta c)$. Imposing the reality condition $|\cos\Theta_\nu|\le 1$ yields
$\beta \ge 1/(n {\cal F}_\nu(\omega))$, or equivalently $v \ge v_{\mathrm{ph},\nu}(\omega)$, which is the condition for the charge to radiate in polarization mode $\nu$. In other words, a real Cherenkov angle exists only when the speed of the moving charge exceeds the phase velocity of the mode, $v \ge v_{\mathrm{ph},\nu}(\omega)$. \\
In the low-frequency limit ($\omega\to 0$) one has $v_{\mathrm{ph},+}(\omega)\to\infty$ and $v_{\mathrm{ph},-}(\omega)\to 0$. Therefore, for any fixed particle speed $v>0$ there exists a sufficiently small frequency range in which $v \ge v_{\mathrm{ph},-}(\omega)$, realizing a threshold-free Cherenkov emission window for the $\nu=-$ mode.  Notably, in the limit $\sigma \to 0$, both angles coalesce into the standard Cherenkov angle $\Theta_{\rm Ch}$, satisfying $\cos \Theta_{\rm Ch}= 1/(\beta n)$. 

In the high-frequency limit ($\omega \to \infty$), one has $\mathcal{F}_\nu \to 1$ and $\Sigma \to 0$. 
In this case, the phase velocity and the Cherenkov angle reduce to their standard forms for each mode,
\begin{equation}
v_{\mathrm{ph},\nu}(\omega) \to \frac{c}{n}, 
\qquad 
\cos\Theta_{\nu} \to \frac{1}{n\beta},
\end{equation}
and the spectral energy density tends to one half of the usual Cherenkov expression, 
\begin{equation}
\lim_{\omega\to\infty} \mathcal{E}_{\nu} = 
\frac{1}{2} 
\frac{q^{2}\omega}{c^{2}}
\left(1 - \frac{1}{n^{2}\beta^{2}}\right),
\end{equation}
so that the sum over both polarization modes reproduces the conventional Frank-Tamm spectrum.
In chiral matter, such as Weyl semimetals or quark-gluon plasma, the permittivity $\epsilon$ depends on the frequency \cite{Hansen:2020irw, Sa:2016cen}, 
and satisfies
\begin{equation}
\lim_{\omega\to\infty} n(\omega)=1.
\end{equation}
Therefore, the phase velocity tends to $v_{\mathrm{ph}}(\omega)\to c$, and Cherenkov radiation becomes kinematically forbidden, since no particle can move faster than $c$. 
Thus, the $\omega$-dependence of $\epsilon$ introduces a natural high-frequency cutoff in the theory, exactly as in the standard Cherenkov case, thereby avoiding any ultraviolet divergence.

Next, we address the second issue in isotropic chiral electrodynamics: whether the radiated energy is positive definite. Although the general expression for energy density in (\ref{CONS}) does not guarantee positivity, it remains to be seen whether positivity holds in specific cases.
 Equation (\ref{THETAC}) enables us to reexpress
 (\ref{ELAMBDA}) as 
\begin{equation}
{\cal E}_\nu=\frac{ q^2 \omega}{2 c^2}\frac{{\cal F}_\nu} {\sqrt{1+\frac{ \Sigma^2}{4}}} 
\sin^2 \Theta_\nu.
\label{ELAMBDA4}
\end{equation} 
This clearly demonstrates that the total radiated energy for each polarization is positive, despite the uncertainties arising from the first equation in (\ref{CONS}).

With each polarization having a physically meaningful and experimentally verifiable interpretation, we now examine in more detail the conditions for radiation occurrence, as outlined in Eq. (\ref{POSDEF1}). As suggested earlier, a  further simplification  arises  since the term in round brackets in this equation  is the perfect $(\sqrt{\sigma^2+4n^2 \omega^2}-\nu \sigma)^2/4c^2$, such that 
\beq
\sqrt{{\cal Q}^2_\nu + k^2}=\frac{1}{2c}\Big( \sqrt{\sigma^2+4n^2 \omega^2}-\nu \sigma\Big),
\label{Q2Pk2}
\eeq
where  $(\sqrt{\sigma^2 +4n^2 \omega^2}-\nu \sigma)=2n \omega {\cal F}_\nu$ is positive definite.
This allows to present Eq. (\ref{POSDEF1}) as 
\beq
\frac{1}{2c}\sqrt{\sigma^2 +4n^2 \omega^2}\geq \frac{\omega}{v} + \frac{\nu \sigma}{2c}.
\label{COND3}
\eeq 
Let us emphasize that this condition is the same as that required for the existence of real  radiation angles $\Theta_\nu$  
defined in Eq. (\ref{THETAC}). Furthermore, a further rewriting Eq. (\ref{COND3}) yields the final condition for radiation.
\beq
\omega (n^2 \beta^2-1)\geq \nu \sigma \beta,
\label{CONDRADFIN}
\eeq
where  $\omega \geq 0$. 
The signs in the inequality (\ref{CONDRADFIN}) are properly taken into account by considering the following two  cases,  which define the contribution of the polarization modes:

\begin{enumerate}[label=(\roman*)]
\item When $n^2\beta^2-1>0$, we have that $v>c/n$, i.e., we are in the high-velocity regime. This condition can  only be satisfied for $n>1$; that is, there is no Cherenkov emission in vacuum ($n=1$) in this regime.
The inequality  (\ref{CONDRADFIN}) yields
 \beq
 \omega > \frac{\nu \sigma \beta}{(n^2 \beta^2-1)}\equiv \nu \, \omega_0.
 \label{CONDVGEQVL}
 \eeq
 Thus the channel  $\nu=+$ opens up when $\omega > \omega_0$ while  it is  always open  for $\nu=-$.  In other words, we always  have one cone, $\nu=-$, together with a second cone provided $\omega>\omega_0$.
From Eq. (\ref{ELAMBDA2}), it is interesting to observe that when both polarization channels are open   the total emitted  energy  
\beq
{\cal E}={\cal E}_+ + {\cal E}_-= \frac{q^2 \omega}{c^2}\Big( 1-\frac{1}{n^2\beta^2} \Big),
\label{ELAMBDA41}
\eeq
turns out to be  independent of $\sigma$,
and coincides with that of the standard CHR in an isotropic media with the same refraction index. 
However, each separate mode  is radiating  at a different   angle with  an explicit dependence on $\sigma$.  
It is a consistency check to verify that the limit
  $\sigma=0$ yields $\omega_0=0$ and $\Theta_+ = \Theta_-= \Theta_{\rm Ch}$. Then, the total radiated energy flows through the standard Cherenkov angle $\Theta_{\rm Ch}$ and the spectral distribution  is given by ${\cal E}$ in Eq. (\ref{ELAMBDA41}), which reproduces  the standard result for CHR when $n>1$.
\item When $n^2\beta^2-1<0$, we have that $v<c/n$, i.e., we are in the low-velocity regime. This condition can be satisfied for both $n>1$ and $n=1$ (vacuum).

For $n>1$, Eq.~(\ref{CONDRADFIN}) demands
\begin{equation}
0 < \omega<-\frac{\nu\,\sigma\,\beta}{1-n^2\beta^2}=-\nu\,|\omega_0|.
\label{CONDVGEQVL1}
\end{equation}
This can only be satisfied for $\nu=-$, and $\sigma\neq 0$. This sector of the model allows for the study of the most sought-after threshold-free CHR \cite{liu2017integrated, zhang2022tunable,hu2020nonlocality,gong2023interfacial}.

For $n=1$ (vacuum), the condition becomes
\begin{equation}
0<\omega<\frac{\sigma\,\beta}{1-\beta^2},
\label{VACCOND0}
\end{equation}
which again applies only to the $\nu=-$ mode.
Observe that for $\sigma=0$, and for any value of $n$, the allowed window for $\omega$ closes and there is no radiation at all, as expected. Note that in this case, for fixed $\omega$, radiation only occurs if $\sigma$ exceeds a minimum value $\sigma_{\min}=\tfrac{\omega}{\beta}(1-n^2\beta^2)$. This threshold depends linearly on $\omega$, so at low frequencies it can be small, allowing radiation even for weak chiral coupling. 
However, as $\beta \to 0$ the $1/\beta$ factor dominates and makes $\sigma_{\min}$ arbitrarily large. Thus, at very low velocities the condition for radiation can only be satisfied if $\sigma$ is also very large. Therefore, a simultaneous expansion of the electromagnetic fields in small $\sigma$ and small $v$, as done in Refs.~\cite{Altschul:2014bba,Schober:2015rya,Altschul:2017xzx}, automatically excludes the physical regime where vacuum radiation could exist. Although taking the limit of smaller and smaller frequencies in Eqs. (\ref{CONDVGEQVL1}) and (\ref{VACCOND0}) is perfectly legitimate, we recall the obvious situation  that when we reach $\omega\equiv 0 $ we fall into  the static domain, where radiation is absent.
\end{enumerate}

\section{The non-relativistic limit}
\label{NRLIMIT}

In this section we give some details of the non-relativistic limit of our
exact electromagnetic fields (\ref{ERHO})-(\ref{BZ}).  We aim to make contact with the
iterative procedure developed in Refs.  \cite{Altschul:2014bba,Schober:2015rya, Altschul:2017xzx} in terms of an expansion
of the fields in powers of $\sigma $\ and $v$ \ in the spacetime domain. The successful
comparison  with their results serves as a partial validation of our
calculations.  Also we discuss the $z$-parity introduced in Refs.
 \cite{Altschul:2014bba,Schober:2015rya, Altschul:2017xzx}, that is the tool through which its authors demonstrate the
absence of  vacuum CHR in this case.

 Our starting point is  a series expansion of the  electromagnetic fields
(\ref{ERHO})-(\ref{BZ}) in powers of $\sigma $,  up to second order, summarized in the
equations (\ref{BRHOS2})-(\ref{EZS2}). The expansion naturally falls in terms of Bessel functions with argument  $(\rho\sqrt{\lambda^2})$, with $\lambda ^{2}=\frac{%
\omega ^{2}}{v^{2}}\left( 1-\frac{n^2 v^{2}}{c^{2}}\right)$. This is consistent with the fact that the limit $\sigma =0$ should reproduce the standard CHR for the material case $\epsilon >1$. Recalling  
that radiation emerges at $\rho \rightarrow \infty $ only when the
argument of the Bessel functions is imaginary we should demand $\lambda ^{2}<0$, which
yields precisely the threshold condition that the velocity of the particle
must be larger than the velocity of light in the medium $c/\sqrt{\epsilon }%
$. Going back to our general condition (\ref{POSDEF1}), we identify $\sqrt{\lambda^2}=i \sqrt{q_\nu}$, where the $\nu $  dependence is absent since we have expanded $q_{\nu
}$ only to zeroth order in $\sigma$. Recall $q_\nu >0$ is the condition for non-zero CHR. In this approximation the two
Cherenkov cones coalesce so that we must add  the polarization contributions
labeled by $\nu $ to obtain the total fields in Eqs. (\ref{BRHOS2})-(\ref{EZS2}).  We verify
that the only non-zero terms are $B_{\phi }$ together with $E_{\rho }\;$and $%
\ E_{z}$. After retaining only the dominant $c^2$ 
contributions in the remaining coefficients, we verify that these fields precisely match those of the original Frank-Tamm calculation in cylindrical coordinates. We comment in passing
that the expansion in Eqs. (\ref{BRHOS2})-(\ref{EZS2}) will not shed  any light  regarding the
vacuum CHR we find in our exact treatment. A perturbative  description of
this radiation would require an expansion of the fields such that the
argument of the Bessel functions $(i \sqrt{q_\nu} \rho)$ includes  the expansion of $q_\nu$ at least to first order in $\sigma$.

The non-relativistic limit we require to make contact with Refs.  \cite{Altschul:2014bba,Schober:2015rya, Altschul:2017xzx}  is given in Eqs. (\ref{ERHOEXP})-(\ref{BZEXP}) in the
space frequency domain. Additionally we need to
invert the Fourier transforms  
\begin{equation}
\tilde{\mbb{F}}_i(\mathbf{x},k)=e^{ikz}\,F_i(\rho ,k)=\int_{-\infty }^{+\infty }dt\,e^{i\omega
t}\,\mbb{F}_i(\mathbf{x},t),\qquad k=\frac{\omega }{v},
\label{RELSTSF}
\end{equation}
for each component $F_i(\rho ,k)$ of the fields obtained  from our
non-relativistic expansion of the fields. The inversion produces

\begin{equation}
\mbb{F}_i(\mathbf{x},t)=v\int_{-\infty }^{+\infty }\frac{\;dk}{2\pi }\,e^{i\;k {Z}%
}\,F_i(\rho ,k),\qquad {Z}=z-vt\;\;\;  \label{FTCOMPONENT}
\end{equation}
In order to make the comparison with Refs. \cite{Altschul:2014bba,Schober:2015rya, Altschul:2017xzx} we still need to transform between the cylindrical components $(V_\rho, V_\phi, V_z)$
and the spherical ones
 ($V_{r},V_{\theta
},V_{\phi }$), according to the transformation 
\begin{eqnarray}
V_{\rho } &=&\sin\theta \, V_{r}+\cos \theta \, V_{\theta },\qquad V_{z}=\cos
\theta \,  V_{r}-\sin \theta \, V_{\theta },
\label{COORDCHANGE}
\end{eqnarray}%
recalling that \ $V_{\phi }$ is unchanged. The inverse transformation is obtained by changing: $V_\rho \rightarrow V_r,\quad  V_z \rightarrow V_\theta, \quad \theta \rightarrow - \theta $.

\subsection{The zeroth order in $\sigma $}

\label{SIGMA0ORD}

The non-relativistic electromagnetic fields are those of a charge that moves
at constant velocity%
\begin{equation}
\mbb{E}=\frac{q}{\epsilon }\frac{\mathbf{\hat{R}}}{R^{2}},\;\;\;\;\mbb{
B}=\epsilon \frac{\mathbf{v}}{c}\times 
\mbb{E},\label{NRAPPRXO}
\end{equation}
written here in the coordinate system where the charge is instantaneously at rest. Here $%
\mathbf{R\;}$is the vector that joins the instantaneous position of the
charge ($\mathbf{v}t$) with the observation point $\mathbf{\ r,\;\;}$and the 
$z$-axis is chosen in direction of the velocity $\mathbf{v.\;}$ Thus 
$|\mathbf{R}|$=R=$\sqrt{%
x^{2}+y^{2}+(z-vt)^{2}}$ and $\theta \;$is the angle between $\mathbf{v\;}
$\ and $\mathbf{R}$. We have $Z=z- vt, \,\, \rho =\sqrt{x^{2}+y^{2}}, \,\, \sin \theta
=\rho/{R}, \,\,\cos \theta ={{Z}}/{R}\;\;$.
The only non-zero components of the electromagnetic
fields  are 
\begin{equation}
\mbb{E}_{r}=\frac{q}{\epsilon }\frac{1}{R^{2}},\qquad\;\mbb{B}_{\phi }=\frac{qv}{c}%
\frac{\rho }{R^{3}},
\label{BSIGMA0}
\end{equation}
yielding 
\begin{equation}
\mbb{E}_{\rho }=\frac{q}{\epsilon }\frac{\rho }{R^{3}},\qquad \mbb{E}_{z}=\frac{q}{%
\epsilon }\frac{{Z}}{R^{3}}.\;  \label{ESIGMA0}
\end{equation}
From the non-relativistic expansions in the Appendix \ref{NRAPP} we have%
\begin{equation}
B_{\phi }=\frac{2q}{c}\,k\;K_{1}\!\left( k\rho \right) ,\qquad E_{\rho }=%
\frac{2q}{\epsilon v}\,kK_{1}(k\rho ),\qquad E_{z}=-\frac{2iq}{\epsilon v}%
k\;K_{0}(k\rho ).
\label{OURAPP}
\end{equation}%
Inserting the Fourier transforms according to Eqs. (\ref{BESSELFT}) we recover the
expressions \ (\ref{BSIGMA0}) and (\ref{ESIGMA0}).
\subsection{Additional terms in the expansion}
\label{ADDTERMSEXP}

The iterative method proposed in Refs. \cite{Altschul:2014bba,Schober:2015rya,Altschul:2017xzx} relies on the following grounds: (1) an expansion of the fields in powers of $v$ and $\sigma$
\beq
\mbb{E}=\sum_{m,n=0}^{\infty} \mbb{E}^{(m,n)}, \qquad 
\mbb{B}=\sum_{m,n=0}^{\infty} \mbb{B}^{(m,n)},
\label{ITERATION}
\eeq 
where each term with superscript $(m,n)$ is  $O(\sigma^m, v^n)$; and  (2) since the fields of a particle moving with constant velocity along the $z$-axis  depend on
$z-vt$, it is possible to set $\partial_z=-v \partial t$ and rewrite  Maxwell's equations as a chain of iterations
\barr
&&\bs{\nabla}\cdot \mbb{E}^{({m,n})}=0, \quad 
\bs{\nabla}\cdot \mbb{B}^{({m,n})}=0, \quad \nonumber \\
&& \bs{\nabla}\times\mbb{E}^{({m,n})}=v \frac{\partial \mbb{B}^{({m,n-1})}}{\partial z},
\qquad
\bs{\nabla}\times\mbb{B}^{({m,n})}= -v \frac{\partial  \mbb{E}^{({m,n-1})}}{\partial z}+ \sigma \mbb{B}^{({m-1,n})}.  
\label{ITERATION1}
\earr
In this way, starting from the zeroth order approximation of the fields given by  (\ref{NRAPPRXO}) the subsequent contributions can be obtained. Some of these iterations  are shown in the first and second  columns of Table \ref{tab:campos} indicating the reference of origin in the last column. Our aim is to compare these terms of the expansion with the non-relativistic expression we obtained in Eqs. (\ref{ERHOEXP})-(\ref{BZEXP}). To this end we calculate the Fourier transform of the space-frequency domain of each contribution, which we present in cylindrical coordinates  in the third column of Table \ref{tab:campos}. Comparison with the corresponding terms of the expansion in the Appendix \ref{Ap-Exp} gives perfect agreement, just  validating our results.

\begin{table*}[htpb]
    \centering
    \scalebox{1.05}{
    \renewcommand{\arraystretch}{2}
    \setlength{\tabcolsep}{10pt}
    \begin{tabular}{c c c c}
        \hline\hline
        \textbf{Field} & \textbf{Function of } $(\mathbf{x},t)$ & \textbf{Function of } $(\mathbf{x},\omega)$ & \textbf{Ref.} \\
        \hline\hline
        $\mathbf{B}^{(0,1)}$ 
        & $ \displaystyle qv \frac{\sin\theta}{R^{2}}\, \mathbf{\hat{\bs{\phi}}}$ 
        & $ \displaystyle 2qk\, e^{ikz} K_{1}(k\rho)\, \mathbf{\hat{\bs{\phi}}}$ 
        & \cite{Altschul:2014bba} \\
        \hline
        $\mathbf{B}^{(1,1)}$ 
        & $\displaystyle \sigma \frac{qv}{2R} \left(2\cos \theta\, \mathbf{\hat{\bs{r}}} - \sin \theta\, \mathbf{\hat{\bs{\theta}}} \right)$ 
        & $\displaystyle
        \left.
        \begin{aligned}
        \sigma q\, e^{ikz} \Big( & \left[2K_{0}(k\rho) - (k\rho) K_{1}(k\rho) \right]\, \mathbf{\hat{k}} \\
        & -i (k\rho) K_0(k\rho)\hat{ \bs{\rho}}\Big)
        \end{aligned}
        \right.
        $
        & \cite{Altschul:2014bba} \\
        \hline
        $\mathbf{B}^{(2,1)}$ 
        & $\displaystyle \sigma^{2} \frac{qv}{2} \sin \theta\, \mathbf{\hat{\bs{\phi}}}$ 
        & $\displaystyle \sigma^{2} q\rho\, e^{ikz} K_{0}(k\rho)\, \mathbf{\hat{\bs{\phi}}}$ 
        & \cite{Schober:2015rya} \\
        \hline
        $\mathbf{E}^{(2,2)}$ 
        & $\displaystyle -\sigma^{2} \frac{qv^2}{4} \left[ \left(\frac{3}{2} \cos^{2} \theta - \frac{1}{2} \right) \mathbf{\hat{r}} - \sin \theta \cos \theta\, \mathbf{\hat{\bs{\theta}}} \right]$ 
        & $\displaystyle
        \left.
        \begin{aligned}
        \sigma^2 q\, e^{ikz}\,\frac{\rho}{4}\Big( & (\omega \rho) K_1(k\rho)\, \hat{ \bs{\rho}} \\
        & +    i v [2K_1(k\rho)-(k \rho)K_0(k \rho)]\, \hat{\mathbf{k}} \Big)
        \end{aligned}
        \right.
        $
        & \cite{Schober:2015rya} \\
        \hline\hline
    \end{tabular}
    }
    \caption{The fields in the space-time domain (second column) are written in spherical coordinates with $\mbf{R}=\mbf{x}- \mbf{v} t, \, R=|\mbf{R}|, \, \cos\theta=(z-vt)/R, \, \sin\theta=\sqrt{x^2+ y^2}/R$. The fields in the space-frequency domain (third column) are obtained by  Fourier  transforming those in the second column and are presented in cylindrical coordinates. The parameter ``k'' in the cited references corresponds to $\sigma/2$ in our conventions. Here we set $c=1$ and $\epsilon=1$.}
    \label{tab:campos}
\end{table*}
As noted in Ref. \cite{Altschul:2014bba}, the introduction of the parameter $\sigma/c$, which has dimensions of one over longitude, enables the possibility of fields decaying slower than $1/R^2$ at infinity.  Unfortunately, the  proposed version of the iterative method does not yield radiation at all. This can be clearly seen after  writing the fields in the coordinate (cylindrical)-frequency domain, as shown  for a few terms in the third column of Table \ref{tab:campos}. The generalization to second order in $\sigma$, carried in the  Appendix \ref{Ap-Exp}, results in an  expansion of the fields in a series of Bessel functions $K_n$ with argument $ k \rho$. Since $k=\omega/v$ and $\rho$ are real, the limit of these functions in the radiation zone ($\rho \to \infty$) is strictly zero due to the  exponential decay for the Bessel functions  indicated in Eq. (\ref{BESSELINFTY}), for all $n$.  Recall that  non-zero radiation  requires  an expansion of the fields in terms of Bessel functions $K_n$ with imaginary argument, as discussed in detail at the beginning of this section. In this sense, the null radiation result reported in Refs. \cite{Altschul:2014bba,Schober:2015rya,Altschul:2017xzx} is correct, but it does not apply to Cherenkov radiation.

\subsection{The $Z$-parity of the fields}
\label{ZPARITY0}

 References \cite{Altschul:2014bba,Schober:2015rya,Altschul:2017xzx} disclosed that each spherical component of the electric and magnetic fields in coordinate space has definite ${ Z}$-parity under the change ${ Z} \rightarrow -{ Z}$, where each component is in coordinate space $(t, \mbf{x})$. These properties are summarized in the left panel of Table~\ref{tab:Zparity} and are correctly determined for the polarization independent contributions to the electromagnetic fields which components  $\{ i \}$ we label with the additional  subindex $a=1$  . When going from spherical to cylindrical coordinates via Eqs.~(\ref{COORDCHANGE}), they translate into the right panel of Table~\ref{tab:Zparity}.
\begin{table}[H]
	\centering
	{
		\renewcommand{\arraystretch}{1.5}
		\setlength{\tabcolsep}{9pt}
		\begin{minipage}{0.3\textwidth}
			\raggedleft
			\begin{tabular}{@{}c ccc@{}}
				\hline\hline
				& \multicolumn{3}{c}{$ Z$-parity} \\ \cline{2-4}
				Field & $r$ & $\theta$ & $\phi$ \\ \hline
				$\mbb{E}_{i1}$ & $+$ & $-$ & $-$ \\
				$\mbb{B}_{i1}$ & $-$ & $+$ & $+$ \\
				\hline\hline
			\end{tabular}
		\end{minipage}
		\hspace{7.5 em}
		\begin{minipage}{0.3\textwidth}
			\raggedright
			\begin{tabular}{@{}c ccc@{}}
				\hline\hline
				& \multicolumn{3}{c}{$ Z$-parity} \\ \cline{2-4}
				Field & $\rho$ & $\phi$ & $z$ \\ \hline
				$\mbb{E}_{i1}$ & $+$ & $-$ & $-$ \\
				$\mbb{B}_{i1}$ & $-$ & $+$ & $+$ \\
				\hline\hline
			\end{tabular}
		\end{minipage}
	}
\caption{The ${ Z}$-parity of the polarization independent contributions to the electric and magnetic fields, labelled by the additional subindex $a=1$, in spherical coordinates (left panel) \cite{Altschul:2014bba,Schober:2015rya,Altschul:2017xzx},  and in cylindrical coordinates (right panel).}
	\label{tab:Zparity}
\end{table} 
 Next we are interested in determining   the ${ Z}$-parity of each field contribution in our series expansion, paying special attention to the polarization  dependent terms, i.e. those proportional to $\nu$, which were  not considered in Refs. \cite{Altschul:2014bba,Schober:2015rya,Altschul:2017xzx}. The   Eq. (\ref{FTCOMPONENT}) tell us  that this parity is fully determined by the $k$ dependence of the function $F(\rho,k)$ of the field. A glance at the non-relativistic expansion in the Appendices \ref{Ap-Exp} and \ref{APPENDIXF} reveals that this dependence has the general form
 \beq
 F(\rho,k)\sim W_{(p,q)}(\rho,k) \equiv k^p K_q(k \rho),
 \label{GENFORMEXP}
 \eeq
where  $p$ and $q$ are  integers and $K_q$ is a modified Bessel function. The Fourier transform  
\barr
W_{(p,q)}({\mbf x},t) &=&= v\, \int_{-\infty}^{+\infty} \frac{ d k}{2 \pi }
\, \,\, e^{ik { Z} }\,W_{(p,q)}(\rho,k)=v\, \int_{-\infty}^{+\infty} \frac{ d k}{2 \pi }
\, \,\, e^{ik { Z} }\,k^p K_q(k \rho) \nonumber \\
&=&  v \frac{\partial  }{i \partial {\bar Z}} \int_{-\infty}^{+\infty} \frac{ d k}{2 \pi }
\, \,\, e^{ik { Z}}\,k^{(p-1)} K_q(k \rho)=  v \frac{\partial  }{i \partial { Z}} W_{((p-1), q)}
({\mbf x},t),
\label{WPQ}
\earr
yields the relation
\beq
{\cal P}\Big( W_{(p,q)}({\mbf x},t) \Big)= -{\cal P}\Big( W_{((p-1),q)}({\mbf x},t) \Big).
\label{PARITYREL}
\eeq 
where we call 
${\cal P}(X)$ the ${ Z}$-parity of the function $X$.  
That is to say, for a given Bessel function ($q$ fixed) we need to read the ${ Z}$-parity  from the explicit Fourier transform of only one combination $W_{(p,q)}$, while the result for the remaining powers of $k$ is obtained from the relation (\ref{PARITYREL}). This allows to read the ${ Z}$-parity directly in the space-frequency domain. Since our expansion includes modified Bessel functions up to order  $q=2$, we read the basic parities from Eqs. (\ref{BESSELFT}) in the Appendix 
\beq
{\cal P}\Big(K_0(k \rho)\Big)=+,\qquad 
{\cal P}\Big( K_1(k \rho)\Big)=-,  \qquad    
 {\cal P}\Big( k  K_2(k \rho) \Big)=-.
\label{FT3} 
\eeq
 The explicit form of 
the fields in the frequency domain are given in the Appendix \ref{Ap-Exp} and we introduce the following notation for any cylindrical component of the electric  field
\beq
[E^{(\nu)}_{ia}]_\alpha(\rho,k), \quad i: \rho, \, \phi, \, z , \quad a=1, 2, \quad  \alpha=0,1,2, 
\label{NOTZPARITY}
\eeq
where $a=1$ denotes the contribution to the field independent of the polarization $\nu$, while $a=2$ refers to the polarization dependent piece. The label $\alpha=0,1,2$ indicates the power of sigma in the component, i.e. its  proportionality to $\sigma^0, \, \sigma$ and $\sigma^2$, respectively. The total electric field component  in the polarization $\nu$ is $ E_i^{(\nu)}= \sum_{a, \alpha} \,  [E^{(\nu)}_{ia}]_\alpha  $ .  Similar notation is adopted  for the magnetic field.
 
Summarizing, just by inspection from the power expansion of the  fields in the Appendix \ref{Ap-Exp}, together with the relation (\ref{PARITYREL}),  plus the basic Fourier transforms (\ref{FT3}) we can determine the ${ Z}$-parity of each contribution. 

We first focus on the $\phi$-components, which receive contributions from both the electric and magnetic fields. In Refs.~\cite{Altschul:2014bba,Schober:2015rya,Altschul:2017xzx}, only the polarization–independent terms ($a=1$) were analyzed, showing a definite ${ Z}$-parity. Our analysis goes further by also including the polarization–dependent contributions ($a=2$). The resulting ${ Z}$-parities for all such components are summarized in Table~\ref{tab:phi}, where one observes that, for instance, the electric field component $E_\phi$ contains three pieces with $\alpha=0,1,2$: while the $a=1$ contribution has odd parity, the  $a=2$ contributions have even parity. A similar pattern occurs for the magnetic components. Consequently, the full $\phi$-components of the fields combine pieces with opposite ${ Z}$-parities, and therefore do not possess a definite ${ Z}$-parity, in contrast with the assignments summarized in Table~\ref{tab:Zparity}.
\begin{table}[H]
	\centering
	{
		\renewcommand{\arraystretch}{1.5}
		\setlength{\tabcolsep}{9pt}
		\begin{tabular}{c c c c}
			\hline\hline
			Field & $ Z$-parity & Field & $ Z$-parity \\ \hline
			$[E^{(\nu)}_{\phi 2}]_{0}$ & $+$ & $[B^{(\nu)}_{\phi 1}]_{0}$ & $+$ \\
			$[E^{(\nu)}_{\phi 1}]_{1}$ & $-$ & $[B^{(\nu)}_{\phi 2}]_{1}$ & $-$ \\
			$[E^{(\nu)}_{\phi 2}]_{2}$ & $+$ & $[B^{(\nu)}_{\phi 1}]_{2}$ & $+$ \\ 
			\hline\hline
		\end{tabular}
	}
\caption{${ Z}$-parity of the $\phi$-components of the electric and magnetic fields.}
	\label{tab:phi}
\end{table}
We find a similar behavior for the remaining components of the fields. 
In this case, the $\rho$- and $z$-components also combine contributions with opposite 
${ Z}$-parities, as summarized in Table~\ref{tab:rho_z}, where the left panel 
corresponds to the $\rho$-components and the right panel to the $z$-components.
\begin{table}[H]
	\centering
	{
		\renewcommand{\arraystretch}{1.5}%
		\setlength{\tabcolsep}{9pt}%
		\begin{minipage}{0.4\textwidth}
			\raggedleft
			\begin{tabular}{c c c c}
				\hline\hline
				Field & $ Z$-parity & Field & $ Z$-parity \\ \hline
				$[E^{(\nu)}_{\rho 1}]_{0}$ & $+$ & $[B^{(\nu)}_{\rho 2}]_{0}$ & $+$ \\
				$[E^{(\nu)}_{\rho 2}]_{1}$ & $-$ & $[B^{(\nu)}_{\rho 1}]_{1}$ & $-$ \\
				$[E^{(\nu)}_{\rho 1}]_{2}$ & $+$ & $[B^{(\nu)}_{\rho 2}]_{2}$ & $+$ \\ 
				\hline\hline
			\end{tabular}
		\end{minipage}
		\hspace{1 cm}
		\begin{minipage}{0.4\textwidth}
			\raggedright
			\begin{tabular}{c c c c}
				\hline\hline
				Field & $ Z$-parity & Field & $ Z$-parity \\ \hline
				$[E^{(\nu)}_{z1}]_{0}$ & $-$ & $[B^{(\nu)}_{z2}]_{0}$ & $-$ \\
				$[E^{(\nu)}_{z2}]_{1}$ & $+$ & $[B^{(\nu)}_{z1}]_{1}$ & $+$ \\
				$[E^{(\nu)}_{z1}]_{2}$ & $-$ & $[B^{(\nu)}_{z2}]_{2}$ & $-$ \\ 
				\hline\hline
			\end{tabular}
		\end{minipage}
	}
	\caption{${ Z}$-parity of the electric and magnetic fields, 
		for the $\rho$-components (left panel) and the $z$-components (right panel).}
	\label{tab:rho_z}
\end{table}
Recapping, the ${ Z}$-parity   for the field components reported  in Refs. \cite{Altschul:2014bba,Schober:2015rya,Altschul:2017xzx}  corresponds to their polarization-independent contribution, resulting  solely in type $a=1$ fields.
Comparison with Table \ref{tab:Zparity} shows that the predicted ${Z}$-parity for   all the   $a=1$ components is correctly  recovered from our field expansions. Nevertheless, the  radiation output is described separately  by each polarization mode, which appears at each characteristic angle. We find that the  total fields $E^{(\nu)}_i=\sum_{a, \alpha} [E^{(\nu)}_{ia}]_\alpha$ ,  $B^{(\nu)}_i=\sum_{a, \alpha} [B^{(\nu)}_{ia}]_\alpha$ have no definite ${ Z}$-parity, since  the $a=2$ components have opposite  ${ Z}$-parity  as compared to the   $a=1$ ones. This invalidates the symmetry argument for the absence of vacuum radiation given in Refs. \cite{Altschul:2014bba,Schober:2015rya,Altschul:2017xzx}.

 We can understand the failure of the proposed iterative method to produce the necessary polarization-dependent contributions due to: (i) the zeroth-order terms in Eqs. (\ref{NRAPPRXO}) are polarization independent, 
and so will be subsequent iterations since there is no polarization parameter that explicitly enters Maxwell's equations.
Certainly, iteration in $\sigma$ produce corrections to the polarization independent contributions, as can be verified in the Appendix \ref{Ap-Exp}. (ii) the chain of  iterative equations (\ref{ITERATION1}) is designed to produce the  total electric and magnetic fields at any order, which in general are the sum of the two polarizations. In fact,  
    starting from our full solutions and then performing a controlled expansion in small $\sigma$ (see  the Appendix \ref{Ap-Exp}), we obtain both polarization-dependent and polarization-independent contributions. 
    However,  when we look for the total fields, the two polarizations must be added together, and we find the cancellation of the polarization-dependent terms in the retained order, leaving only the polarization-independent contributions. In the non-relativistic regime, where no radiation is present anyway, this can be observed  from the expansions in the Appendix \ref{NRAPP}, verifying that we get the same results for some of  the total fields calculated in Refs. [79-81].

\section{ Numerical  Estimations}

\label{APPLIC}

Weyl semimetals are representatives of topological matter  presenting  an effective electromagnetic response described by CFJ electrodynamics \cite{Sa:2016cen,borisenko2019time,lv2015observation,grassano2020influence,zu2021comprehensive}. However, most of the known Weyl materials are characterized by parameters ${\cal B}_0= \sigma$ and $\bs{{\cal B}} \neq 0$. When inversion symmetry is violated, while maintaining time-reversal symmetry, it is in principle possible to  engineer materials to set $\bs{{\cal B}}=0$, yielding a truly  isotropic chiral matter according to our definition. The best possibility for engineering these Weyl points is then by realizing a Weyl semimetal phase in a non-centrosymmetric material (broken inversion symmetry) that preserves time-reversal symmetry, and subsequently using external tuning parameters (like   appliying strain, external magnetic fields, or  specific doping, for example) to precisely control the position of the Weyl nodes relative to the Fermi level. To the best of our knowledge the state of the art in this direction is still very preliminary.  On the other hand, a theoretical possibility is suggested by the work in Ref. \cite{Gomez:2023jyl}, where it was shown that the vector $\bs{{\cal B}} $ can be tuned by introduccing a tilting velocity and an anisotropy matrix at  each of the originally momentum-separated Weyl cones with $\mbf{b} \neq 0$. Dealing with  any of these two possibilities is beyond the scope of the present work. Consequently,  to illustrate the key aspects of Cherenkov radiation for an electron propagating in isotropic chiral matter, we are not in position to choose parameters that  correspond to an existing material. Instead, we select  some  representatives from standard WSMs. 
We take the following order of magnitude values:   $\sigma=0.05$ eV \cite{borisenko2019time} and  $n=3$, the latter corresponding to an average refractive index reported for Weyl semimetals TaAs and NbAs  around  a frequency of 1~eV~\cite{zu2021comprehensive}, yielding a threshold velocity $\beta_{\rm TH} = 1/3$.

 \begin{figure}[h!]
  \centering
 \includegraphics[scale=1]{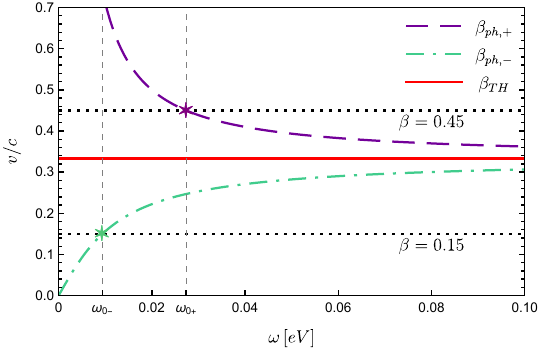} 
  \caption{Plot of the dimensionless phase velocities $v_{\mathrm{ph},\pm}/c \, = \, \beta_{\rm ph,\pm}$ as a function of frequency, for $\sigma=0.05~\mathrm{eV}$ and $n=3$. Horizontal lines at $\beta_{TH}$ and at two example particle speeds ($\beta=0.45$ and $\beta=0.15$) are shown for reference.}
  \label{fig:vph}
\end{figure}

 We begin by discussing the phase velocities $v_{\rm ph  ,\pm}$ as a function of frequency, emphasizing that the condition for non-zero radiation is $\beta> \beta_{\rm ph , \pm}$. For the chosen  parameters, the phase velocities of the two polarization modes are shown in Fig.~\ref{fig:vph}. As 
$\omega\to 0$ one has $\beta_{\mathrm{ph},+}(\omega)\to\infty$, while  $\beta_{\mathrm{ph},-}(\omega)\to 0$. On the other hand  $\beta_{\mathrm{ph},\pm}(\omega)\to \beta_{\rm{TH}}$ when $\omega\to \infty$.
Therefore, for charges moving at high velocities, i.e., 
 $\beta>\beta_{\rm TH}$, one always has $\beta>\beta_{\mathrm{ph},-}(\omega)$ for all $\omega>0$, whereas $\beta>\beta_{\mathrm{ph},+}(\omega)$ only when   $\omega>\omega_{0+}(\beta)$, with $\omega_{0+}(\beta)=\sigma\,\beta/(n^2\beta^2-1)$. In other words, the mode $\nu=+$ radiates only for $\omega > \omega_{0+}$, while the mode $\nu=-$ radiates always when $\beta>\beta_{\rm TH}$ (See Fig. \ref{fig:vph}). By contrast,  when $\beta < \beta_{\rm TH}$, the mode $\nu=+$ is always suppressed. Nevertheless, 
since $\beta_{\mathrm{ph},-}(\omega)\to 0$ when  $\omega\to 0$,  for any fixed $\beta >0$, there always exists a finite low-frequency range $0 < \omega <  \omega_{0-}(\beta)={\sigma\,\beta}/(1-n^2\beta^2)$  in which $\beta >\beta_{\mathrm{ph},-}(\omega)$ such that  radiation occurs.  Moreover, this frequency window narrows as $\beta$ decreases, as shown in Fig.~\ref{fig:vph}. This latter scenario showcases a groundbreaking aspect of CHR, dubbed threshold-free CHR, which offers a significant advancement in the field.

Next, we illustrate the relationship between the frequency range for non-zero {CHR} and the corresponding angles of the Cherenkov cones. The spectral energy distribution for each case is also plotted.
In the high-velocity case ($\beta>1/n$), the Cherenkov angle for the $\nu=-$ mode exists for all $\omega>0$, so this mode can radiate across the entire spectrum. In contrast, the $\nu=1$ mode admits a real angle only above the critical frequency $\omega_{0 +}(\beta)$, in agreement with the phase-velocity analysis. Figure~\ref{CASO1}(a) shows $\cos\Theta_\nu(\omega)$ for both modes, and Fig.~\ref{CASO1}(b) displays the spectral energy distribution for each channel. For $\omega>\omega_{0+}(\beta)$, once the $+$ mode is activated, the sum $\mathcal{E}_{-}(\omega)+\mathcal{E}_{+}(\omega)$ coincides with the standard Cherenkov result, which is consistent with Eq.~(\ref{ELAMBDA41}).
 
\begin{figure}[h!]
  (a)\includegraphics[scale=0.88]{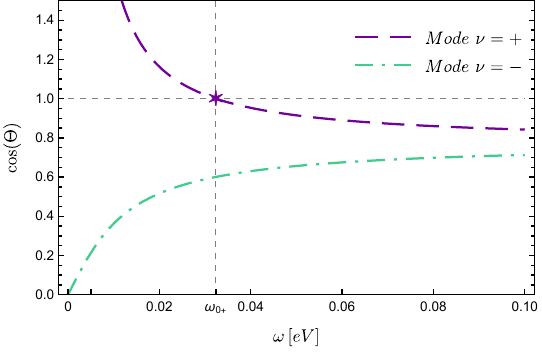}\qquad (b)\includegraphics[scale=0.88]{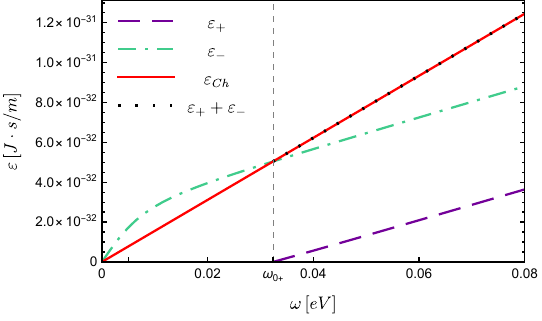}
  \caption{Panel (a): Plot of $\cos \Theta$ for the Cherenkov angles corresponding to  the modes $\nu=\pm$, as a function of the frequency for $\beta > 1/n$. Panel (b): Plot of the spectral energy distribution  for different contributions: each mode ($\mathcal{E}_{+}$ and $\mathcal{E}_{-}$), the sum of both modes ($\mathcal{E}_{+}+\mathcal{E}_{-}$), and the standard CHR case ($\mathcal{E}_{\text{Ch}}$). The parameters are : $n=3$, $\sigma=0.05~\text{eV}$, and  $\beta=0.43$.}
  \label{CASO1}
\end{figure}
For  low velocities ($\beta<1/n$), the Cherenkov angle for the $\nu=+$ mode is nonexistent since $\cos\Theta_{+}\notin[0,1]$, resulting in no radiation. In contrast, for the $\nu=-$ mode the Cherenkov angle exists only within the frequency range  $0<\omega\le \omega_{0-}(\beta)$, i.e., precisely where $0\le \cos\Theta_{-}\le 1$, as shown in Fig.~\ref{CASO2}(a), also  in agreement with the phase-velocity analysis. The spectral energy  distribution $\mathcal{E}_{-}(\omega)$ for  the $\nu=-$ mode is shown in Fig.~\ref{CASO2}(b). This behavior contrasts with the standard  case, where for $\beta<1/n$ no Cherenkov radiation is allowed. This is an equivalent way of showing   that threshold-free CHR is possible in isotropic chiral matter.
\begin{figure}[h!]
 (a)\includegraphics[scale=0.88]{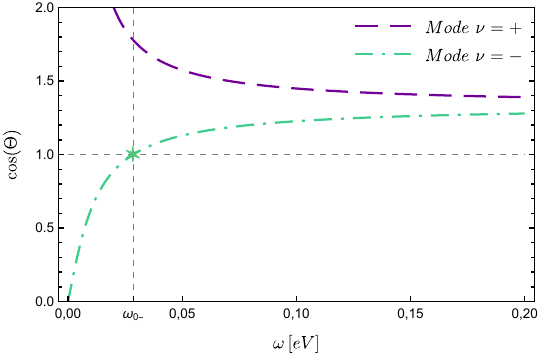}\qquad (b)\includegraphics[scale=0.88]{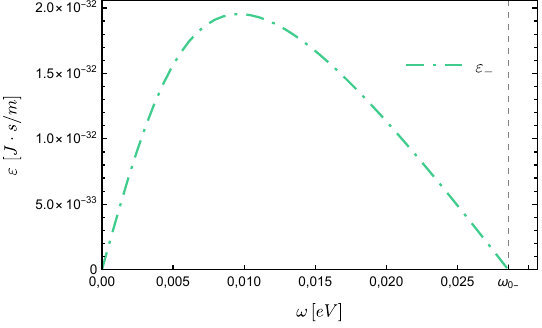}
  \caption{Panel (a): Plot of $\cos \Theta$ for the Cherenkov angles corresponding to  the modes $\nu=\pm$, as a function of the frequency for $\beta <  1/n$. Panel  (b): spectral energy distribution for the $\nu=-$ mode. The parameters: $n=3$, $\sigma=0.05~\text{eV}$,  $\beta=0.25$.}
  \label{CASO2}
\end{figure}

\

An alternative way of looking at the rise of the Cherenkov cones is by rewriting  the condition (\ref{THETAC}), which determines them, as  ${\cal F}_\nu \, \cos\Theta_\nu=1/n\beta$. The left hand side of this equation defines the function $H_\nu(\Theta)$ such that  
\beq
H_\nu(\Theta)\equiv \cos \Theta_\nu\left(\sqrt{1 + \Sigma^2/4} - \nu \Sigma/2\right)=\frac{1}{n\beta}. \label{HTHETA}
\eeq
For a given set of parameters $\omega, n ,\sigma $ each  function $H_\nu(\Theta)$ is plotted as a function of $\Theta$. Then, by sweeping horizontal lines at values $1/n\beta$, the intersection with each  $H_\nu(\Theta)$ curve determines the  allowed number of Cherenkov cones, as shown in Fig. \ref{FIG1}. Moving the horizontal line we identify different regions where we have no Cherenkov radiation, one Cherenkov cone, and two Cherenkov cones. For reference we also plot the Standard Cherenkov radiation function $H_{\rm Ch}(\Theta)=\cos\Theta$ obtained when  $\Sigma =0$. Table \ref{TABLE1} presents the values of the Cherenkov angles for various velocities corresponding to this scenario. 
  
\begin{table}[ht]
\begin{center}
\begin{tabular}{cccc}
\hline 
\hline
{ $\beta$ } \qquad   & { $ \Theta_+(\rm rad) $ } \qquad & { $ \Theta_{\rm Ch}(\rm rad) $ } \qquad  & { $ \Theta_-(\rm rad) $ } \qquad \tabularnewline
\hline 
\hline
$0.45$ \qquad & \qquad $0.0$ \quad & \qquad $0.0$ \quad &\qquad $0.520$ \quad \tabularnewline
$0.60$ \qquad & \qquad $0.0$ \quad & \qquad $0.585$ \quad &\qquad $0.862$ \quad \tabularnewline
$0.75$ \qquad & \qquad $0.547$ \quad & \qquad $0.841$ \quad &\qquad $1.023$ \quad  \tabularnewline
$0.90$ \qquad & \qquad $0.779$ \quad & \qquad $0.981$ \quad &\qquad $1.122$ \quad 
\tabularnewline
\hline
\hline 
\end{tabular}
\caption{Cherenkov angles for $\Sigma=0.5$ and $n=2$, as shown in Fig.~\ref{FIG1}.}
\par
\label{TABLE1}
\end{center}
\end{table}

\begin{figure*}[h!]
\centering
\includegraphics[width=0.65 \textwidth]{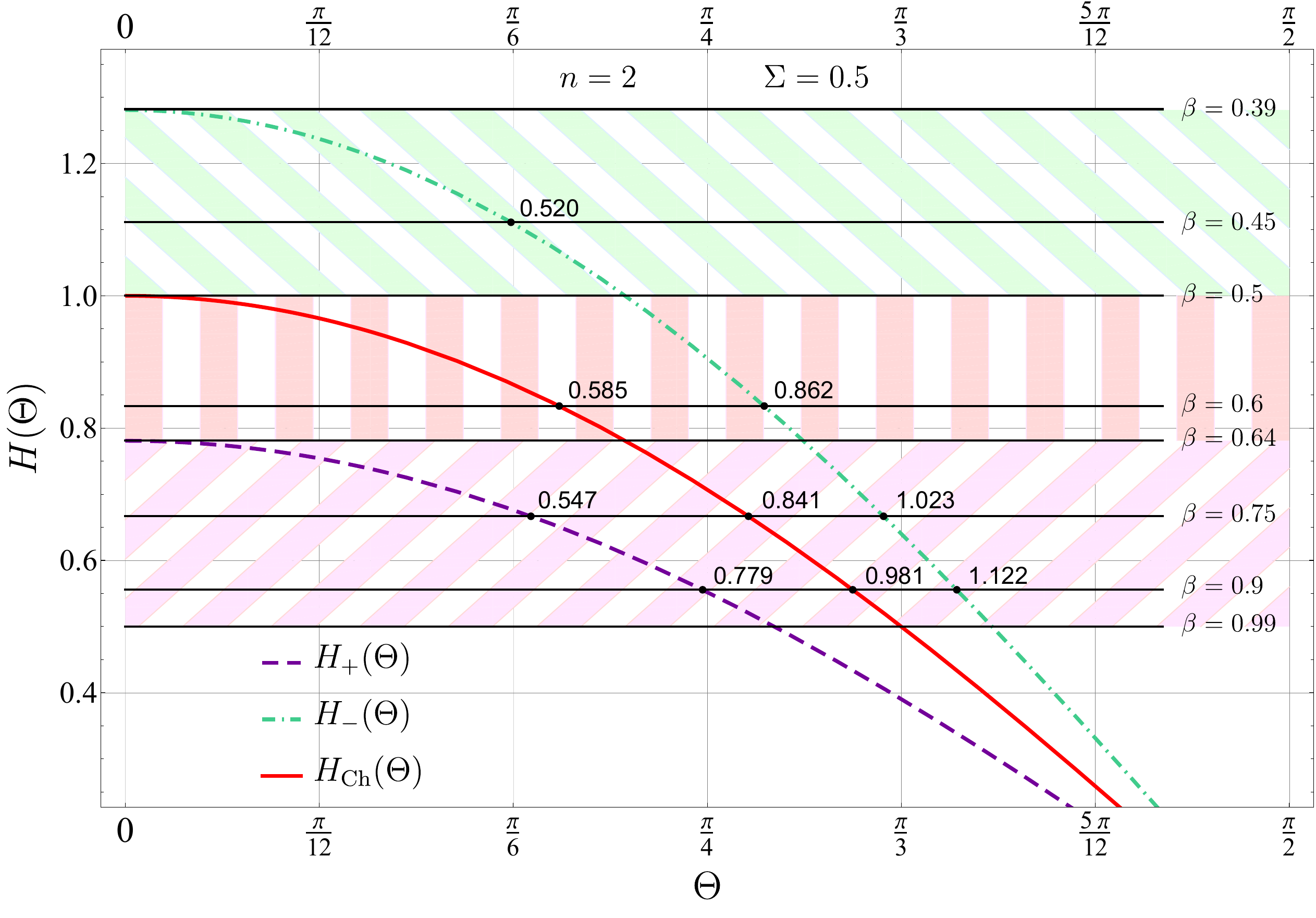}
\caption{Plot of the functions $H_{+}(\Theta)$ (purple dotted line for the $\nu=+$ mode), $H_{-}(\Theta)$ (green dash-dotted line for the $\nu=-$ mode), and $H_{\rm Ch}(\Theta)$ (red solid line for the standard case), for $\Sigma = 0.5, \, n = 2$. The horizontal lines labeled by $\beta$ correspond to $H = 1/(n\beta)$ on the ordinate. The Cherenkov angles are given in radians. }
\label{FIG1}
\end{figure*}
 
 From Eq. (\ref{THETAC}) we verify that in the limit of high frequencies both Cherenkov angles $\Theta_\nu$ converge  into the  standard result 
  $\Theta_{\rm Ch}$, with $\cos \Theta_{\rm Ch}=1/(n\beta) $. We aim to investigate the rate at which this convergence process takes place in order to get an idea of the angular sensitivity  required to distinguish  between these angles, as a function of different parameters of the theory.
 \begin{figure}[h!]
	\centering
	\includegraphics[scale=1]{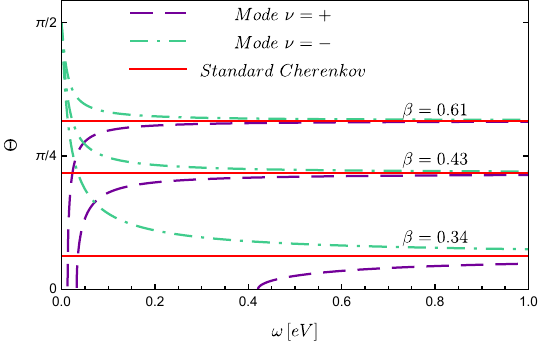}
	\caption{Emission angles $\Theta_\nu$ as a function of the frequency, for different particle velocities. The parameters are $\sigma=0.05 {\rm eV}$ and $n=3$, with $\beta >\beta_{\rm TH}$.}
\label{FIG121}
\end{figure}
The correlation between radiation angles and various particle velocities $\beta> \beta_{\rm TH}$ is shown in Fig.
 \ref{FIG121} as a function of frequency, revealing the onset of the $\nu=+$ mode at the corresponding cut-off frequency $\omega_0$. As the  velocity increases the convergence of both modes occurs more rapidly, thus  diminishing the angular sensitivity. However, this sensitivity can be adjusted by varying either $n$ or $\sigma$. Notably, increasing $n$, while keeping $\beta$  and $\sigma$ fixed decreases the sensitivity, as shown in the right panel of Fig. \ref{FIG122}. In contrast, for fixed $\beta$ and $n$, increasing $\sigma$ enhances the sensitivity. The latter situation is shown in the left panel of Fig. \ref{FIG122}. 
 \begin{figure}[h!]
	\centering
	\includegraphics[scale=0.87]{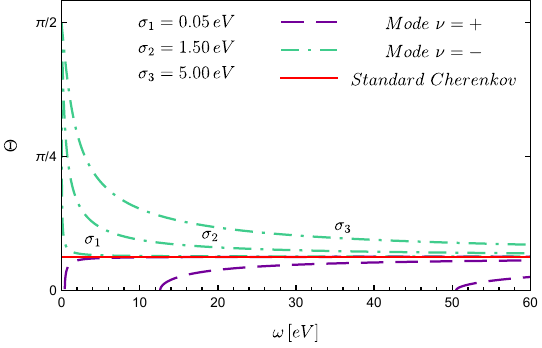}
	\hspace{0.1cm}
	\includegraphics[scale=0.87]{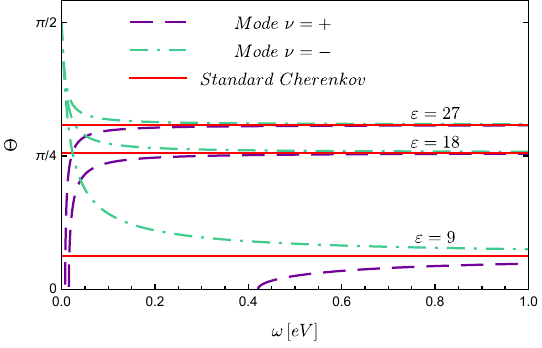}
	\caption{Left panel: Emission angles $\Theta_\nu$ as a function of the frequency, for different values of $\sigma$, with $\beta=0.34$ and $n=3$. Increasing $\sigma$ enhances the angular sensitivity, bringing the emission angles closer to the standard Cherenkov angle. Right panel: Emission angles $\Theta_\nu$ as a function of the frequency for different values of the refraction index  $ n = \sqrt{\epsilon}=3,\ \sqrt{18},\ \sqrt{27}$, with fixed velocity $\beta = 0.34$ and $\sigma = 0.05$ eV. Increasing $n$ reduces the angular sensitivity, bringing the emission angles closer to the standard Cherenkov angle.}
\label{FIG122}
\end{figure}

 A key figure of merit in the threshold-free CHR    ($\beta <  \beta_{\rm TH}) $ is the photon extraction efficiency per unit length for the channel $\nu=-$, defined as \cite{gong2023interfacial, Chen:2022qlr}
    \beq 
    {\tilde \eta}_{-}= d \eta_{-}/dz={\cal E}_{-}/(\hbar \omega \, E_{\rm ch}),
    \label{FEE0}
    \eeq    
which  can be interpreted as the total number of photons emitted per unit length and per unit frequency divided by the  kinetic energy of the charge $E_{\rm ch}= m_{\rm ch} c^2 (1-\gamma_{\rm L})$. Here $\gamma_{\rm L}$ is the standard Lorentz  factor. In our case  we obtain  
\beq
{\tilde \eta}_{-}= f \frac{1}{(\gamma_{\rm L}-1)} \Omega, \qquad  f=\frac{q^2}{c^4\, \hbar \, m_{\rm ch}},
\label{FEE1}
\eeq
where $\Omega$ is defined in Eq. (\ref{ELAMBDA2}) and
the constant factor is  $f=296.8 \,  [{\rm Watt \, m}]^{-1}$. We observe that $\Omega$ is a slowly varying function within the interval  $\omega < \omega_0$, decreasing from $1$ to $0$ in the allowed frequency region, as illustrated in Fig. \ref{FIG11}. Notably, for small velocities, the frequency range is narrower, and the curve rises more sharply to its maximum value of $1$. 

\begin{figure}[h!]
	\centering
\includegraphics{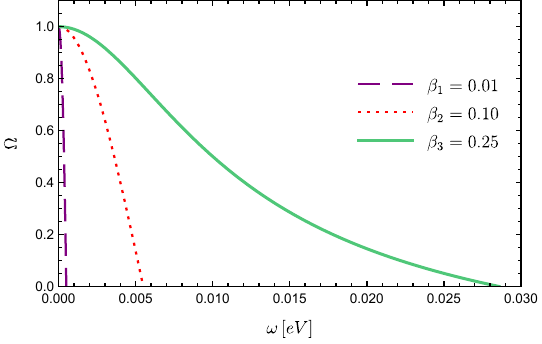}
	\caption{The function  $\Omega(\omega)$ for the charge velocities $\beta_1=0.01$, $\beta_2=0.10$ and $\beta_3=0.25$, with $n=3$ and $\sigma = 0.05$ eV.}
	\label{FIG11}
\end{figure}
In other words, the strongest variations in ${\tilde \eta}_{-}$ arises from the amplifying factor $1/(\gamma_{\rm L}-1)$, as illustrated schematically in Table \ref{T1}. As this factor increases,  CHR emission from low-velocity charges is enhanced, at the cost of a diminishing  allowed frequency range.  A relative enhancement is determined by the ratio of the corresponding values of  ${\tilde \eta}_{-}$   at different velocities within a shared frequency range. To facilitate a meaningful comparison with the values  of 
${\tilde \eta}_{-}$ reported in Fig. 2 of  Ref. \cite{Chen:2022qlr} we observe that  
in our case, when  $\sigma=0.05$ eV, $n=3$  and $\beta= 0.43$,  ${\cal E}_-$ results of the order of 
    $10^{-32}\,$-$\,10^{-31}$ [J-s/m] in the frequency range $0 < \omega  < 0.1$ eV.

\begin{table}[h!]
	\centering
	\begin{tabular}{c c c c}
		\hline 
		\hline
		$\beta$ \quad & \quad $0.01$ \quad &\quad  $0.10$ \quad & \quad $0.25$ \quad \\
		$1/(\gamma_{\rm L}-1) \quad $ & \quad $2 \times 10^4$ \quad &\quad $2 \times 10^2$ \quad & \quad 30 \quad \\
        $\omega_0 \,\,  [{\rm meV}]$  \quad & $0.5 \,  $ & $5.5 \, $ &\quad  $29 \, $    \\
        $\langle{\tilde \eta}_{-}\rangle \,\, [{\rm Watt \, m}]^{-1}$ \quad & \quad $6\times10^{6}$ \quad & $6 \times 10^{4}$ \quad  \quad &  \quad $9 \times10^{3}$ \quad \\
        \hline
		\hline
	\end{tabular}
	\caption{The amplification factor $1/(\gamma_{\rm L}-1)$   for different charge velocities, together with the corresponding maximum allowed frequency $\omega_0$ and the resulting value  of $\langle {\tilde \eta}_{-}\rangle$ given by  ${\tilde \eta}_{-}$ calculated at $\Omega =1$.}
	\label{T1}
\end{table}

\section{Summary and conclusions} 
\label{conclusion}

We examine the Cherenkov  radiation (CHR) of a charge $q$ moving at constant velocity $\mbf{v}= v\hat{\mbf {k}}$ in isotropic chiral matter with $\epsilon \geq 1$, $\mu=1$ and magnetoelectric parameter $\sigma$.
The effective electromagnetic response of chiral matter is a restricted form of Carroll-Field-Jackiw electrodynamics, with an axion angle $\theta(x)= \sigma t$.  This electrodynamics shows unexpected features such as a local energy density which is  not positive definite, accompanied  with an energy-momentum tensor which is not gauge-invariant.  We have carefully addressed these problematic issues in the manuscript. 

As is well known, the local energy-momentum tensor $T^{\mu\nu}$ does not represent physical quantities, as it is subject to redefinitions of the form  
 $\tilde{T}^{\mu\nu}=T^{\mu\nu} +\partial_\rho S^{\mu\rho\nu}$,  with $S^{\mu\rho\nu}$ an arbitrary tensor antisymmetric in its first two indices. The physical quantities that remain invariant are the associated integral charges: the total energy $E$ in space and the total energy flux across a closed surface at infinity. We have demonstrated the gauge invariance of both charges in section \ref{MAXEQSSUB}. Notably, while the Poynting vector $\mbb{S}$ itself is not  gauge-invariant, the total time-integrated energy flux over a closed surface at infinity preserves gauge invariance. This defines the observable spectral energy distribution ${\cal E}$, as given in Eq. (\ref{SED0}), which plays a crucial role in our analysis.

 By solving the modified Maxwell's equations (\ref{MAXW1}) and (\ref{MAXW2}) in the space-frequency domain using cylindrical coordinates, we derive closed-form expressions for the electromagnetic fields (\ref{ERHO}) - (\ref{BZ}) and the potentials in the Coulomb gauge (\ref{TOTALA1}). Each field is a superposition of two circular polarizations, $\nu=+$ and $\nu=-$, corresponding to left- and right-handed helicities respectively, which represent the system's normal modes. The detailed calculations and helicity identification are presented in Appendix \ref{SOLMAXWEQ}. Each circular polarization contributes independently and is experimentally distinguishable through its angular emission pattern.

 We then examine the radiation zone by considering an infinite cylinder with its axis aligned with the particle motion, focusing on the total radiated energy per unit length $\cal{E}$, which depends on the electromagnetic fields and potentials. From its general expression (\ref{FINALSED}), the same values of  $\cal{E}$  is obtained in both the Coulomb gauge Eq.~(\ref{TOTALA}) and the Lorenz gauge Eq. (\ref{ALGAPP}), thus providing an additional proof of gauge invariance.

 Next we impose causality by requiring outgoing waves as $\rho \to \infty$. This condition yields the final constraint for the existence of CHR in each polarization mode, given by Eq. (\ref{POSDEF1}), which determines the allowed frequency range for a given set of parameters. This relation generalizes the standard threshold condition for CHR when $\sigma=0$.

The asymptotic expressions for the fields are given in Eqs. (\ref{EPHICCRAD})-(\ref{AZRAD}), which enable the explicit calculation of the spectral energy distribution $\cal{E}$. Notably, we find that interference effects, codified in ${\cal{E}}_{+,-}$, exactly cancel, as detailed in Appendix \ref{Ap-En}. Consequently, $\cal{E}$ reduces to an independent superposition of the two polarization modes: ${\cal E}=
{\cal E}_+ + {\cal E}_-$.

Effectively, this means we can consider $\mbb{S}= \mbb{S}_+ + \mbb{S}_-$, such that our proof of gauge invariance remains valid for each polarization, providing independent physical significance to each mode. The spectral energy distribution for each polarization $\cal{E}_\nu$
is given in alternative forms in Eqs (\ref{ELAMBDA2}) and (\ref{ELAMBDA}), as closed expressions in the parameters $\sigma$ and $v$.

The expression for the spectral energy distribution of each mode, given by Eq. (\ref{ELAMBDA}) , can be expressed in terms of the angle $\Theta_\nu$ as shown in Eq. (\ref{ELAMBDA4}). This formulation provides a direct proof that both ${\cal E}_\nu$ are positive, thereby resolving the issue posed by the non-positive definiteness of the local energy density in the general case

The threshold condition  (\ref{POSDEF1}) can be further simplified to (\ref{CONDRADFIN}), providing a straightforward basis for determining the contributions of the polarization modes. Based on this relation, we proceed to examine the frequency range for excitation in each mode for a given set of parameters $n, \sigma$ and $v$.

One way to determine the conditions for radiation is to require that the particle velocity $\beta$ exceeds the phase velocity (\ref{VPHASE0}) of the wave. 
This is illustrated in the  Fig. \ref{fig:vph} which shows that  when 
$\omega\to 0$ one has $\beta_{\mathrm{ph},+}(\omega)\to\infty$, 
while  $\beta_{\mathrm{ph},-}(\omega)\to 0$. 
On the other hand, when $\omega\to \infty$, we have both $\beta_{\mathrm{ph},\pm}(\omega)\to \beta_{\rm{TH}}=1/n$.
Therefore, for charges moving at high velocities, i.e., 
 $\beta>\beta_{\rm TH}$, the whole frequency region  $\omega>0$ allows radiation for the polarization $\nu=- $.
On the contrary, for the polarization $\nu=+$,  $\beta>\beta_{\mathrm{ph},+}(\omega)$  is restricted to the interval  $\omega>\omega_{0+}$,  where $\omega_{0+}$ results from the intersection between the growing phase velocity $\beta_{\mathrm{ph},+}(\omega)$ and the charge  velocity. In other words, the mode $\nu=+$ radiates only for $\omega > \omega_{0+}$, while the mode $\nu=-$ radiates always when $\beta>\beta_{\rm TH}$. In the limit $\sigma \to 0$, this set of conditions reduces to the standard Cherenkov threshold, providing a clear consistency check with the standard case.

 On the other hand,  in the low velocity regime,  $\beta < \beta_{\rm TH}$, the mode $\nu=+$ is always suppressed. 
Nevertheless,  there always exists a finite low-frequency
range $0 < \omega < \omega_{0-}$ in which $\beta > \beta_{\mathrm{ph},-}(\omega)$ such that  radiation occurs. 
The frequency $\omega_{0-}$ results from the intersection of the decreasing sector of $\beta_{{\rm ph}, -}(\omega)$ with the charge velocity. Moreover, this frequency window narrows as $\beta$ decreases. This latter scenario showcases a groundbreaking aspect of CHR, dubbed threshold-free CHR, which offers a significant advancement in the field. 

A further argument for the independent contribution of each polarization to radiation lies in their experimental distinguishability, marked by distinct emission angles that reflect the characteristic cones of CHR.
We find at most two cones, determined by the angles (\ref{THETAC}), whose existence  depends on fulfilling the condition  $\cos \Theta_\nu < 1$ in  this equation— a condition that coincidentally matches the requirement for satisfying the dispersion relation (\ref{POSDEF1}). The functions $\cos \Theta_\nu $ in Eq. (\ref{THETAC})  have the following behavior as functions of the frequency: $\cos \Theta_+$ is a decreasing function such that $\lim_{\omega \to 0}\cos \Theta_+ \to \infty$ and 
$\lim_{\omega \to \infty}\cos \Theta_+=1/(n\beta)$. On the other hand
$\cos \Theta_-$ is an increasing function  which starts at zero when $\omega=0$ and goes to the limit $1/(n\beta)$ when $\omega \to \infty$.

Then, an alternative way to determine when emission is allowed for each polarization, is to plot the right-hand side of Eq. (\ref{THETAC}) in terms of the frequency  and look for the range where the curve is below one. For the high-velocity case ($n\beta >1$) this is shown in the left panel of  Fig. (\ref{CASO1}). Again we observe that the polarization $\nu=-$ is allowed for all frequencies, while the mode 
$\nu=+$ is excited only for frequencies larger than $\omega_{0+}$. The right panel of Fig. (\ref{CASO1}) shows the spectral energy distribution for each polarization, together with their sum when both polarizations are excited. This sum equals the spectral energy distribution of standard CHR ($\sigma=0$) with the same remaining parameters. We regard this property as a mathematical curiosity that does not imply any cancellation between the radiation emitted by each cone, which occurs at distinct angles. In other words, in this case we observe two Cherenkov cones when $\omega > \omega_{0+}$  and only one cone when $\omega < \omega_{0+}$. 

The most notable situation arises in the low-velocity limit $\beta n <1 $,  where radiation  is completely forbidden in the standard case with $\sigma=0$.  The polarization $\nu= +$ is also forbidden in this case, since $\cos \Theta_+ >1$ for all frequencies. However, since  $0< \cos \Theta_- < 1/(n \beta)$, with $1/(n \beta) >1$, we can always find a range of 
frequencies
$0 \leq \omega \leq \omega_{0-}$ such that $\cos \Theta_- <1 $, thus opening a frequency window for radiation. 
This is  shown in the left panel of Fig. \ref{CASO2}. 
In other words, this sector  of the model allows for the study of the most sought-after threshold-free CHR arising from slowly-moving charges.  \cite{liu2017integrated,zhang2022tunable, hu2020nonlocality,gong2023interfacial}.  This includes 
the specific scenario with  $n=1$ and $\sigma \neq 0$, which is commonly referred to  as ``vacuum'' CHR in the literature. 
Our calculation confirms that this ``vacuum''  CHR is indeed excited within the frequency range  $0< \omega < \sigma \beta/(1-\beta^2)$. The spectral energy density for the  allowed polarization $\nu=-$ is plotted in the right panel of Fig. \ref{CASO2}.

A two-cone structure in radiation is also known to occur in bi-isotropic chiral media exhibiting optical activity, where the splitting arises from the distinct refractive indices for circular polarizations~\cite{Bolotovskii1963,engheta1990vcerenkov}. While the physical origin in those cases is different from ours, this analogy supports the experimental relevance of our predictions. It would be interesting to explore whether known modifications of the radiation processes in optically active media, such as the fields of a uniformly moving charge, transition radiation, or Cherenkov radiation in chiral waveguides~\cite{barsukov1999vavilov,
Galyamin2013Jul,
Galyamin2017Mar}, also admit counterparts in isotropic chiral matter. In the Appendix \ref{BIREFMEDIA} we include a brief review of electrodynamics in bi-isotropic media emphasizing the similarities and differences  with our approach.

Let us focus now on fixed frequency and rewrite the condition (\ref{THETAC}) as $H_\nu(\Theta)={\cal F}_\nu \cos \Theta_\nu=1/(n \beta)$. In Fig. (\ref{FIG1}). We plot   $H_{\pm}(\Theta)$ together with the standard $H(\rm Ch)$ as a function of the angle. Now, the intersection of these curves with   horizontal lines at  different values of $1/(n \beta)$ determines the allowed Cherenkov angles in each case. This is shown in  Fig. (\ref{FIG1}), and the values of the Cherenkov angles are summarized in Table \Ref{TABLE1}. The pattern of zero, one or two  Cherenkov cones is again apparent. 

As previously noted, both Cherenkov angles $\Theta_\nu$  converge to the standard result $\Theta_{\rm Ch}$ in the high-frequency limit. To estimate the angular sensitivity required to distinguish between these angles as a function of different parameters, it is insightful to examine the rate of this convergence. The correlation between radiation angles and various particle velocities $\beta> \beta_{\rm TH}$ 
is illustrated in Fig. \ref{FIG121} as a function of frequency.
As the  velocity increases the convergence of both modes occurs more rapidly, thus  diminishing the angular sensitivity. However, this sensitivity can be adjusted by varying either $n$ or $\sigma$. Notably, increasing $n$, while keeping $\beta$  and $\sigma$ fixed decreases the sensitivity, as shown in the right panel of Fig. \ref{FIG122}. In contrast, for fixed $\beta$ and $n$, increasing $\sigma$ enhances the sensitivity. The latter situation is shown in the left panel of Fig. \ref{FIG122}. 

A key figure of merit in the threshold-free CHR  ($\beta <  \beta_{\rm TH}) $ is the photon extraction efficiency per unit length ${\tilde \eta}_{-}$ \cite{gong2023interfacial, Chen:2022qlr}, which  can be interpreted as the total number of photons emitted per unit length and per unit frequency divided by the  kinetic energy of the charge.  In our case, this parameter is proportional to the product of the rapidly varying factor $1/(\gamma_{\rm L}-1)$ and the slowly varying factor $\Omega$, where  $\gamma_{\rm L}$ is the Lorentz  factor of the charge. In Fig. \ref{FIG11} we plot $0< \Omega < 1$  in the allowed frequency interval, for different charge velocities.  We observe that  for small velocities, the frequency range is narrower, and the curve rises more sharply to its maximum value of $1$. In other words, the strongest variations in ${\tilde \eta}_{-}$ arise from the amplifying factor $1/(\gamma_{\rm L}-1)$, as illustrated schematically in Table \ref{T1}. As this factor increases, CHR emission from low-velocity charges is enhanced, but at the cost of a reduced allowed frequency range.

Finally, in  section \ref{NRLIMIT} we discuss the non-relativistic limit of our
exact electromagnetic fields (\ref{ERHO})-(\ref{BZ}). From  equations (\ref{BRHOS2})-(\ref{EZS2}) we start by considering the expansion of the fields up to order $\sigma^2$. This yields the argument in  the Bessel functions to be $(\rho \sqrt{\lambda^2})$, where $\lambda^2=\frac{\omega^2}{v^2} (1-\frac{n^2 v^2}{c^2})$ is  of  order zero in $\sigma$, when compared with the exact expression for $\cal{Q}_\nu$ in (\ref{POSDEF1}). Recalling that CHR emerges only when the argument of the Bessel functions is imaginary, we conclude that this expansion, even being $\sigma$ dependent,  cannot tell us anything about $\sigma\neq 0$ CHR. At most  we can obtain approximate expressions for standard CHR demanding  $\lambda^2 <0$, which precisely yield the threshold condition
$v^2> c^2/n^2$. In other words, to obtain some information about $\sigma\neq 0$ CHR we would need to keep an expansion
of $\cal{Q}_\nu$ at least to order $\sigma$ in the argument
of the Bessel functions.  We aim to establish contact with the iterative procedure outlined in Refs. \cite{Altschul:2014bba,Schober:2015rya, Altschul:2017xzx}, which relies on a non-relativistic approximation. This approach expands fields in powers $\sigma $  and $v$ within the spacetime domain. To this end we consider and additional restriction on our previous expansion by keeping only the dominant $c^2$  terms in the coefficients of the Bessel functions and setting  $\lambda= \frac{\omega}{v}=k$
 In this case the argument  $(k \rho)$ of the Bessel functions is positive, thus precluding radiation as discussed in section \ref{ADDTERMSEXP}.
Despite this,  we focus upon the resulting expression for the non-relativistic expressions for the fields that can be found in Eqs. (\ref{ERHOEXP})- (\ref{BZEXP}).  As a partial verification of our results, we compare   with  explicit calculations of the fields performed  in the iterative method.  
We take the expressions for some of the calculated fields in the space-time domain, indicated in the second column of the Table \ref{tab:campos} ( with the corresponding reference in the last column), and calculate the time-Fourier transform shown in the third column of the Table. We then compare these results with our expansion in powers of $\sigma$ obtained in Eqs. (\ref{ERHOEXP})-(\ref{BZEXP}) for the non-relativistic limit, finding perfect  agreement. This consistency check provides partial validation of our calculations.
We also discuss the Z-parity concept introduced in Refs. \cite{Altschul:2014bba,Schober:2015rya, Altschul:2017xzx}, which these authors use to demonstrate the absence of ``vacuum'' CHR. After reformulating the parity transformation rules for cylindrical electromagnetic field components (Table \ref{tab:Zparity}), we propose a straightforward algorithm to directly determine each component's parity in the frequency domain, based on Eqs. (\ref{PARITYREL}) and (\ref{FT3}). The resulting Z-parities for the non-relativistic fields (\ref{ERHOEXP})-(\ref{BZEXP}) are summarized in Tables \ref{tab:phi} and \ref{tab:rho_z}. We observe a critical inconsistency: within a given component, the Z-parity of the polarization-dependent contribution (labelled by the subscript $a=2$) is always opposite to that of the corresponding polarization-independent contribution ($a=1$). Consequently, the non-relativistic field components lack definite Z-parity due to the mixing of $a=1$ and $a=2$ contributions. This finding undermines the criteria used in Refs. \cite{Altschul:2014bba,Schober:2015rya, Altschul:2017xzx} to establish the non-existence of ``vacuum''  CHR, which relies on assigning a definite Z-parity to each field component.


\acknowledgments{We thank Professor R. Potting for useful comments. R.M.v.D. and L.F.U. has been partially supported by DGAPA-UNAM Project No. AG100224. L.F.U.  also acknowledges 
support from  project SECIHTI (M\'{e}xico) CBF-2025-I-1862. R.M.vD. was supported by UNAM Posdoctoral Program (POSDOC). E.B.-A. and M.A.G. acknowledge support from the Priority 2030 Federal Academic Leadership Program, and partial  support from the Foundation for the Advancement of Theoretical Physics and Mathematics ``Basis''. E.B.-A. acknowledges partial support from SECIHTI (México) doctoral scholarship. }

\appendix
\section{Solution of Maxwell's equations} 
\label{SOLMAXWEQ}
\subsection{The electromagnetic fields outside the sources }
\label{EMFOUTS}

In  obvious notation for the components of the fields in cylindrical coordinates we find the following results. 
Next we look for an  equation for the electric field outside the sources,  recalling that $\partial_t=-i\omega$, $\partial_\phi=0$ and $\partial_z=ik$. To this end we take the  curl of Faraday's law, and use  Ampère's law in the resulting  contribution $\bs{\nabla}\times \mbf{B}$, obtaining 
\begin{equation}
	\bs{\nabla} \times (\bs{\nabla} \times \mbb{E}) = -\frac{\epsilon}{c^2} \partial_t^2 \mbb{E} - \frac{\sigma}{c^2} \partial_t \mbb{B}.
	\label{INT1}
\end{equation}
 Finally, applying Faraday's law once again in the last term of Eq. (\ref{INT1}), and going to the frequency-space domain we arrive at
\begin{equation}
	\bs{\nabla} \times (\bs{\nabla} \times \tilde{\mbb{E}}) = \frac{\epsilon \omega^2}{c^2} \tilde{\mbb{E}} + \frac{\sigma}{c} (\bs{\nabla} \times \tilde{\mbb{E}}),
	\label{INT2}
\end{equation}
which decouples the electric field. Expressing the  components of the curl in cylindrical coordinates, and omitting the $z$-dependence in the exponential $e^{ikz}$, it is a direct matter to obtain the following equations for each component of Eq. (\ref{INT2}) in the ($\rho,\omega$) space
\barr
&&    \rho \,\,\,   { \rm component}\, : \quad 
\label{AF1}
		\partial_\rho E_z = \frac{i}{k} \left(k^2 - \frac{\epsilon \omega^2}{c^2} \right) E_\rho - \frac{\sigma}{c} E_\phi, \\
	 &&  \phi \,\,\,   {\rm component} \,: \quad
\label{AF2}
		\frac{\sigma}{c} \partial_\rho E_z = \partial_\rho\left( \frac{1}{\rho} \partial_\rho(\rho E_\phi) \right) - \left(k^2 - \frac{\epsilon \omega^2}{c^2} \right) E_\phi + \frac{i \sigma k}{c} E_\rho, \\
&& z \,\,\,  {\rm   component} \, : \quad 
	\frac{1}{\rho} \partial_\rho \left( i k \rho E_\rho - \rho \partial_\rho E_z - \frac{\sigma}{c} \rho E_\phi \right) = \frac{\epsilon \omega^2}{c^2} E_z.
    \label{AMPCOMP}
\earr

Next  we eliminate $\partial_\rho E_z$ from Eqs.~(\ref{AF1}) and (\ref{AF2}) finding
a first equation that couples $E_\rho$ and 
$E_\phi$
\begin{equation} 	\left( \rho^2 \partial_\rho^2 + \rho \partial_\rho - \left(1 + \rho^2 \alpha^2 \right) \right) E_\phi = -\frac{i \epsilon \omega^2 \sigma}{k c^3} \rho^2 E_\rho,
    \label{EPHIERHO1}
\end{equation}
where we denote
\begin{equation}
	\alpha^2 = k^2 - \frac{\epsilon \omega^2}{c^2} - \frac{\sigma^2}{c^2}.
    \label{ALPHA2}
\end{equation}
 Going back to Gauss’s law we express $E_z$ as a function of $E_\rho$
\begin{equation}\label{EZ1}
	E_z = \frac{i}{k} \frac{1}{\rho} \partial_\rho(\rho E_\rho),
\end{equation}
which is  subsequently substituted into   Eq.~(\ref{AF1}), yielding 
\begin{equation} 
	\left( \rho^2 \partial_\rho^2 + \rho \partial_\rho - \left(1 + \rho^2 \gamma^2 \right) \right) E_\rho = \frac{i k \sigma}{c} \rho^2 E_\phi, 
    \label{ERRO1}
\end{equation}
with
\begin{equation}
	\gamma^2 = k^2 - \frac{\epsilon \omega^2}{c^2}.
    \label{GAMMA2}
\end{equation}
The next task is to solve the coupled system of Eqs.(\ref{EPHIERHO1}) and (\ref{ERRO1}). 
The operators in the left hand side of both equations  are identified as those corresponding to a modified Bessel function of order one, which motivates us to take $E_\phi$ and $E_\rho$ as linear combinations of the  independent solutions  $K_1$ and $I_1$ \cite{abramowitz1965handbook}. The boundary conditions
at $\rho \to 0$ as well as those at  $\rho \to \infty$
forces us to include only  $K_1$ as part of the solution and to reject the contribution of $I_1$. In this way  we take
\beq
E_\phi= X K_1(Q\rho), \qquad E_\rho= Y K_1(Q\rho).
\label{EPHISEPERHO}
\eeq
Substituting in Eqs. (\ref{EPHIERHO1}) and (\ref{ERRO1}) and using the relations among the derivatives of the modified Bessel functions  we obtain the matrix equation
\begin{equation}
	\begin{bmatrix}
		Q^2 - \alpha^2 & \frac{in^2 \omega^2 \sigma}{kc^3} \\
		-\frac{ik\sigma}{c} & Q^2 - \gamma^2
	\end{bmatrix}
	\begin{bmatrix}
		X \\
		Y
	\end{bmatrix}
	=
	\begin{bmatrix}
		0 \\
		0
	\end{bmatrix},
    \label{MATRIXEQ}
\end{equation}
which provides the dispersion relation 
\begin{eqnarray}
Q^4-(\alpha^2+\gamma^2)Q^2+\alpha^2\gamma^2-\frac{n^2 \omega^2 \sigma^2}{c^4}&=&0,
\label{DISPRELAPP0}
\end{eqnarray}
yielding  the allowed momenta $Q$ of the solution. This is a quartic equation with solutions
 \begin{equation}
	Q^2_\nu(\omega)=\frac{\omega^2}{v^2}-\frac{n^2 \omega^2}{c^2}-\frac{\sigma^2}{2c^2}+\nu \frac{\sigma}{2c^2} \sqrt{\sigma^2+4n^2 \omega^2}, \qquad \nu= \pm,
	\label{DISPREL}
\end{equation}
where we used $k=\omega/v$. In our conventions, the modes $\nu= + \, (\nu=-) $ describe left (right) circular polarization in the plane perpendicular to the wave vector.

Having imposed the zero determinant condition in (\ref{MATRIXEQ}) we obtain the relation 
$
Y_\nu=i \Omega_\nu  X_\nu$ between  the coefficients 

$X_\nu$ and $Y_\nu$ ,  with 
\beq
\Omega_\nu=
{k \sigma}/{(Q^2_\nu-\gamma^2)c}.
\label{RELCOEF}
\eeq
In this way the electric field results in 
\barr
&& E_\phi=\sum_{\nu=\pm } X_\nu \, K_1(Q_\nu \rho), \qquad E_\rho= i\sum_{\nu=\pm }\Omega_\nu\, X_\nu \, K_1(Q_\nu \rho), \nonumber \\
&&  \hspace{1.5cm} E_z=\frac{1}{k}\sum_{\nu=\pm }\Omega_\nu\, Q_\nu\, X_\nu \, K_0(Q_\nu \rho),
\label{TOTALE}
\earr
where the expression for $E_z$ is obtained from Eq. (\ref{EZ1}).
The magnetic field components are given in terms of the electric field via Faraday's law
\beq
 B_\rho= -\frac{c}{v} E_\phi, \qquad  B_z=-\frac{ic}{k \rho v}\partial_\rho(\rho E_\phi), \qquad   B_\phi=\frac{c}{v}E_\rho +\frac{ic}{k v}\partial_\rho E_z. 
\label{B}
\eeq
Equations  (\ref{B}) for  the magnetic field give
\begin{eqnarray}
	&& \hspace{1.3cm} B_\phi=i\frac{c}{v}\sum_{\nu=\pm } \left( 1 -  \frac{Q^2_\nu}{k^2} \right) \,X_\nu \Omega_\nu K_1(Q_\nu \rho), \nonumber \\
	&&  B_\rho=-\frac{c}{v}\sum_{\nu= \pm }X_\nu K_1(Q_\nu \rho), \quad B_z = i\frac{ c}{\omega}\sum_{\nu=\pm } Q_\nu X_\nu  K_0(Q_\nu\rho). 
	\label{TOTALB}
\end{eqnarray}

\subsection{The boundary conditions} 
\label{BOUNDC}

The last step in the calculation of the electromagntic fields is to determine the coefficients $X_\nu$. We do this by  imposing   the boundary conditions at $\rho \rightarrow 0$, where the sources are located.  To this end  we apply the integral form of  Gauss and 
 Ampère's laws in the space-frequency domain. 

Starting  from the Gauss law, we have
\begin{equation}
	\epsilon \oint_{\partial V} \tilde{\mbb{E}} \cdot \hat{\mbf{n}} \, dS = 4\pi \int_V \bar{\rho} \, d^3x.
	\label{APGL}
\end{equation}
We consider a Gaussian surface consisting of a cylinder of radius $\rho=a \rightarrow 0$ and length $h$ along the $z$-axis, so the surface $\partial V $ is divided into  the mantle of the cylinder and  two circular caps: cap 1 to the left and cap 2 to the right. The splitting of the integral (\ref{APGL}) yields
\begin{eqnarray}
	- \int_{\text{cap 1}} \tilde{\mbb{E}}_z \rho\, d\rho\, d\phi 
	+ \int_{\text{cap 2}} \tilde{\mbb{E}}_z \rho\, d\rho\, d\phi 
	+ \int_{\text{mantle}} \tilde{\mbb{E}}_\rho R\, d\phi\, dz &=& 4\pi \int d\rho\, d\phi\, dz\, \rho \frac{q}{2\pi v \rho} \delta(\rho) e^{i\frac{\omega}{v}z} \nonumber \\
	 &=&\frac{4\pi q}{v \epsilon} \int dz\, e^{i\frac{\omega}{v}z}.
     \label{APGLDETAIL}
\end{eqnarray}
Note that the integrals over the caps are not trivial since $\tilde{\mbb{E}}_z=E_z e^{i\frac{\omega}{v}z} $. In fact, from the last Eq. (\ref{TOTALE}) we have in general 
\begin{eqnarray}
	\int \tilde{\mathbb{E}}_z \, \rho\, d\rho\, d\phi 
	&=& \frac{2\pi e^{i\frac{\omega}{v}z}}{k} \sum_{\nu=\pm} \Omega_\nu Q_\nu X_\nu \int_0^a K_0(Q_\nu \rho) \rho\, d\rho, \\
	&=& \frac{2\pi e^{i\frac{\omega}{v}z}}{k} \sum_{\nu=\pm} \Omega_\nu Q_\nu X_\nu \left[ \frac{1 - Q_\nu a K_1(Q_\nu a)}{Q_\nu^2} \right].
    \label{APGLDETAIL1}
\end{eqnarray}
Since $a \rightarrow 0$ and
$
	\lim_{a \rightarrow 0} K_1(Q_\nu a) = 
	{1}/{(Q_\nu a)}
$
we find $\int \tilde{\mathbb{E}}_z \rho\, d\rho\, d\phi=0$ in each cap.
Thus the remaining contribution from the  Gauss's law is only from the mantle where $\rho=a$, yielding 
\begin{eqnarray}
	a \int E_\rho(\rho) \,  e^{i\frac{\omega}{v}z} \, d\phi\, dz &=& 2 i \pi a \sum_{\nu =\pm}\Omega_\nu X_\nu K_1(Q_\nu a) \int dz\, e^{i\frac{\omega}{v}z} = \frac{4\pi q}{v \epsilon} \int dz\, e^{i\frac{\omega}{v}z}, \\
	&&2 i \pi a \sum_{\nu=\pm} \Omega_\nu X_\nu \frac{1}{Q_\nu a} = \frac{4\pi q}{v \epsilon}.
    \label{APGLDETAIL2}
\end{eqnarray}
In the above  we have substituted the expression for $E_\rho(\rho,\omega)$ from  Eq. (\ref{TOTALE}) together with the $a\rightarrow 0$ limit of $K_1(Q_\nu \, a)$. The result provides 
a first relation between the coefficients $X_1$ and $X_2$. 
\begin{equation}
\label{LG}
	i \frac{\Omega_+ X_+}{Q_+} + i \frac{\Omega_- X_-}{Q_-} = \frac{2q}{v \epsilon}.
\end{equation}
A second equation to determine the coefficients  is obtained from Ampère's law
\begin{equation}
	\oint_{\partial S} \tilde{\mbb{B}} \cdot d\mbf{l} = \frac{4\pi}{c} \int_{S} \tilde{\mbb{J}} \cdot \hat{\mbf{n}}\, dS.
	\label{AMBC}
\end{equation}
We consider the Amperian loop as a circle of radius $\rho=a \rightarrow 0$ perpendicularly centered on the trajectory of the moving particle, such that 
$
d\mbf{l} = a \, d\phi\, \hat{\bs{\phi}}$ and $
	\hat{\mbf{n}}\, dS = \hat{\mbf{z}}\, \rho\, d\rho\, d\phi.
$
Then equation (\ref{AMBC}) reads
\begin{eqnarray}
	\int \tilde{\mbb{B}}_\phi\,  a \, d\phi= e^{i\frac{\omega}{v}z} 2\pi a B_\phi &=& \frac{4\pi}{c} \int \frac{q}{2\pi \rho} \delta(\rho) e^{i\frac{\omega}{v}z} \rho\, d\rho\, d\phi = \frac{4\pi q}{c} e^{i\frac{\omega}{v}z}, \nonumber \\
	a B_\phi &=& \frac{2q}{c}.
    \label{AMBC1}
\end{eqnarray}
From  the expression for $B_\phi$ in (\ref{TOTALB}) together with the zero limit  for  $K_1$, we get
\begin{eqnarray}
	a \frac{i c}{v \omega^2} \sum_{\nu=\pm} X_\nu \Omega_\nu \left( \omega^2 - v^2 Q_\nu^2 \right) \frac{1}{Q_\nu a} &=& \frac{2q}{c}
    \label{AMBC2}
\end{eqnarray}
which yields the second required equation 
\begin{equation}
\label{LA}
	i X_+ \Omega_+ \frac{ \omega^2 - v^2 Q_+^2 }{Q_+} + i X_- \Omega_- \frac{ \omega^2 - v^2 Q_-^2 }{Q_-} = \frac{2q v \omega^2}{c^2}.
\end{equation}
The system (\ref{LG})-(\ref{LA}) yields the  
solutions
\begin{equation}
	X_+ = -\frac{i\omega}{c}\, \Gamma \,Q_+, \quad  X_- = \frac{i\omega}{c}\, \Gamma\, Q_-,  \qquad \Gamma={2q}/{\sqrt{\sigma^2+4n^2 \omega^2}}.
    \label{RESCOEFAPP}
\end{equation}
The final expressions for the electric and magnetic fields are obtained by substituting these coefficients in Eqs. (\ref{TOTALE})  and (\ref{TOTALB}) with the results
\begin{eqnarray}
E_\rho&=&\frac{\omega\Gamma}{c}\left( Q_+\Omega_+  K_1(Q_+\rho)-Q_-\Omega_-  K_1(Q_-\rho)\right), \\
	E_\phi&=&-\frac{i\omega\Gamma}{c}\left(Q_+ K_1(Q_+ \rho) - Q_- K_1(Q_- \rho)\right), \\
	E_z&=&-\frac{iv\Gamma}{c}\left(\Omega_+ Q^2_+ K_0(Q_+ \rho)-\Omega_- Q^2_-  K_0(Q_- \rho)\right),\\
	B_\rho&=&\frac{i\omega\Gamma}{v}\left(Q_+ K_1(Q_+ \rho)-Q_- K_1(Q_- \rho)\right), \\
	B_\phi&=&\frac{\Gamma}{v\omega}\left( Q_+ \Omega_+\left( \omega^2  - v^2Q^2_+ \right)K_1(Q_+ \rho)- Q_- \Omega_- \left( \omega^2  - v^2Q^2_- \right)K_1(Q_- \rho)\right), \\
	B_z&=&\Gamma\left( Q^2_+ K_0(Q_+\rho)-  Q^2_- K_0(Q_-\rho)\right).
    \label{TOTALEBAPP}
\end{eqnarray}

\subsection{The electromagnetic potentials}
\label{EMPOT}
To evaluate the local energy flux $\mbf{S}$ we require the electromagnetic potentials $\Phi$ and $\mbb{A}$. Starting from the relation between the electric field and the potentials
$
\mbb{E} = -\bs{\nabla} \Phi - \frac{1}{c} \partial_t \mbb{A},
$
we next compute the scalar potential in the Lorenz gauge 
$
\bs{\nabla} \cdot \mbb{A} = -\frac{\epsilon}{c} \partial_t \Phi. 
$
Imposing Gauss law leads to the following equation for the scalar potential
\begin{equation}
-\nabla^2 \Phi + \frac{\epsilon}{c^2} \partial_t^2 \Phi = \frac{4\pi}{\epsilon} \rho.
\label{LG1}
\end{equation}
Passing to the $(\omega,\mbf{k})$ Fourier space, we obtain
\begin{equation}
\left(k^2 - \frac{\epsilon \omega^2}{c^2} \right) \Phi(k, \omega) = \frac{2 q}{\epsilon}  \delta(\omega - k_z v).
\label{LG2}
\end{equation}
Transforming  back to frequency   space we write

\begin{equation}
\Phi (\rho ,z,\omega )=\frac{2q}{\epsilon }\int \frac{d^{2}k_{\perp }\,dk_{z}%
}{(2\pi )^{3}}\frac{\delta (\omega -k_{z}v)}{\mathbf{k}_{\perp
}^{2}+k_{z}^{2}-\frac{\epsilon \omega ^{2}}{c^{2}}}e^{i\mathbf{k}_{\perp
}\cdot \bs{\rho }}e^{ik_{z}z}= \frac{2q}{\epsilon v}e^{i\frac{\omega}{v}z}\int \frac{d^{2}k_{\perp
}\,}{(2\pi )^{3}}\frac{1}{\mathbf{k}_{\perp }^{2}+\gamma ^{2}}e^{i\mathbf{k}%
_{\perp }\cdot \bs{\rho }},  \label{LG3}
\end{equation}%
where we have integrated over $k_z$ and defined 
\begin{equation}
\gamma ^{2}=\frac{\omega ^{2}}{v^{2}}%
\left( 1-\frac{\epsilon v^{2}}{c^{2}}\right). 
\end{equation}%
This is a real quantity that can be either positive or negative. Before dealing with this issue we can still proceed in general by doing the angular integration in (\ref{LG3}) recalling that
\begin{equation}
J_{0}(p\rho )=\int_{0}^{2\pi }\frac{d\phi }{2\pi }e^{ip\rho \cos \phi }.
\end{equation}%
We obtain
\begin{equation}
\Phi (\rho ,z,\omega )=\frac{2q}{\epsilon v}e^{i\frac{\omega}{v}z}\int_{0}^{\infty }\frac{%
p dp\,}{(2\pi )^{2}}\frac{J_{0}(p \rho )}{p^{2}+\gamma ^{2}},
\label{PHI1}
\end{equation}%
with $p=|\mbf{k}_\perp|$.  Next we manage to enlarge the integration range of $p$ from $-\infty$ to $+\infty$. To this end we consider the following identities involving the Hankel functions $H_0^{(1,2)} (x)$, 
\begin{equation}
J_{0}(x)=\frac{1}{2}\left( H_{0}^{(1)}(x)+H_{0}^{(2)}(x)\right)
,\qquad H_{0}^{(1)}(e^{i\pi }x)=-H_{0}^{(2)}(x),
\end{equation}%
which allows to write $J_{0}$ as 
\begin{equation}
J_{0}(p\rho)=\frac{1}{2}\left( H_{0}^{(1)}(p\rho)-H_{0}^{(1)}(-p\rho)\right). 
\label{J0DESCOMP}
\end{equation}
Substituting back in (\ref{PHI1}) and making the change of variables $p \rightarrow -p$ in the second contribution from  (\ref{J0DESCOMP}) we arrive at 
\begin{equation}
\;\Phi (\rho ,z,\omega )=\frac{q}{\epsilon v}e^{ikz}\int_{-\infty}^{+\infty}\frac{%
p dp\,}{(2\pi )^{2}}\frac{H_{0}^{(1)}(p \rho )}{p^{2}+\gamma ^{2}}.
\label{PHI2}
\end{equation}
Since $H_{0}^{(1)}(z)$ has a logarithmic branch point singularity at $z=0$ it is necessary to determine an adequate integration path. 
Guided by the Sommerfeld identity (See for example Refs. \cite{sommerfeld1949partial,chew1999waves} 
we choose the so called Sommerfeld integration path $\Gamma_S$, depicted in Fig. \ref{path} , which takes proper care of causality.

\begin{figure}[h!]
  \centering
 \includegraphics[scale=0.8]{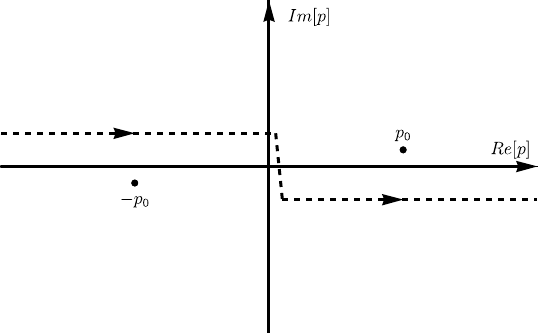} 
  \caption{Sommerfeld integration path $\Gamma_S$.}
  \label{path}
\end{figure}

In addition  we close the integration path in the upper-half plane because  we have 
\begin{equation}
H_{0}^{(1)}(z)\rightarrow \sqrt{\frac{2}{\pi z}}%
e^{i\left( z-\frac{\pi }{4}\right) },
\end{equation}
when $|z|\rightarrow \infty $. Now we are in conditions to deal with the integral (\ref{PHI2}) using Cauchy´s theorem in the circuit determined by $\Gamma_S$, having already made sure that the semicircle at infinity does not contribute.

Next we consider separately the two cases $\gamma^2=s \, \delta^2$, where $s$ is the sing of $\gamma^2$ and $\delta=+\sqrt{ |\gamma^2|} >0 $.

\subsubsection{Case $s=+, \,\,  \gamma^2 >0$}

The integral we have to consider is 
\begin{equation}
I_1=\oint_{\Gamma _{S}}\frac{p dp\,}{(2\pi )^{2}}\frac{H_{0}^{(1)}(p \rho )}{%
p^{2}+ \delta ^{2}},
\end{equation}%
where we introduce the decomposition 
\begin{equation}
\frac{1}{p^{2}+\delta ^{2}}=\frac{1}{2p} \left( \frac{1}{%
p+i\delta }+\frac{1}{p-i\delta }\right) 
\end{equation}%
\qquad 
We identify $p=i\delta$ as the only pole contributing to the Cauchy's theorem 
so that 
\beq
I_1 =\frac{i}{4\pi }\;\;H_{0}^{(1)}(i\delta \rho )
\eeq
This yields the potential
\begin{equation}
\Phi_1 (\rho ,z,\omega )=\frac{i}{%
4\pi }\left( \frac{q}{\epsilon v}e^{i\frac{\omega}{v}z}\right) \;H_{0}^{(1)}(i\delta \rho )= 
\frac{1}{2\pi ^{2}}\left( \frac{q}{\epsilon v}%
e^{i\frac{\omega}{v}z}\right) K_{0}(\delta \rho ),
\label{PHI11}
\end{equation}%
where the last expression is obtained from the identity
\begin{equation}
K_{0}(z)=\frac{i\pi }{2}H_{0}^{(1)}\left( iz\right) ,\;\;\;\;\;\;-\pi <\arg
z\leq \frac{\pi }{2}.
\end{equation}%
We observe that the result (\ref{PHI11}) coincides with a direct calculation of this integral which proceeds straightforwardly due to the absence of poles in the real axis $p$.

\subsubsection{ Case $ s= -, \,\,  \gamma^2 < 0$}

In this case, the integral to be considered is 
\begin{equation}
I_2=\oint_{\Gamma _{S}}\frac{%
p dp\,}{(2\pi )^{2}}\frac{H_{0}^{(1)}(p\rho )}{p^{2}-\delta^{2}}
\end{equation}
and we proceed in complete analogy to the previous case. The decomposition now yields 
\begin{equation}
\frac{1}{p^{2}-\delta^{2}}=\frac{1}{2p}\left( \frac{1}{p+\delta}+\frac{1}{%
p-\delta}\right) 
\end{equation}%
which singles out the only contribution of the pole at $p=+ \, \delta$. Now, the result for the potential is
\begin{eqnarray}
\Phi_2 (\rho ,z,\omega ) &=&\frac{i}{4\pi}\left(\frac{q}{%
\epsilon v}e^{i\frac{\omega}{v}z}\right)\; H_{0}^{(1)}(\delta \rho )= \frac{1\,}{2\pi ^{2}}\left(\frac{q}{\epsilon v}%
e^{i\frac{\omega}{v}z}\right)K_{0}(-i\delta \rho )
\label{PHI22}
\end{eqnarray}%
Expressions (\ref{PHI11}) and (\ref{PHI22})  can be unified such that the  potential for arbitrary real $\gamma^2$ reads
\begin{equation}
\Phi (\rho ,z,\omega )=\frac{q}{\epsilon v}e^{i\frac{\omega }{v}%
z}\oint_{\Gamma _{S}}\frac{kdk\,}{(2\pi )^{2}}\frac{H_{0}^{(1)}(k\rho )}{%
k^{2}+\gamma ^{2}}=\frac{1}{2\pi ^{2}}\left( \frac{q}{\epsilon v}%
e^{i\frac{\omega}{v}z}\right) K_{0}\left( \gamma _{s}\rho \right) ,
\label{FINPOTAUX}
\end{equation}%
with the notation
\begin{equation}
\gamma _{s}=s^{3/2}\sqrt{|\gamma ^{2}|}, \qquad e^{i\pi}=-1.
\end{equation}%

\subsection{The vector potential}

Knowing the scalar potential and the components of the electric field, we can determine the components of the vector potential through the relation 
$ \tilde{\mbb{A}} =-i\frac{c}{\omega}(\tilde{\mbb{E}}+\bs{\nabla} \Phi ).  
$
We obtain
\barr
&&A_\rho = -\frac{i c}{\omega} E_\rho + \frac{i c q \gamma_s }{2 \epsilon \omega v \pi^2} K_1(\gamma_s \rho),\label{ARHO1} \qquad
A_\phi = -\frac{i c}{\omega} E_\phi, \label{APHI1}
\qquad A_z = -\frac{i c}{\omega} E_z + \frac{c k q}{2 \epsilon \omega v \pi^2} K_0(\gamma_s \rho), \quad k=\frac{\omega}{v}.
\label{AZ1}
\label{ALGAPP}
\earr
To simplify the expressions for the electromagnetic potentials, we perform a  gauge transformation with parameter
\begin{equation}
\Lambda = \frac{i q c}{2 \epsilon \pi^2 v \omega} K_0(\gamma_s\rho)= \frac{ic}{\omega} \Phi,
\label{GT1APP}
\end{equation}
which is designed to obtain $\Phi'=0$. This  as can be immediately verified via the standard 
transformations 
\begin{eqnarray}
\Phi' &=& \Phi - \frac{1}{c} \partial_t \Lambda,\qquad  A'_i = A_i + \partial_i \Lambda.
\label{GT2APP}
\end{eqnarray}
The remaining components of the vector potential reduce to 
\begin{eqnarray}
A'_\rho &=& A_\rho + \partial_\rho \Lambda = -\frac{i c}{\omega} E_\rho, \qquad 
A'_\phi = A_\phi= -\frac{ic}{\omega} E_\phi, \qquad 
A'_z = A_z + \partial_z \Lambda = -\frac{i c}{\omega} E_z.
\label{GT3APP}
\end{eqnarray}
Summarizing, in the new gauge we have $\Phi'=0$
and $A'_i=-\frac{ic}{\omega}E_i$.
Therefore   $\partial_i A'_i = 0$ outside the sources ($\rho >0$), as adequate for the radiation regime.  We recognize that now we are  working in the Coulomb gauge.
For simplicity, we drop the primes in the potentials and express them in terms of the previously obtained solutions for the electromagnetic fields. The final results are  
\begin{eqnarray}
A_\rho &=& -i\Gamma \sum_{\nu=\pm } \nu \, Q_\nu \Omega_\nu K_1(Q_\nu \rho), \qquad 
A_\phi = -\Gamma  \sum_{\nu=\pm } \nu \, Q_\nu K_1(Q_\nu \rho), \nonumber \\
A_z &=& - \frac{v\Gamma}{\omega} \sum_{\nu=\pm } \nu \, \Omega_\nu Q^2_\nu K_0(Q_\nu \rho), \qquad 
\Phi = 0.
\label{TOTALA}
\end{eqnarray}
Let us emphasize that the scalar potential (\ref{FINPOTAUX}) ultimately  plays no role in the calculation, but we have presented its detailed  derivation to show that the choice of the Coulomb gauge rest  on a firm basis.

\subsection{The polarization of each mode}

\label{POL}

Let us start by recalling 
the components of the electric field for each polarization $\nu$ 
\beq
    E_{\rho,\nu}  = \nu \frac{\omega \Gamma}{c} Q_\nu \Omega_\nu K_1(Q_\nu \rho), \qquad 
    E_{\phi,\nu}  = -i\nu \frac{\omega \Gamma}{c} Q_\nu  K_1(Q_\nu \rho), \qquad 
    E_{z,\nu}  = -i \nu \frac{\omega \Gamma}{ck} Q^2_\nu \Omega_\nu K_0(Q_\nu \rho),
\eeq
where we are not writing the factor $e^{ikz}$ which goes together with each component. Here we have
\begin{align}
    \Gamma & = \frac{2q}{\sqrt{\sigma^2 + 4 n^2 \omega^2}}, & \Omega_\nu &= \frac{k\sigma}{c(Q_\nu^2-\gamma^2)}, & \gamma^2 &= k^2\Big(1 - \frac{\epsilon v^2}{c^2}\Big).
\end{align}
In the asymptotic region \( \rho \to \infty \), the modified Bessel functions behave as
\begin{equation}
    K_{0,1}(Q_\nu \rho) \sim \sqrt{\frac{\pi}{2 Q_\nu \rho}}\, e^{-Q_\nu \rho}.
\end{equation}
For causal propagation, we have chosen  $ Q_\nu = -i \mathcal{Q}_\nu $, with $ \mathcal{Q}_\nu > 0 $ real.  In cylindrical coordinates  the asymptotic electric field then reads
\begin{equation}
\tilde{\mbb{E}}_\nu(\mathbf{x},\omega)= \nu \frac{\omega \Gamma }{c}\sqrt{\frac{i\pi \mathcal{Q}_{\nu }}{%
2\rho }}\,\left( \hat{\bs{\phi}}-i \frac{k\sigma}{c(%
\mathcal{Q}_{\nu }^{2}+\gamma ^{2})  
} \mathbf{P}_\nu  \right)\exp{i \mathcal{Q}_\nu \rho + i k z} 
\label{ERADPOL}
\end{equation}
where we introduce the transverse vector
\begin{equation}
    \mathbf{P}_\nu = \hat{\boldsymbol{\rho}} - \frac{\mathcal{Q}_\nu}{k} \hat{\mathbf{z}},\qquad 
    \mathbf{P}_\nu =  \frac{\sqrt{\mathcal{Q}_{\nu }^{2}+k^{2}}}{k} \, \hat{\mathbf{P}}_\nu,
    \label{TRANSV}
\end{equation}
which satisfies \( \mathbf{K}_\nu \cdot \mathbf{P}_\nu = 0 \), where the propagation vector is \( \mathbf{K}_\nu = \mathcal{Q}_\nu \hat{\boldsymbol{\rho}} + k \hat{\mathbf{z}} \). 
Then we have 
\begin{equation}
\tilde{\mbb{E}}_\nu(\mathbf{x},\omega)= \nu \frac{\omega \Gamma }{c}\sqrt{\frac{i\pi \mathcal{Q}_{\nu }}{%
2\rho }} \left( \hat{\bs{\phi}}-i \frac{k\sigma}{c(%
\mathcal{Q}_{\nu }^{2}+\gamma ^{2})  
} \frac{\sqrt{\mathcal{Q}_{\nu }^{2}+k^{2}}}{k} \, 
\hat{\mathbf{P}}_\nu  \right)\exp{i \mathcal{Q}_\nu \rho + i k z}. 
\label{ERADPOL1}
\end{equation}
From Eq.(\ref{POSDEF1}) together with the definition of $\gamma$ we readily obtain 
\begin{equation}
(\mathcal{Q}_{\nu }^{2}+\gamma ^{2})=\frac{\sigma }{2c^{2}}\left( \sigma
-\nu \sqrt{\sigma ^{2}+4n^{2}\omega ^{2}}\right). 
\label{Q2PGAMMA2}
\end{equation}%
Recalling  Eq. (\ref{Q2Pk2}) we have 
\beq
\sqrt{{\cal Q}^2_\nu + k^2}=-\frac{\nu }{2c}\left( \sigma
-\nu \sqrt{\sigma ^{2}+4n^{2}\omega ^{2}}\right) ,
\label{Q2Pk2APP}
\eeq
which yields
\begin{equation}
\frac{k\sigma }{c(\mathcal{Q}_{\nu }^{2}+\gamma ^{2})}\frac{\sqrt{\mathcal{Q}%
_{\nu }^{2}+k^{2}}}{k}=-\nu 
\label{IMPRELPOL}
\end{equation}
Finally,  
the electric field becomes 
\begin{equation}
\tilde{\mbb{E}}_\nu(\mathbf{x},\omega)= i\frac{\omega \Gamma }{c}\sqrt{\frac{i\pi \mathcal{Q}_{\nu }}{%
2\rho }}\left( \hat{\mathbf{P}}_\nu -i \nu \hat{\bs{\phi}}   \right)\exp{i \mathcal{Q}_\nu \rho + i k z} 
\label{POLCOND}
\end{equation}
We verify that 
\beq
\hat{\mathbf{P}}_{\nu } \times \hat{\boldsymbol{\phi}}=
\hat{\mathbf{K}}_{\nu },
\eeq
such that  $\hat{\mathbf{P}}_{\nu }, \,\,  \hat{\boldsymbol{\phi}}, \,\,
\hat{\mathbf{K}}_{\nu }$ constitutes a right handed orthogonal set of unit vectors. The plane $(\hat{\mathbf{P}}_{\nu }$, $\hat{\boldsymbol{\phi}})$ is perpendicular to the propagation vector so that Eq. (\ref{POLCOND}) indicates  
that  each mode $\nu =\pm 1$ corresponds to a
circularly polarized electric field.
Following the conventions of
\cite{Jackson:1998nia} we identify the mode  $\nu=+ $ as having
 negative helicity, which  corresponds to a right circular polarization in the optics convention. Conversely, the mode   
$\nu=- $ is identified as having positive helicity, being described as left circular polarization in the optics convention. 
\section{ Calculation of the spectral distribution of the total radiated energy per unit length $\cal{E}$}
\label{Ap-En}
 We start from the expression (\ref{FINALSED}) for the spectral energy distribution
\begin{equation}
	\mathcal{E} =  \lim_{\rho \to \infty} \frac{\rho}{2\pi} \, \mathrm{Re} \left[ c E^*_\phi B_z - c E^*_z B_\phi + \frac{\sigma}{2}( E^*_z A_\phi  - E^*_\phi A_z ) \right],
	\label{CALE1}
\end{equation}
and 
calculate the far-field limit $\mathcal{Q}_\nu \rho \rightarrow \infty$ of the required fields using the asymptotic behavior
\begin{equation}
	K_{0,1}(-i \mathcal{Q}_\nu \rho) \sim \sqrt{\frac{\pi}{-2 i \mathcal{Q}_\nu \rho}} e^{i \mathcal{Q}_\nu \rho},
\end{equation}
with ${\cal Q}_\nu$  real and positive. 
Due to the trivial dependence of the fields in 
$e^{ikz}$,  the $z$-dependence cancels out and in Eq. (\ref{CALE1}) an we  just write the $\rho$ - dependent factor for simplicity
\begin{align}
	E^*_\phi &= -\frac{ \omega \Gamma}{c}  \sqrt{\frac{\pi}{{2i}\rho}}\left(  \sqrt{\mathcal{Q}_{+}} e^{i \mathcal{Q}_{+} \rho} -  \sqrt{\mathcal{Q}_{-}} e^{i \mathcal{Q}_{-} \rho} \right), \label{EPHICCRAD}\\
	E^*_z &= -\frac{i v \Gamma}{c} \sqrt{\frac{\pi}{{2i}\rho}} \left( \Omega_+ \mathcal{Q}_+ \sqrt{\mathcal{Q}_+} e^{i \mathcal{Q}_+ \rho} - \Omega_- \mathcal{Q}_- \sqrt{\mathcal{Q}_-} e^{i \mathcal{Q}_- \rho} \right), \label{EZCCRAD}\\
	B_\phi &= -\frac{i \Gamma}{v \omega} \sqrt{\frac{\pi}{{-2i}\rho}} \left(  \Omega_+ (\omega^2 + v^2 \mathcal{Q}_+^2) \sqrt{\mathcal{Q}_+} e^{-i \mathcal{Q}_+ \rho} -  \Omega_- (\omega^2 + v^2 \mathcal{Q}_-^2) \sqrt{\mathcal{Q}_-} e^{-i \mathcal{Q}_- \rho} \right),\label{BPHIRAD} \\
	B_z &= -\Gamma 
	\sqrt{\frac{\pi}{{-2i} \rho}}\left( \mathcal{Q}_+ \sqrt{\mathcal{Q}_+} e^{-i \mathcal{Q}_+ \rho} - \mathcal{Q}_- \sqrt{\mathcal{Q}_-} e^{-i \mathcal{Q}_- \rho} \right),\label{BZRAD} \\
	A_\phi &= i \Gamma  \sqrt{\frac{\pi}{{-2i} \rho}}\left(  \sqrt{\mathcal{Q}_+} e^{-i \mathcal{Q}_+ \rho} -  \sqrt{\mathcal{Q}_-} e^{-i \mathcal{Q}_- \rho} \right), \label{APHIRAD}\\
	A_z &= \frac{v \Gamma}{\omega} \sqrt{\frac{\pi}{{-2i} \rho}} \left( \Omega_+ \mathcal{Q}_+ \sqrt{\mathcal{Q}_+} e^{-i \mathcal{Q}_+ \rho} - \Omega_- \mathcal{Q}_- \sqrt{\mathcal{Q}_-} e^{-i \mathcal{Q}_- \rho} \right).
	\label{AZRAD}
\end{align}
Each term in (\ref{CALE1}) contains products exponentials of the form $e^{\pm i \mathcal{Q}_\lambda \rho}$. When the modes coincide, these contributions combine to one, while  for different modes, the product yields  a factor $e^{\pm i(\mathcal{Q}_1 - \mathcal{Q}_2)\rho}$. Moreover, calculating the required products reveals that the only surviving complex factors are in the exponentials. In this way, when taking the real part we find that all mixed terms appear with a cosine factor. The resulting products in Eq. (\ref{CALE1}) are 
\begin{align}
\mathrm{Re}[c E^*_\phi B_z] &= {\frac{\pi \Gamma^2}{2\rho }} \,\, \omega \left( \mathcal{Q}_+^2 + \mathcal{Q}_-^2 -  \sqrt{\mathcal{Q}_+ \mathcal{Q}_-} (\mathcal{Q}_+ + \mathcal{Q}_-) \cos[(\mathcal{Q}_+ - \mathcal{Q}_-) \rho] \right), \label{PROD1}\\
\mathrm{Re}[-c E^*_z B_\phi] &= {\frac{\pi\Gamma^2 }{2\rho }} \,\,  \frac{1}{\omega} \left(  \frac{}{}\Omega_+^2 \mathcal{Q}_+^2 (\omega^2 + v^2 \mathcal{Q}_+^2) + \Omega_-^2 \mathcal{Q}_-^2 (\omega^2 + v^2 \mathcal{Q}_-^2) \right. \nonumber \\
	& \left. -  \Omega_+ \Omega_- \sqrt{\mathcal{Q}_+ \mathcal{Q}_-} (\mathcal{Q}_+ (\omega^2 + v^2 \mathcal{Q}_-^2) + \mathcal{Q}_- (\omega^2 + v^2 \mathcal{Q}_+^2)) \cos[(\mathcal{Q}_+ - \mathcal{Q}_-)\rho] \frac{}{} \right), \label{PROD2} \\	
\mathrm{Re}\left[\frac{\sigma}{2} A_\phi E^*_z\right]  &= \frac{\pi \Gamma^2}{2\rho} \frac{\sigma v }{2 c} \left( \Omega_+ \mathcal{Q}_+^2 + \Omega_- \mathcal{Q}_-^2 -  \sqrt{\mathcal{Q}_+ \mathcal{Q}_-} (\Omega_+ \mathcal{Q}_+ + \Omega_- \mathcal{Q}_-) \cos[(\mathcal{Q}_+ - \mathcal{Q}_-) \rho] \right), \label{PROD3}\\
\mathrm{Re}\left[-\frac{\sigma}{2} A_z E^*_\phi\right] &= \frac{\pi \Gamma^2}{2\rho} \frac{\sigma v }{2 c} \left( \Omega_+ \mathcal{Q}_+^2 + \Omega_- \mathcal{Q}_-^2 -  \sqrt{\mathcal{Q}_+ \mathcal{Q}_-} (\Omega_+ \mathcal{Q}_+ + \Omega_- \mathcal{Q}_-) \cos[(\mathcal{Q}_+ - \mathcal{Q}_-) \rho] \right)
\label{PROD4}
\end{align}
Let us observe that all the products are symmetric under the interchange $(+ \leftrightarrow -)$.
 We observe that the equality between the equations   
(\ref{PROD3}) and (\ref{PROD4}) is a direct consequence of our choice of potentials in the Coulomb gauge. In fact we have taken  $A_\phi=-\frac{ic}{\omega} E_\phi $ and   
$A_z=-\frac{ic}{\omega} E_z $. Substituting the electromagnetic potentials in the corresponding left-hand sides, we have 
\beq
\mathrm{Re}\left[\frac{\sigma}{2} A_\phi E^*_z\right]
= -\mathrm{Re}\left[\frac{\sigma}{2}\frac{ic}{\omega} E_\phi E^*_z\right], \qquad
\mathrm{Re}\left[-\frac{\sigma}{2} A_z E^*_\phi\right]= \mathrm{Re}\left[\frac{\sigma}{2} \frac{ic}{\omega}E_z E^*_\phi\right].
\label{COMPEQ}
\eeq
The equality arises by recalling that $\mathrm{Re}(z)
=\mathrm{Re}(z^*)$ in the right-hand side of the second equation in (\ref{COMPEQ}), which reproduces the left-hand side of the first equation.

Taking the sum of Eqs. (\ref{PROD1}) to (\ref{PROD4}) yields ${\cal E}= {\cal E}_+ +{\cal E}_- +{\cal E}_{(+,-)}$ from Eq. (\ref{CALE1}). Isolating the term proportional to  $\cos[(\mathcal{Q}_+ - \mathcal{Q}_-) \rho]$ we can read each of the polarization contributions ${\cal E}_\nu$ together with the interference term ${\cal E}_{(+,-)}$. The results are 
\barr
\hspace{-1cm} {\cal E}_{(+,-)}&=& - {\frac{\Gamma^2}{4 }}   \sqrt{\mathcal{Q}_+ \mathcal{Q}_-} \cos[(\mathcal{Q}_+ - \mathcal{Q}_-) \rho] \times \nonumber \\
&&\Big[{\omega }(\mathcal{Q}_+ + \mathcal{Q}_-) + \frac{1}{\omega }  \Omega_+ \Omega_-  \Big(\mathcal{Q}_+ (\omega^2 + v^2 \mathcal{Q}_-^2) + \mathcal{Q}_- (\omega^2 + v^2 \mathcal{Q}_+^2)\Big) + 
\sigma {\frac{v}{c} }  \, (\Omega_+ \mathcal{Q}_+ + \Omega_- \mathcal{Q}_-)\Big], \label{MIXEDCALE}\\
\hspace{-1cm}{\cal E}_\nu &=& {\frac{\Gamma^2}{4 }} \Big( {\omega} \mathcal{Q}_\nu^2  + \frac{1}{\omega }\Omega_\nu^2 \mathcal{Q}_\nu^2 (\omega^2 + v^2 \mathcal{Q}_\nu^2) +  v\sigma \Omega_\nu \mathcal{Q}_\nu^2 \Big),
\label{INDCALE}
\earr
respectively.  The next step is to simplify the expressions 
(\ref{MIXEDCALE}) and (\ref{INDCALE}) substituting the explicit expressions for $\Gamma, \Omega_\nu,   {\cal Q}_\nu$   
which we recall again at this stage
\barr
&&\Gamma= \frac{2q}{\sqrt{\sigma^2+4n^2 \omega^2}} ,\qquad \Omega_\nu=- \frac{k \sigma}{({\cal Q}_\nu^2+ \gamma^2)\, c} , \qquad \gamma^2=k^2 \Big( 1-\frac{n^2 v^2}{c^2}\Big), \\
&& {\cal Q}^2_\nu= \Big(\frac{n^2 \omega^2}{c^2}+\frac{\sigma^2}{2c^2}-\nu \frac{\sigma}{2c^2} \sqrt{\sigma^2+4n^2 \omega^2}\Big) -\frac{\omega^2}{v^2} \geq 0.	 
\earr
Let us collect some additional relations steming from the above definitions  which will be useful in the following.         

We  observe  that the round bracket in ${\cal Q}_\nu^2$ is a perfect square
\begin{eqnarray}
\Big(\frac{n^2 \omega^2}{c^2}+\frac{\sigma^2}{2c^2}-\nu \frac{\sigma}{2c^2} 
\sqrt{\sigma^2+4n^2 \omega^2}\Big)&&=\frac{1}{4c^2} \Big( \sqrt{\sigma^2+4n^2 \omega^2}- \nu \sigma \Big)^2,
\label{ROUNDBRACKET}
\end{eqnarray}
as  can be immediately verified. Recalling  $k=\omega/v$ we obtain  
\begin{eqnarray}
(\omega^2+ v^2 \mathcal{Q}^2_\nu)= v^2\Big(\mathcal{Q}^2_\nu+ k^2 \Big)= 
\frac{v^2}{4c^2} \Big( \sqrt{\sigma^2+4n^2 \omega^2}- \nu \sigma \Big)^2.
\label{COMB11}
\end{eqnarray}
Next we consider the combination $({\cal Q}_\nu^2+\gamma^2)$ which simplifies to 
\begin{eqnarray}
\mathcal{Q}^2_\nu+ \gamma^2 &&= \frac{\sigma}{2c^2} \Big( \sigma-\nu \sqrt{%
\sigma^2+4n^2 \omega^2} \Big).
\label{COMB12}
\end{eqnarray}
This provides  a direct calculation of the product  
\begin{eqnarray}
(\mathcal{Q}^2_++ \gamma^2)(\mathcal{Q}^2_- + \gamma^2)&&= -\frac{\sigma^2
n^2 \omega^2}{c^4},
\label{PROD11}
\end{eqnarray}
together with
\begin{eqnarray}
\Omega_+ \Omega_-=  -\frac{c^2}{n^2 v^2}
\label{PROD21}
\end{eqnarray}
Inserting the required expressions into Eq. (\ref{MIXEDCALE}) and using MATHEMATICA 
 yields the result ${\cal E}_{(+,-)}=0$ ,
 indicating that no mixing occurs between the polarization modes $\nu$. This cancellation is not due to averaging on fast oscillations, but arises from the exact algebraic structure. Consequently, the spectral distribution of the total radiated energy per unit length  separates cleanly into the contributions from each individual mode
\begin{equation}
	{\cal E} = {\cal E}_+ + {\cal E}_-.
\end{equation}
Since this result is of great relevance  for the correct interpretation of our results we provide  an algebraic proof of ${\cal E}_{(+,-)}=0$. We start from Eq. (\ref{MIXEDCALE}) and focus on the square bracket, which  we rewrite as the sum of two factors $F_\pm $  of ${\cal Q}_\pm/\omega $
\begin{eqnarray}
F_+ \frac{{\cal Q}_+}{\omega}+ F_- \frac{{\cal Q}_-}{\omega}\equiv && \Big[\omega^2 + \Omega_+ \Omega_- \Big( \omega^2 + v^2 \mathcal{Q}_-^2 %
\Big) + {\sigma \frac{v}{c} \omega } \, \Omega_+ \Big]\frac{\mathcal{Q}_+}{\omega} \nonumber \\ &&+ \Big[\omega^2 + \Omega_+ \Omega_- \Big( \omega^2 + v^2 \mathcal{Q}_+^2 %
\Big) + {\sigma \frac{v}{c} \omega } \, \Omega_- \Big]\frac{\mathcal{Q}_-}{\omega}. 
\label{CALC1APP}
\end{eqnarray}
Next we separately calculate each coefficient $F_\pm$. Let us start  with $F_+$ and  let us follow the indicated steps, where we use most of the previous definitions and relations between Eqs. (\ref{ROUNDBRACKET}) and Eq. (\ref{PROD21})
\begin{eqnarray}
F_+&&=\Big[\omega^2 + \Omega_+ \Omega_- \Big( \omega^2 + v^2 \mathcal{Q}_-^2 %
\Big) + {\sigma \frac{v}{c} \omega } \, \Omega_+ \Big], \nonumber \\
&&= \omega^2 + \Big(-\frac{c^2}{n^2 v^2} \Big) \Big( \frac{v^2}{4c^2} \Big( 
\sqrt{\sigma^2+4n^2 \omega^2}+ \sigma \Big)^2\Big) + {\sigma \frac{v}{c} \omega } \, %
\Big(- \frac{k \sigma}{(\mathcal{Q}_+^2 + \gamma^2)c} \Big), \nonumber \\
&&= \omega^2 - \frac{1}{4 n^2 } \Big( \sqrt{\sigma^2+4n^2 \omega^2}+ \sigma %
\Big)^2 - \frac{\sigma^2 \omega^2}{c^2} \,\frac{1}{(\mathcal{Q}_+^2 + \gamma^2)}, \nonumber \\
&&= \omega^2 - \frac{1}{4 n^2 } \Big( \sqrt{\sigma^2+4n^2 \omega^2}+ \sigma %
\Big)^2 - {2 \sigma c^2 \omega^2} \,\frac{1}{\Big( \sigma- \sqrt{%
\sigma^2+4n^2 \omega^2} \Big)},\nonumber \\
&&= \omega^2 - \frac{1}{4 n^2 } \Big( \sqrt{\sigma^2+4n^2 \omega^2}+ \sigma %
\Big)^2 - \frac{2 \sigma c^2 \omega^2}{c^2} \,\frac{1}{\Big( \sigma- \sqrt{%
\sigma^2+4n^2 \omega^2} \Big)} \frac{\Big(\sigma+ \sqrt{\sigma^2+4n^2
\omega^2}\Big)}{\Big(\sigma+ \sqrt{\sigma^2+4n^2 \omega^2}\Big)}, \nonumber  \\
&&= \omega^2 - \frac{1}{4 n^2 } \Big( \sqrt{\sigma^2+4n^2 \omega^2}+ \sigma %
\Big)^2 - {2 \sigma  \omega^2} \,\frac{\Big(\sigma+ \sqrt{\sigma^2+4n^2
\omega^2}\Big)}{\Big( -4n^2 \omega^2 \Big)}, \nonumber \\
&& = \omega^2 - \frac{1}{4 n^2 } \Big( \sqrt{\sigma^2+4n^2 \omega^2}+
\sigma \Big)^2 + \,\frac{\sigma^2}{2n^2}+ \frac{\sigma}{2n^2}\sqrt{%
\sigma^2+4n^2 \omega^2}.
\label{CALC2APP}
\end{eqnarray}
Expanding the square in the last equation we obtain $F_+=0$. The analogous calculation shows $F_-=0$, thus yielding the final result ${\cal E}_{(+,-)}=0 $. 
An analogous calculation for Eq. (\ref{INDCALE}) produces
\begin{equation}
	{\cal E}_\nu = \frac{ q^2 \omega}{2c^2}\left( 1-\frac{1}{n^2 \beta^2}-\nu\frac{\sigma}{\sqrt{\sigma^2+4n^2 \omega^2}}\left(1+\frac{1}{n^2 \beta^2}\right)\right),
	\label{ELAMBDA5}
\end{equation}
where $\beta=v/c$ is the ratio of the velocity of the moving  charge over the velocity of the light in vacuum. 

Finally we rewrite Eq. (\ref{ELAMBDA5}) such that the positivity of ${\cal E}_\nu$ is evident.
On one hand we consider the following chain of steps starting from (\ref{ELAMBDA5})
\begin{eqnarray}
	{\cal E}_\nu 
	&=&\frac{q^2 \omega}{2c^2} \left( 1-\nu\frac{\sigma}{\sqrt{\sigma^2+4n^2 \omega^2}}-\frac{1}{n^2 \beta^2}\left(1+\nu\frac{\sigma}{\sqrt{\sigma^2+4n^2 \omega^2}}\right) \right), \nonumber \\
	&=&\frac{  q^2 \omega}{2c^2} \left( 1-\nu\frac{\sigma}{\sqrt{\sigma^2+4n^2 \omega^2}}\right)\left[1-\frac{1}{n^2 \beta^2}\frac{\left(1+\nu\frac{\sigma}{\sqrt{\sigma^2+4n^2 \omega^2}}\right)}{ \left(1-\nu\frac{\sigma}{\sqrt{\sigma^2+4n^2 \omega^2}}\right)}\right], \nonumber \\
		&=&\frac{ q^2 \omega}{2c^2} \left( 1-\nu\frac{\sigma}{\sqrt{\sigma^2+4n^2 \omega^2}}\right)\left[1-\frac{1}{n^2 \beta^2}\frac{\left(\sqrt{\sigma^2+4n^2 \omega^2}+\nu\sigma\right)}{ \left(\sqrt{\sigma^2+4n^2 \omega^2}-\nu\sigma\right)}\frac{\left(\sqrt{\sigma^2+4n^2 \omega^2}-\nu\sigma\right)}{ \left(\sqrt{\sigma^2+4n^2 \omega^2}-\nu\sigma\right)}\right], \nonumber \\
		&=& \frac{  q^2 \omega}{2 c^2} \left( 1-\nu\frac{\sigma}{\sqrt{\sigma^2+4n^2 \omega^2}}\right)\left[1-\frac{q}{n^2 \beta^2}\frac{4n^2 \omega^2 }{ \left(\sqrt{\sigma^2+4n^2 \omega^2}-\nu\sigma\right)^2}\right], \nonumber \\
	&=&\frac{  q^2 \omega}{2 c^2} \left( 1-\nu\frac{\sigma}{2n \omega\sqrt{\frac{\sigma^2}{4n^2 \omega^2}+1}}\right)\left[1-\frac{1}{n^2 \beta^2}\frac{1}{ \left(\sqrt{\frac{\sigma^2}{4n^2 \omega^2}+1}-\nu\frac{\sigma}{2n\omega}\right)^2}\right].
    \label{CALC3APP}
\end{eqnarray}
Recalling the definition  $\Sigma = \frac{\sigma}{n \omega}$, we rewrite
\begin{equation}
{\cal E}_\nu	=\frac{ q^2 \omega}{2 c^2} \left( 1-\nu\frac{\Sigma/2}{\sqrt{\frac{\Sigma^2}{4}+1}}\right)\left[1-\frac{1}{n^2 \beta^2}\frac{1}{ \left(\sqrt{\frac{\Sigma^2}{4}+1}-\nu\frac{\Sigma}{2}\right)^2}\right].
\label{CALC4APP}
\end{equation}
On the other hand, from Eqs. (\ref{MODK}) and (\ref{THETAC}) we recall the expression
\beq
	\cos \Theta_\nu =\frac{2\omega}{\beta \left( \sqrt{\sigma^2+4 n^2 \omega^2}-\nu \sigma \right)} =
	\frac{1}{ n  \beta}\frac{1}{\left( \sqrt{\frac{\Sigma^2}{4 }+1}-\nu\frac{\Sigma}{2} \right) },
    \label{RELCOSTHETAAPP}
\eeq
in terms of which we can replace the square bracket in Eq. (\ref{CALC4APP}) by  $1-\cos^2 \Theta_\nu= \sin^2 \Theta_\nu$ yielding the final result 
\begin{equation}
	{\cal E}_\nu=\frac{  q^2 \omega}{2c^2} \left( 1-\nu \frac{\Sigma/2}{\sqrt{\frac{\Sigma^2}{4}+1}}\right)\, \sin^2\Theta_\nu.
	\label{FINALCALEAPP}
\end{equation}
The numerator in the round bracket of Eq. (\ref{FINALCALEAPP}) is of the form $(\sqrt{1+ z^2}-\nu z)$ , with $z=\Sigma/2$  real, which is positive for all $z$. Then, it becomes evident that ${\cal E}_\nu$ is always positive  in spite the doubts  sparkling  from the initial expressions (\ref{CONS}).
\section{Expansion of the fields in powers of  $\sigma$}
\label{Ap-Exp}
\subsection{General expansion  }
\label{Ap-Exp1}
We present the expansion to order $\sigma^2$ of our general expressions for the electric and magnetic fields in each polarization $\nu$. Taking
\beq
\lambda^2=\Big( \frac{\omega^2}{v^2} - \frac{\epsilon \omega^2}{c^2} \Big)
\label{PARAMEXPBESSEL}
\eeq
we obtain 
\begin{eqnarray}
B_{\rho \nu} &=& i \nu \frac{q\,\omega}{v\sqrt{\epsilon\,\omega^{2}}}\,
\lambda\;
K_{1}\!\left(\rho\,\lambda\right)\nonumber\\
&& - \frac{i\,q\,\rho\,\omega}{2c^{2}v}\,
K_{0}\!\left(\,\rho\,\lambda\right)\,\sigma \nonumber\\
&& + i \nu  \frac{q\,\omega^{3}}{8c^{4}v^{3}(\epsilon\,\omega^{2})^{3/2}}
\left[-
\frac{c^{4}+3c^{2}v^{2}\epsilon - v^{2}\epsilon^{2}\rho^{2}\omega^{2}}
{\lambda}\,
K_{1}\!\left(\rho\,\lambda\right)
+ 2\,c^{2}v^{2}\epsilon\,\rho\,K_{2}\!\left(\rho\,\lambda\right)
\right]\,\sigma^{2},
\label{BRHOS2}
\earr
\barr
B_{\phi \nu} &=& \,\frac{ q}{c}\,\lambda \;
	K_{1}\!\left(\rho\,\lambda \right) \nonumber\\
	&& + \nu \frac{q}{2c^{3}}\sqrt{\epsilon\,\omega^{2}}\,
	\left[
	-\frac{(c^{2}-3v^{2}\epsilon)}{\epsilon\,v^{2}\,\lambda}\,
	K_{1}\!\left(\rho\,\lambda\right)
	- \rho\,K_{2}\!\left(\rho\,\lambda\right)
	\right]\,\sigma \nonumber\\
	&& + \frac{q}{8c^{5}}\,
	\left[
	-\frac{(8c^{2}-\epsilon\,\rho^{2}\omega^{2})}{\lambda}\,
	K_{1}\!\left(\rho\,\lambda \right)
	+ 4c^{2}\rho\,K_{2}\!\left(\rho\,\lambda\right)
	\right]\,\sigma^{2},
    \label{BPHIS2}
\earr
\barr
	B_{z \nu} &=& \nu \frac{q\,(c^{2}-v^{2}\epsilon)\,\sqrt{\epsilon\,\omega^{2}}}
	{c^{2}v^{2}\epsilon}\,
	K_{0}\!\left(\rho\,\lambda\right) \nonumber\\
	&& + \frac{q}{2c^{2}}\left[
	2\,K_{0}\!\left(\rho\,\lambda \right)
	-\rho\,\lambda \;
	K_{1}\!\left(\rho\,\lambda\right)
	\right]\,\sigma \nonumber\\
	&& + \nu \frac{q}{8c^{4}v^{2}\sqrt{\epsilon\,\omega^{2}}}
	\left[
	\left(-3c^{2}v^{2} - c^{4}/\epsilon + v^{2}\epsilon\,\rho^{2}\omega^{2}\right)\,
	K_{0}\!\left(\rho\,\lambda \right)
	+ \frac{2\,(c^{2}-2v^{2}\epsilon)\,\rho\,\omega^{2}}
	{\lambda}\,
	K_{1}\!\left(\rho\,\lambda \right)
	\right]\,\sigma^{2},
     \label{BZS2}
\earr
\barr
	E_{\rho \nu} &=& \frac{ q}{v\,\epsilon}\,\lambda \;
	K_{1}\!\left(\rho\,\lambda \right) \nonumber\\
	&& + \nu\frac{q}{2c^{2}v^{3}\epsilon^{2}}\sqrt{\epsilon\,\omega^{2}}\,
	\left[
	\frac{\,(c^{2}+v^{2}\epsilon)}{\lambda}\,
	K_{1}\!\left(\rho\,\lambda\right)
	- v^{2}\epsilon\,\rho\,K_{2}\!\left(\rho\,\lambda \right)
	\right]\,\sigma \nonumber\\
	&& +  \frac{\,q\,\rho^{2}\omega^{2}}{8c^{4}v\,\lambda }\,
	K_{1}\!\left(\rho\,\lambda \right)\,\sigma^{2},
     \label{ERHOS2}
\earr
\barr
	E_{\phi \nu} &=& -i\,\nu \frac{q\,\omega}{c\,\sqrt{\epsilon\,\omega^{2}}}
	\,\lambda\;
	K_{1}\!\left(\rho\, \lambda \right) \nonumber\\
	&& + \frac{i\,q\,\rho\,\omega}{2c^{3}}\,
	K_{0}\!\left(\rho\,\lambda \right)\,\sigma \nonumber\\
	&& +i\nu  \frac{q\,\omega^{3}}{8c^{5}(\epsilon\,\omega^{2})^{3/2}}
	\left[
	\frac{\,(c^{4}+3c^{2}v^{2}\epsilon - v^{2}\epsilon^{2}\rho^{2}\omega^{2})}
	{v^2\lambda}\,
	K_{1}\!\left(\rho\,\lambda \right)
	- 2\,c^{2}\epsilon\,\rho\,K_{2}\!\left(\rho\,\lambda \right)
	\right]\,\sigma^{2},
    \label{EPHIS2}
\earr
\barr
	E_{z \nu} &=& -\,\frac{i q(c^{2}-v^{2}\epsilon)\,\omega}{c^{2}v^{2}\epsilon}\,
	K_{0}\!\left(\rho\,\lambda\right) \nonumber\\
	&& + i\nu \frac{q\,\omega}{2c^{2}\sqrt{\epsilon\,\omega^{2}}}
	\left[
	-\frac{\,(c^{2}+v^{2}\epsilon)}{\epsilon\,v^{2}}\,
	K_{0}\!\left(\rho\,\lambda \right)
	+ \rho\,\lambda \;
	K_{1}\!\left(\rho\,\lambda \right)
	\right]\,\sigma \nonumber\\
	&& + \frac{iq\,\rho\,\omega}{8c^{4}}
	\left[
	-\,\rho\,K_{0}\!\left(\rho\,\lambda\right)
	+ \frac{2}{\lambda}\,
	K_{1}\!\left(\rho\,\lambda \right)
	\right]\,\sigma^{2}.
    \label{EZS2}
\end{eqnarray}

\subsection{The non-relativistic approximation}
\label{NRAPP}
 Looking back, we can understand  from the outset why the iterative method employed in Refs. \cite{Altschul:2014bba,Schober:2015rya, Altschul:2017xzx}  produces a null result for radiation, regardless of the fact that the polarization terms were  missed. In fact, after expanding the theory  in powers of $\sigma$ in the space-frequency domain, we are left with $k=\frac{\omega}{v}$  as the only parameter with dimensions of one over length,  that inevitable appears as the combination $k \rho$  in the argument of the Bessel functions. This is indeed verified  
by keeping the dominant $c^2$ terms in the coefficients of the Bessel functions and setting $\lambda= \frac{\omega}{v} =k$ in the  expansion indicated in the previous section \ref{Ap-Exp1}. This yields
\barr
    E^{(\nu)}_\rho &=& \frac{ q }{\epsilon v}\, k K_1(k\rho)+ \nu\, \frac{q}{2c^2v^2\sqrt{\epsilon^3}}\sigma\left( {c^2}K_1(k\rho)-v^2\epsilon \rho k K_2(k\rho) \right)+\frac{q \rho^2 v }{8c^4}\sigma^2 k K_1(k\rho),\label{ERHOEXP}\\
    E_\phi^{(\nu)} &=&-\nu \frac{iq}{c\sqrt{\epsilon}} k K_1(k\rho)+\frac{i q \rho v}{2c^3} \sigma k K_0(k\rho) +\nu \,  \frac{iq}{8c^3\sqrt{\epsilon^3}}\sigma^2\left(\frac{c^2}{v^2}\frac{1}{k}K_1(k\rho)-2\epsilon\rho K_2(k\rho) \right),
    \label{EPHIEXP}\\
E_z^{(\nu)} &=&-\frac{iq}{\epsilon v} k K_0(k\rho)-\nu \, \frac{iq}{2c^2\sqrt{\epsilon}}\sigma\left(\frac{c^2}{v^2\epsilon}K_0(k\rho) - \rho k K_1(k\rho) \right)-\frac{i q \rho v }{8c^4}\sigma^2\left(\rho k K_0(k\rho)-2 K_1(k\rho) \right),  \label{EZEXP} \\
B^{(\nu)}_\rho &=&\nu\, \frac{iq}{v\sqrt{\epsilon}}{k}K_1(k\rho)-\frac{i q \rho }{2c^2}\sigma  k K_0(k\rho)-\nu \,  \frac{iq}{8c^2v^3\sqrt{\epsilon^3}}\sigma^2\left( c^2 \frac{1}{k}K_1(k\rho)-2v^2\epsilon \rho K_2(k\rho)\right),
 \label{BRHOEXP}\\
B^{(\nu)}_\phi &=& \frac{kq}{c}K_1(k\rho)-\nu\, \frac{q v\sqrt{\epsilon}}{2c^3}\sigma\left( \frac{c^2}{v^2\epsilon}K_1(k\rho)-\rho  k K_2(k\rho) \right)+\frac{q \rho}{2c^3}\sigma^2 K_0(k\rho)
\label{BPHIEXP} \\
    B^{(\nu)}_z &=&\nu\, \frac{q}{v\epsilon} k K_0(k\rho)+\frac{q}{2c^2}\sigma\left(2 K_0(k\rho)-\rho k K_1(k\rho) \right)-\nu\, \frac{q}{8c^2 v \sqrt{\epsilon^3}}\sigma^2\left(
    \frac{c^2}{v^2} \frac{1}{k}K_0(k\rho)-2\rho  \epsilon K_1(k\rho) \right) \label{BZEXP},  
\earr

These expressions are consistent with the results obtained in Table \ref{tab:campos}. Noting that $k$ is real and recalling the asymptotic form of the Bessel functions for  real arguments
\beq
\lim_{\rho \to \infty} K_n(k \rho)=\sqrt{\frac{\pi}{2k\rho}} e^{-k \rho}
\label{BESSELINFTY}
\eeq
we realize that all fields in the above expansion decay exponentially to zero in the far-away region yielding zero radiation, even though the polarization-dependent terms are  included .

\section{The Fourier transform of 
the Bessel functions}
\label{APPENDIXF}

The following integral representation of the Bessel functions 
\begin{eqnarray}
&&\int_{-\infty }^{+\infty }du\frac{e^{-iku}}{(\rho ^{2}+u^{2})^{1/2}}%
=2K_{0}(k\rho ), \label{FT1APP}\\
&&\int_{-\infty }^{+\infty }du\frac{u\,e^{-iku}}{(\rho ^{2}+u^{2})^{1/2}}%
=-2i\rho K_{1}(k\rho ),  \label{FT2APP}\\
&&\int_{-\infty }^{+\infty }du\frac{e^{-iku}}{(\rho ^{2}+u^{2})^{3/2}}=2%
\frac{k}{\rho }K_{1}(k\rho ), \label{FT3APP} \\
&&\int_{-\infty }^{+\infty }du\frac{u\,e^{-iku}}{(\rho ^{2}+u^{2})^{3/2}}%
=-2ikK_{0}(k\rho ),  \label{FT4APP} \\
&&\int_{-\infty }^{+\infty }du\frac{u^{2}\,e^{-iku}}{(\rho ^{2}+u^{2})^{3/2}}%
=2K_{0}(k\rho )-2(k\rho )K_{1}(k\rho ), \label{FT5APP} \\
&&\int_{-\infty }^{+\infty }du\frac{u^{3}\,e^{-iku}}{(\rho ^{2}+u^{2})^{3/2}}%
=2i\rho \Big(-K_{1}(k\rho )+(k\rho )K_{0}(k\rho ))\Big),
 \label{FT6APP}
\end{eqnarray}
together with
the recurrence relation%
\begin{equation}
\Psi _{(n-1)}(z)-\Psi _{(n+1)}(z)=\frac{2n}{z}\Psi _{(n)}(z),\;\;\;\Psi
_{(n)}(z)=e^{i\pi n}K_{n},
 \label{RECRELAPP}
\end{equation}
allows a direct calculation of  the Fourier transforms required in the
manuscript, according to the expression 
\begin{equation}
F(\rho ,t)=v\,\int_{-\infty }^{+\infty }\frac{dk}{2\pi }\,\,\,e^{ik {Z}%
}\,F(\rho ,k).
 \label{FTGENAPP}
\end{equation}
Some results for the Fourier transform according to Eq. (\ref{FTCOMPONENT}) are 
\begin{eqnarray}
 K_{0}(k \rho )\quad &\rightarrow &\quad \frac{v}{2}\frac{1}{R}, \qquad 
kK_{0}(k\rho ) \quad \rightarrow \quad \frac{iv}{2}\frac{{Z}}{R^{3}}, \qquad 
K_{1}(k\rho )\quad \rightarrow \quad \frac{iv}{2}\frac{{Z}}{\rho R} \nonumber \\
kK_{1}(k\rho )\quad &\rightarrow &\quad \frac{v}{2}\frac{\rho }{R^{3}},\qquad 
kK_{2}(k\rho ) \quad \rightarrow \quad iv\left( \frac{1}{2R^{3}}-\frac{1}{\rho ^{2}R}%
\right) {Z}.
\label{BESSELFT}
\end{eqnarray}

\section{A brief on the electrodynamics of bi-refringent chiral media}
\label{BIREFMEDIA}

 As noted in Eq.(\ref{CONSTRELTHETA}), Maxwell equations for isotropic chiral matter (\ref{MAXW1}) and (\ref{MAXW2}) can be derived from the constitutive relations
\beq
\mbb{D}= \epsilon \mbb{E} + \theta(x) \mbb{B}, \qquad
\mbb{H}= \epsilon \mbb{B} - \theta(x) \mbb{E},  
\label{CONSTRELAPP}
\eeq
with  $\theta(x)=\sigma t$. At first sight they are very similar to those of a 
 bi-isotropic chiral  media, also known as Pasteur-type medium, which has been extensively studied by the  electromagnetic community \cite{lakhtakia1989time, lindell1994electromagnetic}. In this Appendix we aim to highlight the similarities and differences between both electrodynamics.
 
 To begin with, let us mention that chiral media are modeled by calculating the electromagnetic response of an assembly of macroscopic chiral objects (those  that cannot be brought into congruence with their mirror image by tranlation and rotation) randomly embedded in a dielectric \cite{jaggard1979electromagnetic}. The resulting constitutive relations  in frequency space, valid only  for harmonic time-dependent fields are 
\begin{eqnarray}
\tilde{\mbb{D }}&=&\epsilon \tilde{\mbb{ E}}+i\xi_c \tilde{\mbb{H}},\qquad \tilde{\mbb{B}}={\mu }\tilde{\mbb{H}}
-i\xi_c \tilde{\mbb{E}}, 
\label{CONSTRELBIREF}
\end{eqnarray}
in the by now standard notation of Refs. 
\cite{lindell1994electromagnetic,sihvola1991bi}.
Taking $\mathbf{E}$ and $\mathbf{B}$ as the fundamental fields we can reexpress them as
\begin{equation}
\tilde{\mbb{D}}=\Big(\epsilon -\xi_c^2 \Big) \tilde{\mbb{E}}+i\xi_c 
\tilde{\mbb{B}},\qquad 
\tilde{\mbb{H}}=\tilde{\mbb{B}}+ i\xi_c \tilde{\mbb{E}},
\label{chiralconstitutiverel}
\end{equation}
where $\xi_\text{c}$ is a real parameter typically dependent on the frequency.
This is distinct from chiral matter described in Eq.~(\ref{CONSTRELAPP}), which involves a constant frequency independent parameter $\sigma$.

Our use of the term “isotropic chiral matter” refers not to the bi-isotropic chiral medium of classical electrodynamics, but to   matter which electromagnetic response is  described by the CFJ model  according to Eqs. 
(\ref{MAXW1}) and (\ref{MAXW2}). CFJ electrodynamics  was  introduced also  in the context of the Standard Model Extension (SME) \cite{PhysRevD.55.6760,PhysRevD.58.116002} and later studied in effective models of Weyl semimetals and other systems exhibiting the chiral anomaly.
In high-energy physics, the term chiral  matter does not refers to media exhibiting optical chirality, but
typically denotes matter which displays  a chiral anomaly. As briefly sketched in section 
\ref{MAXEQSSUB}, CFJ electrodynamics emerges after integrating the fermions in a Dirac-like action, coupled to the electromagnetic field, that models the  microscopic description  of electrons in the material, and  it is valid  for arbitrary spacetime dependent  fields. Furthermore, the parameter $\sigma$ is obtained from the lattice characteristics of the medium.  This leads to fundamentally different field dynamics and dispersion relations, which we have explored analytically in the space-frequency domain.    

Isotropic chiral matter behaves similarly to an optical chiral medium only under specific plane-wave conditions, and  also exhibits additional nonreciprocal effects and admits unstable modes, leading to physical consequences fundamentally different from those of standard bi-isotropic chiral media.  

In the following we present a relation between such electrodynamics.  It is only possible to perform a bridge between both theories using the constitutive relations (\ref{CONSTRELAPP}),
 if we go to the frequency domain and consider only monochromatic plane-wave excitations, together with the condition $\xi_c < < \epsilon$. We start from Ampere's law  (\ref{MAXW2}), which takes the form 
 \beq
\nabla \times \tilde{\mbb{B}}(\mbf{x}, \omega) 
+ \frac{i \omega}{c}  \epsilon \tilde{\mbb{E}}(\mbf{x}, \omega) 
- \frac{1}{c}  
 \sigma \tilde{\mbb{B}}(\mbf{x}, \omega)
= \frac{4\pi}{c} \tilde{\mbb{J}}(\mbf{x}, \omega),
\label{AMPEREAPP}  
\eeq
 in frequency space. Next, inspired by the constitutive relation (\ref{chiralconstitutiverel}) for bi-isotropic media, we seek for solutions of the form
 \beq
\tilde{\mbb{D}}(\mbf{x}, \omega)=\epsilon \tilde{\mbb{E}}(\mbf{x}, \omega)+ i\xi_c(\omega) \tilde{\mbb{B}}(\mbf{x}, \omega), \qquad
\tilde{\mbb{H}}(\mbf{x}, \omega)=\tilde{\mbb{B}}(\mbf{x}, \omega)+ i\xi_c(\omega) \tilde{\mbb{E}}(\mbf{x}, \omega)
\label{CR1}, 
\eeq
with the  parameter $\xi_c(\omega)$ to be determined. Substituting in $\nabla \times \tilde{\mbb{H}}+ (i \omega/c) \tilde{\mbb{D}}= (4 \pi/c)\tilde{\mbb{J}}$, using Faraday's law  and comparing with Eq. (\ref{AMPEREAPP}) we obtain
\begin{equation}
    \xi_\text{c} = \frac{1}{2} \frac{\sigma}{\omega}.
    \label{MAP}
\end{equation}

Even though the map (\ref{MAP}) exists, the underlying physics and effects in each medium are different. For example, the dispersion relation for a plane wave excitation in  CFJ elecrodynamics  is ($\epsilon =\mu=1$)
\begin{equation}
    \omega^2 = |\mathbf{k}|^2 + \nu \,  \frac{\sigma}{c} \,  |\mathbf{k}|,
\label{dispersionrelationcompletek0}
\end{equation}
where is easy to see that one of the polarization, $\nu=-1$, is unstable as it could yield imaginary frequencies that could potentially lead to exponentially growing modes. However, we consider this situation irrelevant for our purposes since plane waves are not the normal modes of the Cherenkov radiation we are investigating.   Nonetheless, this property of plane  waves in CFJ electrodynamics is in stark difference with the stable dispersion relation in bi-isotropic chiral media
\begin{equation}
    \omega = \nu \, \xi_\text{c} k\pm k\sqrt{1+\xi_\text{c}^2}.
    \label{DISPRELPLWCHIRAL}
\end{equation}
 Another important difference is that the chiral response $\xi_\text{c}$ of bi-isotropic chiral media is intrinsically frequency-dependent  and should  vanish in the static limit, whereas $\sigma$ is a constant parameter depending on the microscopic structure of the medium as in  effective models of Weyl semimetals, for example. Besides, the relation (\ref{MAP}) implies $\xi_c \to \infty$ in the static case, so that the connection between the theories in (\ref{MAP}) is at most formal. Moreover,  the  response of isotropic chiral matter given in (\ref{CONSTRELAPP}) can be viewed as arising from a time-dependent Tellegen media \cite{Hehl:2004tk}, which  induces nonreciprocal behavior at the boundaries. For instance, while bi-isotropic chiral media do not produce polarization rotation upon reflection (see Chapter 1 of \cite{jaggard1990recent}), time-varying slabs of isotropic chiral matter can yield cross-polarized reflection due to nonreciprocal time-interface effects (see Supplemental Material of \cite{2rrr-glyn}). 

Finally, while the Lagrangian formalism provides a powerful and concise framework to capture the underlying electrodynamics, it also has known limitations: incorporating realistic features such as material losses or strong dispersion quickly becomes cumbersome. Indeed, early works on axion electrodynamics often employed the Lagrangian approach \cite{Essin:2008rq}, whereas more detailed treatments, especially in condensed matter systems, tend to adopt effective Hamiltonians or phenomenological constitutive models \cite{Ahn:2022dpk}.

Summarizing, our restricted version of CFJ electrodynamics  does not rely on constitutive relations with spatial or frequency dispersion. Instead, the modified terms appears directly in Maxwell’s equations as  an effective current derived from the action. 

\bibliography{references3.bib}

\end{document}